\documentclass{ptephy}

\def\bra{\langle}
\def\ket{\rangle}
\def\rmd{{\rm d}}

\def\Im{{\rm Im}~}

\def\Jpsi{{J\!/\!\psi}{}}
\def\X{\text{$X$(3872)}}

\def\ubar{\overline{{u}}}

\def\cbar{\overline{{c}}}
\def\qbar{\overline{{q}}}

\def\Dbar{\overline{{D}}{}}

\def\DDbarz{$D^0\Dbar^{*0}$}
\def\DDbarpm{$D^+D^{*-}$}
\def\ccbar{$c\cbar$}
\def\DDbar{$D\overline{D}{}^*$}

\newcommand{\xbld}[1]{\mbox{\boldmath $#1$}}

\def\vecr{\xbld{r}}

\def\veck{\xbld{k}}
\def\vecp{\xbld{p}}

\def\TinP{T^{(P)}}
\def\GinP{G^{(P)}}
\def\ii{{\rm i}}
\def\rmd{{\rm d}}

\def\Belle{\text{Belle}}
\def\BABAR{\text{\it{\hspace{-.1em}B\hspace{-.1em}A\hspace{-.1em}B\hspace{-.1em}A\hspace{-.1em}R\hspace{.1em}}}}

\def\Vcoul{V_\text{coul}}
\def\Vconf{V_\text{conf}}
\def\Vcmi{V_\text{CMI}}

\begin{document}

\setlength{\baselineskip}{18pt}

\title{On the origin of the narrow peak and the isospin symmetry breaking of the X(3872)}
\date{\today}

\author{\name{Sachiko Takeuchi}{1,4}, \name{Kiyotaka Shimizu}{2}, and \name{Makoto Takizawa}{3,4}}
\address{\affil{1}{Japan College of Social Work, Kiyose, Tokyo 204-8555, Japan}
\affil{2}{Department of Physics, Sophia University, Chiyoda-ku, Tokyo 102-8554, Japan}
\affil{3}{Showa Pharmaceutical University, Machida, Tokyo 194-8543, Japan}
\affil{4}{Theoretical Research Division, Nishina Center, RIKEN, Saitama 351-0198, Japan}
\email{s.takeuchi@jcsw.ac.jp}}

\begin{abstract}
The \X\ formation and decay processes in the $B$-decay
are investigated
by a $c\bar c$-two-meson hybrid model.
The two-meson state consists of the
\DDbarz, \DDbarpm,
$\Jpsi\rho$, and 
$\Jpsi\omega$ channels.
The energy-dependent decay widths of
the $\rho$ and $\omega$ mesons are introduced.
The $D$-$\Dbar^*$ interaction is
taken to be consistent with a lack of the $B\overline{B}{}^*$ bound state.
The coupling between the $D\Dbar^*$ and $\Jpsi\rho$ or the  $D\Dbar^*$ and $\Jpsi\omega$  channels 
is obtained from a quark model.
The \ccbar-\DDbar\ coupling 
is taken as a parameter to fit the \X\ mass.
The spectrum is calculated up to 4 GeV.

It is found that very narrow $\Jpsi\rho$ and  
$\Jpsi\omega$ peaks appear around the \DDbarz\ threshold.
The size of the $\Jpsi\pi^3$ peak we calculated is 1.27--2.24 times as large as 
that of 
$\Jpsi\pi^2$.
The isospin symmetry breaking in the present model comes from the mass
difference of the charged and neutral $D$ and $D^*$ mesons, which gives 
a sufficiently large isospin mixing to explain the experiments.
It is also found that values of the ratios of the transfer strengths
can give the information on the \X\ mass 
or the size of the \ccbar-\DDbar\ coupling.
\end{abstract}
\subjectindex{B69, D32}

\maketitle

\section{Introduction}

The \X\ peak has been found first by \Belle\ \cite{Choi:2003ue}
in the $\Jpsi \pi\pi K$ observation from the $B$ decay.
Its existence was confirmed by
various experiments \cite{Acosta:2003zx, Abazov:2004kp, Aubert:2004ns, Aaij:2011sn}.
The mass of \X\ is found to be 3871.69$\pm$0.17 MeV,
which is very close to or even corresponds to the \DDbarz\ threshold, 
3871.80$\pm$0.12 MeV, within the experimental errors \cite{Agashe:2014kda}.
Whether it is a resonance or a bound state 
has not been determined by the experiments yet.
The \X\ full width is less than 1.2 MeV \cite{Choi:2011fc}, 
which is very narrow for such a highly excited resonance.
The CDF group performed 
the helicity amplitude analysis of the \X\ $\rightarrow \Jpsi \pi^+\pi^-$ 
decay and concluded that the state is $J^{PC}$=$1^{++}$
or $2^{-+}$  \cite{Abulencia:2006ma}. 
Recently, LHCb experiments determined that
its quantum numbers are $J^{PC}=1^{++}$,
ruling out the possibility of $2^{-+}$ \cite{Aaij:2013zoa}.

The \X\ is observed first in the $\Jpsi \pi^n$ spectrum from the $B$ decay.
Later, the peak in the final \DDbarz\ states is also found. 
The experiments for the ratio of the partial decay width of \X\ in 
the \DDbarz\ channel to that in the $\Jpsi\pi^2$ channel, $r_\text{\DDbarz}$, however, are
 still controversial:
the Belle results give 8.92 $\pm$ 2.42 for this value \cite{Choi:2011fc,Adachi:2008sua}
while the \BABAR\ results give 19.9 $\pm$ 8.05 \cite{Aubert:2007rva,Aubert:2008gu}.
These values are taken from the charged $B$ decay experiments
because the \X\ peak in the $B^0$ decay is still vague.

Let us mention an exceptional feature of the \X, which we will discuss in this paper extensively.
It is found that the \X\ decays both to the $\Jpsi\rho$ and to the $\Jpsi\omega$ states.
According to the experiments \cite{Abe:2005ix,delAmoSanchez:2010jr}, 
the decay fraction of \X\ into $\pi^+\pi^-\Jpsi$
is comparable to that into $\pi^+\pi^-\pi^0 \Jpsi$:
\begin{eqnarray}
{Br(X\rightarrow \pi^+\pi^-\pi^0 \Jpsi)
\over 
Br(X\rightarrow \pi^+\pi^-\Jpsi)}
&=&1.0 \pm 0.4 \pm 0.3 ~~~(\Belle)
\label{eq:eq1}\\
&=&0.8\pm 0.3 ~~~(\BABAR).
\label{eq:eq2}
\end{eqnarray}
This isospin mixing is very large comparing to the usual one.
For example, the size of the breaking in the $D^+$-$D^0$ mass difference is 0.003.


Many theoretical works are being reported since the first observation of  \X.
The 1$^{++}$ channel is investigated 
by the lattice QCD 
\cite{Chiu:2006hd,Prelovsek:2013cra}.
It is reported that the $\chi_{c1}(1P)$, \X, and the \DDbar\ scattering states
are found
\cite{Prelovsek:2013cra}.
It seems, however, that the present lattice calculation still has difficulty
in dealing with a very shallow bound state or a resonance 
near the complicated thresholds with $m_u\ne m_d$.
One has to wait for future works to obtain the realistic \X\ configuration 
on the lattice QCD.
There are many phenomenological models to describe the \X.
Most of the them
%
can be classified into four types: the ones that take
the \ccbar\ charmonium picture, 
the tetraquark picture, 
the two-meson hadronic molecule picture, 
and the charmonium-two-meson hybrid picture,
which are summarized in the review articles \cite{Brambilla:2010cs,Swanson:2006st,Godfrey:2008nc}.
The existence of \ccbar(2P) at 3950 MeV was predicted by the quark model which reproduces the 
meson masses below the open charm threshold very accurately \cite{Godfrey:1985xj}.
This \ccbar$(2P)$ seems to be a robust state, 
because the quark model with the screened confinement force 
also predicts its existence
with a slightly lighter mass, 3901 MeV \cite{Li:2009zu}.
The charmonium options for the
\X\ has been carefully studied 
in refs.\ \cite{Barnes:2003vb,Barnes:2005pb,Butenschoen:2013pxa}.
In order to explain the production rate of \X\ 
in the high energy proton-(anti)proton collision experiments
 by the Tevatron 
 or the LHC,
 a configuration small in size is favored.
Also, 
the observed  rate of the
 \X\ radiative decay to $\psi(2S)\gamma$ 
is comparable to that
to $\Jpsi\gamma$ \cite{Aubert:2008ae,Bhardwaj:2011dj,Aaij:2014ala},
which strongly suggests that the \X\ has the \ccbar(2P) component
because such result 
is difficult to explain by the hadronic molecule picture \cite{Swanson:2004pp}.
On the other hand, however, it
is difficult to explain the \X\ properties by assuming a simple $1^{++}$ \ccbar\ state
 \cite{Brambilla:2010cs,Swanson:2006st,Godfrey:2008nc}.
The $\chi_{c1}(2P)$ mass predicted by the quark models is much heavier than the
observed  \X\ mass.
The spectrum of the final pions suggests that there is the $\Jpsi\rho$ component in \X.
The diquark-antidiquark or the tetraquark structure of \X\ has been studied in refs.\
\cite{Maiani:2004vq,Matheus:2006xi,Maiani:2007vr,Vijande:2007fc,Dubnicka:2010kz}.
%
The tetraquark state may be described by coupled two-meson states which are
closely bound with the attraction arising from the quark degrees of freedom.
Moreover, as seen in Table \ref{tbl:threshold-mass}, 
there are four two-meson thresholds which are very close to the \X\ mass.
It is natural to assume that \X\ has a large amount of these two-meson components.
The possibility of \X\ being the hadronic molecular structure has been widely discussed  
\cite{Swanson:2004pp,Close:2003sg,Voloshin:2003nt,Wong:2003xk,Swanson:2003tb,Tornqvist:2004qy,
Voloshin:2004mh,AlFiky:2005jd,Fleming:2007rp,Braaten:2007dw,Braaten:2007ft,Liu:2008fh,
Canham:2009zq,Stapleton:2009ey,Lee:2009hy,Gamermann:2009uq,Wang:2013kva}. 
%
%
Thus, as a model which has both of the above strong points,
the charmonium-hadronic molecule hybrid structure has been proposed for \X\
\cite{Kalashnikova:2005ui,Suzuki:2005ha,Barnes:2007xu,Zhang:2009bv,Matheus:2009vq,
Kalashnikova:2009gt,Ortega:2010qq,Danilkin:2010cc,Coito:2010if,Coito:2012vf,Ferretti:2013faa,
Chen:2013pya,Takizawa:2012hy}.
%
%

\begin{table}[tdp]
\caption{The masses and  widths of mesons  and the \X\ thresholds,
 and their energy difference (in MeV) \cite{Agashe:2014kda}.}
\begin{center}
\begin{tabular}{ccccccccccc}\hline
$m_{D^0}$ & $m_{D^{*0}}$ & $m_{D^+}$ & $m_{D^{*+}}$\\\hline
1864.84$\pm$0.07 & 2006.96$\pm$0.10 & 1869.61$\pm$0.10 & 2010.26$\pm$0.07
\\ \hline
$m_\Jpsi$ & $m_{\rho^0}$ &$\Gamma_{\rho^0}$ & $m_\omega$ & $\Gamma_\omega$ \\\hline
3096.916$\pm$0.011 & 775.26$\pm$0.25 & 147.8$\pm$0.9 & 782.65$\pm$0.12 &8.49$\pm$0.08 \\\hline
\end{tabular}
\\[3ex]
\begin{tabular}{ccccccccccc}\hline
$m_{D^0}+m_{D^{*0}}$ & $m_\Jpsi+m_\rho$ & $m_\Jpsi+m_\omega$ & $m_{D^+}+m_{D^{*+}}$& $m_\X$ &  $\Gamma_\X$ \\\hline
3871.80$\pm$0.12 & 3872.18$\pm$0.25 & 3879.57$\pm$0.12 & 3879.87$\pm$0.12& 3871.69$\pm$0.17
&$<$1.2\\ 
- & 0.38 & 7.77 & 8.07&$-$0.11\\\hline
\end{tabular}
\label{tbl:threshold-mass}
\end{center}
\end{table}%

In this work 
we also employ the charmonium-two-meson hybrid picture.
We argue that the \X\ is a hybrid state of the \ccbar\ and the two-meson molecule:
a superposition of the \DDbarz,  \DDbarpm, $\Jpsi\rho$ and $\Jpsi\omega$
molecular states
and the $c\bar c(2P)$ quarkonium.
In our previous work, where only the \DDbar\ channels are included for the two-meson components,
it has been found that this picture
 explains many of the observed properties of the \X\ in a quantitative way
 \cite{Takizawa:2012hy}:
the \X\ can be a shallow bound state 
(or an $S$-wave virtual state),
absence of the charged $X$, and absence of the $\chi_{c1}(2P)$ peak in the $J^{PC}=1^{++}$ spectrum.
%
Since the quark number is not conserved in QCD,
taking the \ccbar\ and \DDbar\ as orthogonal base 
is an approximation. 
In the low-energy QCD, however,
the light quarks get the dynamical masses 
because of the spontaneous chiral 
symmetry breaking.
Also, since adding a $q\bar q$ pair without changing the parity 
requires the change of the angular momentum of the systems,
the charm quark configuration in the \ccbar$(2P)$ state and that in the 
$1^{++}$ two-meson state
can be very different from each other.
Here we assume that the bare \ccbar$(2P)$ exists as the quark model predicts, 
which couples to the two-meson states.
%
%

In this article, we investigate the 
$J^{PC}=1^{++}$ mass spectra up to 4 GeV
observed in the $B$ decay
as well as the \X.
For this purpose, we employ the hadron model with 
the $\Jpsi\rho$ and $\Jpsi\omega$ as the two-meson states 
(denoted by the $\Jpsi V$ channels in the following)
as well as \DDbarz\ and \DDbarpm.
The source of \X\ is assumed to be
 the \ccbar$(2P)$ state, which is created from the $B$ meson by the weak decay 
as $B\rightarrow c\cbar +K$.
In order to clarify the mechanism 
how the large decay widths of the $\rho$ and $\omega$ mesons 
give rise to the very narrow peak of \X,
the energy dependent decay widths of the $\rho$ and $\omega$ mesons are introduced into the 
meson propagators.
The size of
 the isospin symmetry breaking 
seen in Eqs.\ (\ref{eq:eq1}) and (\ref{eq:eq2})
corresponds to the relative strength of $\Jpsi\rho$ and $\Jpsi\omega$ final states.
The isospin symmetry breaking in the present model originates from the
difference in the charged and neutral $D$ and $D^*$ meson masses.
We will demonstrate that two kinds of ratios of the decay modes reflect the size of 
the \ccbar-\DDbar\ coupling
and that the ratio of the \DDbarz\ to $\Jpsi \rho$ changes
largely as the binding energy of \X.
These ratios can be calculated because 
the present model includes  the relevant two-meson states  dynamically
and because the bound state and the mass spectrum are
calculated simultaneously.
A part of this work is discussed in \cite{Takeuchi:2014mma}.

Among the heavy quarkonia,
\X\ seems a very interesting object in a sense that 
the relevant two-meson thresholds exist closely below the $Q\bar Q$ state.
It has an advantage that the state is well investigated both from the experimentally 
and theoretically.
In this article, we focus our attention 
to \X\ and discuss the genuine exotic resonances such as 
$Z_b(10610)^{0,\pm}$, $Z_b(10650)^{\pm}$ or $Z_c(3900)^\pm$ elsewhere.
The study of the \X\ gives us the information of the size of the interaction 
between $D$ and $\overline{D}{}^*$,
 and therefore that between $B$ and $\overline{B}{}^*$ through the 
heavy quark symmetry.  
It will help us to understand the 
structures of these genuine exotic 
states.
The present work also gives us the information on the \ccbar-\DDbar\ coupling, which is
a clue to understand the light $q\overline{q}$ pair creation and annihilation processes.

We will discuss the method in section 2. The models and parameters are explained in 
section 2.1. The transfer strength is derived in section 2.1, while the 
derivation of the 
$\Jpsi V$-\DDbar\ transfer potential from the quark model is explained in section 2.2.
The results and the discussions are given in section 3.
The bound state we obtained is discussed in section 3.1.
The transfer strength by various parameter sets are shown in section 3.2.
The ratios of the decay modes are discussed in section 3.3.
The features of the present work are compared to the preceding works in section 3.4. 
The summary is given in section 4.

\section{Method}
\subsection{Model Space and Model Hamiltonian}

Our picture of \X\ is a superposition of 
the two-meson state and the \ccbar\ quarkonium.
The two-meson state consists of 
the \DDbarz,
\DDbarpm,
$\Jpsi\omega$, and 
$\Jpsi\rho$ channels.
The \ccbar\ quarkonium, which couples to the \DDbar\ channels,
is treated as a bound state embedded in the continuum (BSEC)
\cite{Takeuchi:2008wc,Newton}.
In the following formulae, we denote the two-meson state by $P$, and the \ccbar\ quarkonium
by $Q$.

The wave function is written as 
\begin{align}
\Psi &= \sum_{i=1}^4\;c_i\;\psi^{(P)}_i+c_0\;\psi^{(Q)}~.
\end{align}
We assume that the state is $J^{PC}=1^{++}$, but do not specify the isospin.
The wave function of each two-meson channel in the particle basis is
\begin{align}
\psi^{(P)}_1 &= {1\over \sqrt{2}} (D^0\overline{D}{}^{*0}+D^{*0}\overline{D}{}^0)
\\
\psi^{(P)}_2 &= {1\over \sqrt{2}} (D^+D^{*-}+D^{*+}D^-)
\\
\psi^{(P)}_3 &= \Jpsi\,\omega
\\
\psi^{(P)}_4 &= \Jpsi\,\rho~.
\end{align}

The model hamiltonian, $H=H_0+V$, 
 can be written as:
 \begin{eqnarray}
 H&=&
 \left(\begin{array}{cc}
 H^{(P)} & V_{PQ}\\
 V_{QP}&E^{(Q)}_0
 \end{array}
 \right)
 \end{eqnarray}
 with
 \begin{align}
 H_0&=\left(\begin{array}{cc}
 H^{(P)}_0 & 0\\
 0&E^{(Q)}_0
 \end{array}
 \right)
 &\text{and}&&
 V&=\left(\begin{array}{cc}
 V_{P} & V_{PQ}\\
 V_{QP}&0
 \end{array}
 \right)~,
 \end{align}
 where $H^{(P)}$ is the Hamiltonian for the two-meson systems,
 $V_{PQ}$ and $V_{QP}$ are the transfer potentials between the two-meson systems and the \ccbar\ quarkonium.
 $E^{(Q)}_0$ is a $c$-number and
 corresponds to the bare BSEC mass,
the mass  before the coupling to the $P$-space  is switched on.
 
 Since the concerning particles are rather heavy and 
 the concerning energy region is close to the threshold, 
 the nonrelativistic treatment is enough for this problem.
 For the free hamiltonian for the $P$-space, we have
 \begin{align}
 H^{(P)}_0&=\sum_i  \left(M_i+m_i+ {k_i^2\over 2 \mu_i}
 \right)~,
 \end{align}
 where $M_i$ and $m_i$ are the masses of the two mesons of the $i$-th channel, 
 $ \mu_i$ is their reduced mass,
 $k_i$ is their relative momentum.
 Because of the same reason, the system will not depend
 much on the details of the interaction.
 So, we employ 
 a separable potential for the interaction between the two mesons, $V_{P}$.
 The potential $V_P$ between the $i$th and $j$th channels  
 is written as 
 \begin{eqnarray}
 V_{P;ij}(\vecp,\vecp')&=&
 v_{ij}~
 f_\Lambda(p)f_\Lambda(p')
 \;
 Y_{00}(\Omega_p)Y^*_{00}(\Omega_{p'})~~~~~\text{with}~~~~~
 f_\Lambda(p)={1\over \Lambda}{\Lambda^2\over p^2+\Lambda^2}~,
 \label{eq:g}
 \end{eqnarray}
 where $v_{ij}$ is the strength of the two-meson interaction.
 We use a typical hadron size for the  value
 of the cutoff, $\Lambda$, and use the same value
 for all the channels.
 The transfer potential $V_{QP}$ between the $Q$ space and
 the $i$th channel of the $P$ space is taken to be
 \begin{eqnarray}
 V_{QP;i}(\vecp) 
 &=&
 {g_i}~\sqrt{\Lambda}\; f_\Lambda(p)\;Y^*_{00}(\Omega_p)~,
 \end{eqnarray}
 where the factor $g_{i}$ stands for the  strength of the
 transfer potential.
 We use the same function, $f_\Lambda$, in eq.\ (\ref{eq:g})
 also for the $V_{PQ}$  for the sake of simplicity.

 The channel dependence of $v_{ij}$ and $g_i$ is assumed to be
 \begin{eqnarray}
 \{v_{ij}\} &=& \left\{ \begin{array}{rrrr}
 v& 0&  ~~u& u \\
 0& v&   u& -u \\
 u&u & v'&0 \\
 u&-u   & 0&v' \\
 \end{array}\right\}
 \text{~~~and~~~}
 \{g_i\}= \left\{ \begin{array}{rrrr}
 g&  ~g&  ~0& ~0
 \end{array}\right\}
 \label{eq:vijgi}
 \end{eqnarray}
 for the \DDbarz, \DDbarpm, $\Jpsi\omega$, and $\Jpsi\rho$ channels, respectively.

 \begin{table}[tbp]
 \caption{Model parameters for the interaction.
 The interaction strength, $v$, $v'$, $u$, and $g$,
 are defined by Eq.\ (\ref{eq:vijgi}).
 The $g_0=0.0482$ is the strength of the \ccbar-\DDbar\ coupling 
 which gives the correct \X\ mass when $v=v'=0$, and $u=0.1929$. (See text.)
 For all the parameter sets, $\Lambda=500$ MeV, and $E_0^{(Q)}=3950$ MeV.
 }
 \begin{center}
 \begin{tabular}{lccccccc}
 \hline
 &$v$ & $v'$ & $u$ & $g$ & $(g/g_0)^2$  \\
 \hline
 A &$-$0.1886 &  0      & 0.1929 & 0.0390 &0.655\\ 
 B &$-$0.2829 &  0      & 0.1929 & 0.0331 &0.472\\
 C &$-$0.1886 &  0      & 0.2894 & 0.0338 &0.491\\
 QM &   \phantom{$-$}0.0233 &$-$0.2791& 0.1929 & 0.0482 &1.003\\
 \hline
 \end{tabular}
 \label{tbl:param}
 \end{center}
 
 \end{table}%
 
 As for the size of the attraction between the two mesons,
 we have tried four sets of parameters, A, B, C and QM.
 The parameters
 of each parameter set are listed in Table \ref{tbl:param}.
 The following assumptions are common to all the parameter sets:
 (1) the attraction in the \DDbarz\ channel is 
 the same as that of \DDbarpm, (2) there is no direct mixing between these \DDbarz\ and \DDbarpm\ channels,
 (3) the interaction between the $\Jpsi$ and the $\omega$ meson
 is the same as that of the $\Jpsi$ and the $\rho$ meson,
 and (4) there is no transfer potential between the two $\Jpsi V$ channels. 
 These assumptions mean that 
 the interaction between the two mesons in the $I(J^{PC})=1(1^{++})$ state
 is the same as  that of $0(1^{++})$.
 The interaction strength in the $\Jpsi V$ channels, $v'$, however,
 can be different from the one for the \DDbar\ channels, $v$.
 The size of the coupling between the \DDbar\ and the $\Jpsi V$ channels, $u$,
 is derived from the quark model, 
 which we will explain later in this section.

 As for the \ccbar\ quarkonium mass, $E_0^{(Q)}$,
 we use the $\chi_{c1}(2P)$ mass obtained by the quark model \cite{Godfrey:1985xj}.
 As for the strength of the transfer potential, $\{g_i\}$,
 we assume 
 that \DDbarz\ and \DDbarpm\ couple to the \ccbar\ quarkonium directly 
 whereas the $\Jpsi V$ channels do not.
 It is because the former coupling occurs by the one-gluon exchange while the latter coupling 
 is considered to be small because of the OZI rule.
 Since the annihilation terms which cause the \ccbar-\DDbarz\ and \ccbar-\DDbarpm\ couplings 
 are considered to be the same,
 we assume these two channels have the same $g$.
 The $g$ is taken as a free parameter 
 in each parameter set
 to reproduce the \X\ peak at the observed energy.
 
 Suppose both of $v$ and $v'$ are equal to zero, the 
 coupling $g$ has to be 0.0482
 to give the correct \X\ mass,
 which we denote $g_0$ in the following.
 The rough size of the \ccbar\ quarkonium contribution 
 to the attraction to bind the \X\ can be 
 expressed by $(g/g_0)^2$.
 When $(g/g_0)^2$ is close to 1, the attraction 
 comes mainly from the \ccbar-\DDbar\ coupling, whereas the attraction 
 comes largely from the 
 two-meson interaction when
 $(g/g_0)^2$ is smaller.
 The size of $g_0$ in the present work
 is somewhat smaller than but not very different from 
 the corresponding value in the previous work, 0.05110,
 where the $\Jpsi V$ channels were not introduced yet 
 \cite{Takizawa:2012hy}.
 It seems that the effect of the $\Jpsi V$ channels on the \X\ mass is not large.
 As we will show later, its effect on the transfer spectrum in the higher energy region is
 not large, either.
 Introducing the $\Jpsi V$ channels, however, 
 changes the phenomena at the \DDbarz\ threshold drastically.
 
 For a single channel problem with the Lorentzian separable interaction,
 the binding energy, $E_B$, can be obtained analytically:
 \begin{align}
 -v\mu  &=  {(\alpha+\Lambda)^2 \over \Lambda }~~~
 \text{with}~~\alpha=\sqrt{2\mu E_B}~.
 \label{eq:sepawf}
 \end{align}
 For the $B^0 \overline{B}{}^{*0}$ system, 
 the condition to have a bound state is $v <- 0.1886$ with $\Lambda=500$ MeV.
 In the parameter set A, we assume this value,  $- 0.1886$, for 
 the strength of the interaction between the $D$ and $\overline{D}^*$ mesons.
 Namely, the $D$-$\overline{D}^*$ attraction is 
 taken as large as possible 
 on condition 
 that
 there is 
 no bound state
 in the $B^0\overline{B}{}^{*0}$ systems
 if the attraction of the same size is applied
 \cite{Takizawa:2012hy}.
 Since it requires $v<-0.5173$  for the \DDbarz\ channel 
 to have a bound state  only by the $D^0$-$\Dbar{}^{*0}$ attraction, 
 this assumption means that here we assume
 only 0.36 of the required attraction 
 comes from the \DDbarz;
the rest  is provided by the \ccbar-\DDbar\ coupling.
 We also assume that the interaction between $\Jpsi$ and $\rho$ or $\Jpsi$ and $\omega$  is taken to be zero,
 $v'=0$, for the parameter set A.
 
 In the parameter set B [C], 
 we use $v$ [$u$] 1.5 times as large as that of the parameter set A to see the 
 parameter dependence.
 We use the one from the quark model also for 
 the diagonal part, $v$ and $v'$, in the parameter set QM.

 We have introduced the width into the $\Jpsi V$ channels,
 which represents the decays to $\Jpsi \pi^n$.
 In the present model, the source of the isospin symmetry breaking 
 is the charged and neutral $D$ and $D^{*}$ 
 meson mass difference.
 The couplings and the 
 two-meson interactions mentioned above conserve the isospin symmetry.

\subsection{The Lippmann-Schwinger equation and the transfer strength}
 
 We solve the  Lippmann-Schwinger (LS) equation
 to investigate the \X.
 Let us show some of its formulae for the case with the BSEC.
 The LS equation for the $T$-matrix and the full propagator $G$ can be written as 
 \begin{eqnarray}
 T&=&V+VG_0T
 \\
 G&=&G_0+G_0VG=G_0+GVG_0
 \end{eqnarray}
 with
 \begin{align}
 G_0&=\left(E-H_0+{\rm i}\varepsilon\right)^{-1}
 ~=~{\mathcal P} \left(E-H_0\right)^{-1}-{\rm i}\pi\delta\left(E-H_0\right) 
 \\
 G&=\left(E-H+{\rm i}\varepsilon\right)^{-1}
 ~,
 \end{align}
 where ${\mathcal P}$ indicates that the principal value should be taken for the integration of this term.
 
 Suppose there is no $Q$-space, then the 
 `full' propagator solved within the $P$-space, $\GinP$, 
 can be obtained as
 \begin{align}
 \GinP &=
 \left(E-H_P+{\rm i}\varepsilon\right)^{-1}
 \\
 &= \GinP_0\left(1+V_P\GinP\right)
 = \left(1+\GinP V_P\right)\GinP_0~.
 \label{eq:GP}
 \end{align}
 
 When the coupling to the $Q$-space is introduced,
 the full propagator for that state becomes
 \begin{eqnarray}
 G_Q&=&\left(E-E_Q^{(0)}-\Sigma_Q\right)^{-1}
 ~,
 \end{eqnarray}
 where $\Sigma_Q$ is the self energy of the $Q$-space,
 \begin{eqnarray}
 \Sigma_Q&=&V_{QP}\GinP V_{PQ}~.
 \label{eq:SigQ}
 \end{eqnarray}

 Since $\Sigma_Q$ is the only term which has an imaginary part in $G_Q$, we have
 \begin{eqnarray}
 \Im G_Q&=& \Im G_Q^* \Sigma_Q G_Q\\
 &=& \Im G_Q^* V_{QP}\GinP V_{PQ} G_Q\\
 &=& \Im G_Q^* V_{QP}\GinP \GinP{}^{*-1}\GinP{}^* V_{PQ} G_Q
 ~.
 \end{eqnarray}
 Using $\Im \GinP{}^{*-1} = \Im \GinP_0{}^{*-1}$
 and Eq.\ (\ref{eq:GP}), the above equation can be rewritten as
 \begin{eqnarray}
 \Im G_Q
 &=&\Im G_Q^* V_{QP}(1+\GinP V_P) \GinP_0(1+V_P\GinP{}^*) V_{PQ} G_Q~.
 \label{eq:24}
 \end{eqnarray}
 In the actual calculation we use the following relation with the $T$-matrix within the 
 $P$-space, $\TinP$, 
 \begin{align}
 V_P\GinP&=\TinP\GinP_0 ,~~~~ 
 \GinP V_P =\GinP_0 \TinP
 \\
 \TinP &= \left(1-V_{P}\GinP_0\right)^{-1}V_{P}~.
 \end{align}
 
 \begin{figure}[tbp]
 \begin{center}
 \includegraphics[scale=0.6]{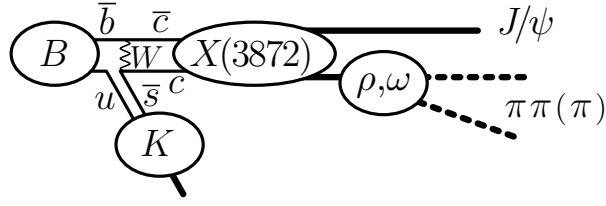}
 
 \caption{The \X\ formation process with the final $\Jpsi V$ channel
 in the $B$ meson decay.}
 \label{fig:b2x}
 \end{center}
 \end{figure}

 It is considered that the 
 \X\ state is produced via the
 \ccbar\ quarkonium (Fig.\ \ref{fig:b2x}).
 Thus the transfer strength from the \ccbar\ quarkonium to the final meson states
 corresponds to the observed mass spectrum
 with a certain factor of the weak interaction as well as the formation factor of the \ccbar\ quarkonium,
 which we do not consider in this work.
 In the following, we explain how we calculate the transfer strength.
 Notations of the kinematics are summarized in Appendix A.1.
 
 First, we derive the strength without the $\rho$ or $\omega$meson widths.
 The transfer strength from the \ccbar\ to the two-meson state, $W$,
 becomes
 \begin{eqnarray}
 {\rmd W\over \rmd E} &=& 
 -{ 1\over \pi} \Im \bra \text{\ccbar}|G_Q | \text{\ccbar}\ket~,
 \label{eq:17}
 \end{eqnarray}
 where $E$ is the energy of the system when the center of mass of \DDbarz\  is at rest.
 In order to obtain the strength to each final two-meson state separately,
 we rewrite the Eq.\ (\ref{eq:17}) as
 \begin{eqnarray}
 {\rmd W\over \rmd E} &=& 
 -{ 1\over \pi} 
 \Im 
 \bra \text{\ccbar}|
 G_Q{}^* V_{QP}(1+\GinP V_P) \GinP_0(1+V_P\GinP{}^*) V_{PQ} G_Q
 | \text{\ccbar}\ket
 \label{eq:20b}
 \\
 &=&
 -{ 1\over \pi} 
 \Im \sum_f\int \rmd^3{\veck}
 \bra f;\veck | \GinP_0(E)| f;\veck\ket 
 \left| \bra f; \veck | 
 (1+V_P\GinP{}^*) V_{PQ} G_Q
 | \text{\ccbar}\ket\right|^2 
 ~,
 \label{eq:20}
 \end{eqnarray}
 where 
 the summation is taken over all the final two-meson channels, $f$, with the
 momentum $\veck$,
 which is denoted by $| f; \veck\ket$.
 Eq.\ (\ref{eq:20}) is derived 
 by using the fact that
 the imaginary part of rhs of  Eq.\ (\ref{eq:20b}) arises 
 only from the imaginary part of 
 $\GinP_0$ in the middle of the matrix element.
 The free propagator $\GinP_0$ can be rewritten as
 \begin{eqnarray}
 \bra f;\veck | \GinP_0(E)| f;\veck\ket 
 &=&
 \frac{1}{E-(M_f+m_f+\frac{k^2}{2\mu_f})+\ii \varepsilon} 
 =
 \frac{2\mu_i}{k_f^2-k^2+\ii \varepsilon}
 \\
 &=&2\mu_f
 \frac{{\mathcal P}}{k_f^2-k^2}-\ii\pi\mu_f
 {\delta(k-k_f) \over  k_f}
 ~,
 \label{eq:g0}
 \end{eqnarray}
 where the $k_f$ is the size of the  relative momentum of the two-meson system 
 in the $f$th channel, $E = M_f+m_f+k_f^2/(2\mu_f)$. 
 Thus we have
 the strength for the open channel $f$  as
 \begin{eqnarray}
 {\rmd W(\text{\ccbar}\rightarrow f)\over \rmd E}
 &=&
 \mu_{f}k_f 
 \left| \bra f;k_f|(1+V_P\GinP{}^*)V_{PQ}G_Q |\text{\ccbar} \ket \right|^2~.
 \end{eqnarray}
  Since the out-going wave function solved in the $P$ space, $|f;k_f\ket\!\ket$, can be expressed by the plane wave  as 
 $|f;k_f\ket\!\ket = (1+\GinP V_P)|f;k_f\ket$, the above equation can be rewritten 
 as 
 $\mu_{f}k_f 
 \left| \bra\!\bra f;k_f|V_{PQ}G_Q |\text{\ccbar} \ket \right|^2
$.

 
 Next, we introduce the $\rho$ and $\omega$ decay widths.
 For this purpose, we modify the free propagator in the 
 $P$-space, $\GinP_0$, as
 \begin{eqnarray}
 \bra f;k|\GinP_0(E)|f;k\ket\rightarrow \bra f;k|\tilde{G}^{(P)}_0(E)|f;k\ket
 &=& 
 \left(E-H^{(P)}_0 +{\ii \over 2} \Gamma_V\right)^{-1}
 ~.
 \end{eqnarray}
 The width comes from the imaginary part of the self energy of the $\rho$ or $\omega$ mesons which couple to the $\pi^n$ states.
 The real part of the self energy 
 is taken care of by using the observed masses  in the denominator. 
 The width of the mesons,  $\Gamma_V$, depends on the energy of the $n\pi$ final state, $E_{n\pi}$.
 We use the energy dependent decay width which produces
 the observed  $\rho$ or $\omega$ width.  (see Appendix A.2.)
Here we neglect the $D^{*}$ meson width 
 because
they are  small compared to the $\rho$ or $\omega$ width: 
83.4 $\pm$ 1.8 keV for $D^{*+}$(2010) and less than 2.1 MeV for $D^{*0}$(2007). 
 
 By the above substitution, the full propagator, $\GinP$ and $G_Q$, the
 self energy $\Sigma_Q$ are also modified as
 \begin{align}
 \tilde{G}^{(P)} &= \left(E-H^{(P)} +{\ii \over 2} \Gamma_V\right)^{-1} = \tilde{G}^{(P)}_0(1-V_{PP} \tilde{G}^{(P)}_0)^{-1}
 \\
 \tilde\Sigma_Q &= V_{QP}\tilde{G}^{(P)}V_{PQ}
 \\
 \tilde G_Q &=\left(E-E_Q^{(0)}-\tilde{\Sigma}_Q\right)^{-1}~.
 \end{align} 
 Thus  the strength for the open channel $f$ becomes
 \begin{eqnarray}
 {\rmd W(\text{\ccbar}\rightarrow f)\over \rmd E}
 &=&
 {2\over \pi} \mu_{f} 
 \int {k^2 \rmd k~\mu_f\Gamma_f
 \over
 \big(k_f^2-k^2\big)^2 +\big(\mu_f\Gamma_f\big)^2}
 \left| \bra f;k|(1+V_P\tilde{G}^{(P)}{}^*)V_{PQ}\tilde G_Q |\text{\ccbar} \ket \right|^2~,
 \nonumber\\
 \label{eq:eq32}
 \end{eqnarray}
where $\Gamma_f$ is the width of the $f$th channel.
 The width of the $\Jpsi V$ channels 
 depend both on  $k$ and on $k_f$
 through  $E_{n\pi}$.
 The above strength is normalized  as
 \begin{align}
 \sum_f \int_0^\infty \rmd E {\rmd W(c\overline{c}\rightarrow f)\over \rmd E}
 &=1
 \label{eq:energy_sum_rule}
 \end{align}
 when energy-independent widths are employed.
For the energy-dependent widths small deviation appears:
it becomes 0.990 for the parameter set A.
 
 In order to see the mechanism to have a peak,
 we factorize the transfer strength as
 \begin{align}
 {\rmd W(\text{\ccbar}\rightarrow f)\over \rmd E}
 &=
 \Delta_f(E)\; D_{PQ}(E)\; |\bra\text{\ccbar} |\tilde G_Q(E)|\text{\ccbar} \ket|^2
 \label{eq:eq40factors}
 \\
 \Delta_f(E)&=
 {2\over \pi }
 \int {k^2 \rmd k~\mu_f\Gamma_f
 \over
 \big(k_f^2-k^2\big)^2 +\big(\mu_f\Gamma_f\big)^2}
 \;{f_\Lambda(k)^2\over f_\Lambda(k_f)^2}
 \label{eq:eq40delta}
 \\
 D_{PQ}(E)&=
 \mu_f  \; \left| \bra f;k_f|(1+V_{P}\tilde{G}^{(P)}{}^*)V_{PQ}|\text{\ccbar} \ket \right|^2
 ~,
 \label{eq:eq40dpq}
 \end{align}
 where
 $\Delta_f(E)\rightarrow k_f$ as $\Gamma_f\rightarrow 0$.
 For the energy around the \DDbarz\ threshold, 
 the integrand of the factor $\Delta_{\Jpsi\rho}(E)$
 has the maximum at around $k\sim$ 1.26 fm, 
 which corresponds to $E_{2\pi}\sim$ 670 MeV.
 There, $\Gamma_{\Jpsi \rho}$ is 0.89 times as large as that of the energy independent value,
 147.8 MeV. 
 On the other hand, 
 since the $\omega$ meson width is much smaller than that of $\rho$ meson,
 $E_{3\pi}$ which gives main contribution 
 is much closer to the peak: $E_{3\pi}\sim$ 762 MeV. 
 There, the width also reduces to 0.89 times of the energy independent value 8.49 MeV.

\subsection{The  $\Jpsi\omega$- and $\Jpsi\rho$-$D\overline{D}{}^*$ transfer potential from the quark model}

 In this subsection we explain 
 how we obtain the transfer potential between the $\Jpsi\omega$- and $\Jpsi\rho$-\DDbar\
 channels from a quark model.
 For this purpose, we employ the model of ref.\ \cite{Godfrey:1985xj},
 where they found the $q\qbar$ meson masses as well as their decays 
 are reproduced reasonably well.
 Since the results of the present work 
 do not depend much on the model detail as we will discuss later,
 we simplify the quark model in order to apply it to multiquark systems as follows:
 (1) we remove the smearing from the gluonic interaction,
 (2) we remove the momentum dependence of the strong coupling constant ($\alpha_s$) 
 but let it depend on the flavors of the interacting quarks,
 (3) we only use a single gaussian orbital configuration for each mesons,
 each of whose size parameters 
 corresponds to the matter root mean square radius ($rms$) of the original model solved without the
 spin-spin term,
 and (4) we remove the energy dependence from the spin-spin term and
 multiply the term by a parameter ($\xi$) to give a correct hyperfine splitting.

 The quark hamiltonian consists of the kinetic term, $K_q$,
 the confinement term, $\Vconf$, the color-Coulomb term, $\Vcoul$, and 
 color-magnetic term, $\Vcmi$: 
 \begin{align}
 H_q&=K_q+\Vconf+\Vcoul+\Vcmi \label{eq:49}\\
 K_q &= \sum_i K_i \text{~~~with~} K_i=\sqrt{m_i + p_i^2}\\
 \Vcoul &= \sum_{i<j} {(\lambda_i\cdot\lambda_j)\over 4}{\alpha_{s\,ij}\over r_{ij}}
 \\
 \Vconf &= \sum_{i<j} {(\lambda_i\cdot\lambda_j)\over 4}\Big(-{4\over 3}\Big)^{-1}({b \,r_{ij}+c})
 \\
 \Vcmi &= -\sum_{i<j} {(\lambda_i\cdot\lambda_j)\over 4}{(\sigma_i\cdot\sigma_j)}{2\pi\over 3}\alpha_{s\,ij}{\xi_{ij}\over m_im_j}
 \delta^3(r_{ij}) 
 ~,
 \label{eq:53}
 \end{align}
 where $m_i$ and $p_i$ are the $i$th quark mass and momentum, respectively,
 $r_{ij}$ is the relative distance between the $i$th and $j$th quarks,
 $\alpha_{s\,ij}$ is the strong coupling constant which depends on the 
 flavor of the interacting the $i$th and $j$th quarks,
 $b$ is the string tension, $c$ is the overall shift.

 The parameters are summarized in Table \ref{tbl:qm}.
 The obtained meson masses and the components are listed in Table \ref{tbl:qm2}.
 We use the values for the quark masses and the confinement parameters,
 $m_q$, $b$, and $c$, in ref.\ \cite{Godfrey:1985xj} as they are.
 Each $q\qbar$ system has three other parameters: $\alpha_s$, $\xi$, 
 and  the size parameter of the wave function, $\beta$.
 The values of $\alpha_s$ and the $\xi$ are
 taken so that the model gives the observed masses of the spin 0 and 1 mesons:
 $D$, $D^*$, $\eta_c$, $\Jpsi$, $\omega$ (the underlined entries in Table \ref{tbl:qm2}).
 We do not use the $\eta$ meson mass for the fitting
 because the mass difference between 
 $\omega$ and its spin partner $\eta$ cannot be considered as a
 simple hyperfine splitting.
 Instead, we use the $\omega$ mass obtained from the original model
 without the spin-spin term, $M_0$,  as a guide. 
 
 \begin{table}[btp]
 \caption{Quark model parameters.
 The $u$ and $c$ quark masses, $m_u$ and $m_c$,
 the string tension $b$ and the overall shift $c$ are taken from 
 ref.\ \cite{Godfrey:1985xj}. As for the $\alpha_s$, $\xi$ and $\beta$,
 see text.
 }
 \begin{center}
 \begin{tabular}{cccccc} \hline
 $m_u$(MeV) & $m_c$(MeV) & $b$(GeV$^2$) & $c$(MeV) \\\hline
 220 & 1628 & 0.18 & $-$253\\ \hline
 \end{tabular}
 \\[1ex]
 \begin{tabular}{cccccc} \hline
 & $\alpha_s$ & $\xi$ & $\beta$(fm) \\ \hline
 $u\ubar$ & 0.9737 & 0.1238 & 0.4216 \\
 $u\cbar$, $\ubar c$ or $uc$ & 0.6920 &0.2386  & 0.3684   \\
 $c\cbar$ & 0.5947 & 0.5883 & 0.2619 \\
 \hline
 \end{tabular}
 \label{tbl:qm}
 \end{center}
 \end{table}%
 \begin{table}[btp]
 \caption{Meson masses and the components of the quark potentials.
 All entries are in MeV. 
 $\bra K\ket$, $\bra \Vcoul\ket$, $\bra \Vconf\ket$, and
 $\bra \Vcmi\ket$ are the expectation values of the kinetic, the color-Coulomb,
 the confinement, and the color-magnetic terms, respectively.
 $M_0$ is the summation of the first three terms.
Since we fit the meson masses, $M_0+\bra \Vcmi\ket$
is equal to 
the observed mass, $M_{obs}$,
which (and the hyperfine splitting, {\it hfs})
 are taken from ref.\ \cite{Beringer:1900zz}$^\dag$.
 The underlined entries are used for the fitting.
 The values in the parentheses are
 the results of the original model with no spin-spin interaction.
}
 \begin{center}
 \begin{tabular}{cccccrrrrrrrr} \hline
 &&$\bra K\ket$ & $\bra \Vcoul\ket$ & $\bra \Vconf\ket$ 
 &  $M_0$ & $\bra \Vcmi\ket$ & $M_{obs}$ & {\it hfs}\\ \hline
 simplified& $D$      &2402.9&$-$557.6 & 126.2 &1971.5 &$-$106.6& \underline{1864.9} &142.1
 \\
 & $D^*$    &2402.9&$-$557.6 & 126.2 &1971.5 &    35.5& \underline{2007.0}
 \\
 & $\eta_c$ &3726.1&$-$674.1 &  16.6& 3068.6 & $-$84.9&\underline{2983.7}&113.2
 \\
 & $\Jpsi$  &3726.1&$-$674.1 &  16.6& 3068.6 &    28.3&\underline{3096.9}
 \\
 &$\omega$  &1159.2&$-$685.6 & 181.0&  \underline{654.6} &   128.0& \underline{782.7}&-
 \\\hline
 original
 &$\omega$ &(1198.7)&($-$721.0)&(176.9)& (654.6)& 
 & 771.3
 \\ \hline
 \end{tabular}\\
 \label{tbl:qm2}
\begin{flushleft}{\footnotesize $^\dag$The values cited here are different from those of the 
 2014 version [6] 
  by no more than 0.1 MeV.}\end{flushleft}
 \end{center}
 \end{table}%

 As seen in Table \ref{tbl:qm}, 
 the $\alpha_s$ becomes smaller as the interacting quark masses become
 larger.
 The size parameter of the orbital gaussian is small for the $c\cbar$
 system, and larger for the $u\ubar$ system.
 The factor for the CMI, $\xi$, varies widely from 0.1238 to 0.5883.
 These values, however, are reasonable because
 $({m\over E})^2\sim ({ m_u\over \bra K \ket_\omega/2})^2=0.144$ and
 $({m_c\over \bra K \ket_{\Jpsi}/2})^2=0.764$.
 In the following, we will explain how we derive the potential between the hadrons from the quark model.
 The obtained potential, however, is mostly determined by the observables as seen from Table
 \ref{tbl:qm2}. It does not depend much on
 the detail of the quark model, 
 except for the color-spin dependence of the quark potential
 and the meson size parameters, $\beta$'s.

 We use the following base functions
 to extract the two-meson interaction.
 \begin{align}
 \psi^{(1)}_i &= |\Dbar{}_1 D^*{}_1\ket \;\phi(\beta_{uc},r_{14})\phi(\beta_{uc},r_{23})\phi(\beta_i,r_{14-23})
 \\
 \psi^{(2)}_i &= | V_1\Jpsi{}_1\ket \; \phi(\beta_{uu},r_{13})\phi(\beta_{cc},r_{24})\phi(\beta_i,r_{13-24})
 \\
 \phi(\beta,r) &= (\pi \beta^2 )^{-3/4}\exp[-{r^2\over 2\beta^2}]~,
 \end{align}
 where $|\Dbar{}_1 D^*{}_1\ket$ [$ | V_1\Jpsi{}_1\ket$] corresponds to the spin-flavor-color part of the 
 wave function in which the $u\cbar$ [$u\ubar$] quark pair is in the color singlet state
 (see appendix B).
 As for the orbital part, we use a single gaussian function for the internal meson wave function
 and gaussian base for the relative wave function  of the two mesons.
 These base functions are not orthogonal to each other.
 Their normalization becomes
 \begin{align}
 {\mathcal N} &= \bra \psi^{(c)}_i|\psi^{(c')}_j\ket = 
 \begin{pmatrix}N &\frac{1}{3}\nu\\ \frac{1}{3}{}^t\nu&N\\ \end{pmatrix}\\
 N_{ij} &= \int 4\pi r^2 \rmd r \phi(\beta_i,r)\phi(\beta_j,r) = \left({2\beta_i\beta_j\over \beta_i^2+\beta_j^2}\right)^{3/ 2}\\
 \nu_{ij}&=\int \prod_{all~ \boldmath{r} 's} \rmd \vecr~\phi(\beta_{uc},r_{14})\phi(\beta_{uc},r_{23})\phi(\beta_i,r_{14-23})\phi(\beta_{uu},r_{13})\phi(\beta_{cc},r_{24})\phi(\beta_j,r_{13-24})~.
 \end{align}
 The $\nu$ vanishes as ${\mathcal O}(\beta_i^{-3})$ when the $\beta_i\sim \beta_j$ becomes large, whereas the
 $N$ becomes one when $\beta_i=\beta_j$.
 
 The normalization can be `diagonalized'  by 
 \begin{align}
 {\mathcal B} {\mathcal N} {}^t{\mathcal B}&=
 \begin{pmatrix}N &0\\0 & N\\ \end{pmatrix}
 \\
 {\mathcal B} &= 
 \begin{pmatrix}\displaystyle\sqrt{N} {1\over \sqrt{{\tilde N}}} &\displaystyle-\frac{1}{3}\sqrt{N} {1\over \sqrt{{\tilde N}}}\nu {1\over N}\\[3ex]0 & 1\\ \end{pmatrix}
 \\
{\tilde N} &= N -{1\over 9}\nu {1\over N}\,{}^t\nu~.
 \end{align}
 The transfer matrix ${\mathcal B}$ is not unique.
 We choose the above ${\mathcal B}$ so that 
 the base functions become
 $| V_8\Jpsi{}_8 \ket$
 and $| V_1\Jpsi{}_1 \ket$
 rather than
 $|\Dbar{}_1 D^*{}_1\ket$
 and $| V_1\Jpsi{}_1 \ket$
 at the short distance.
 By choosing this and by adding the width in 
 the $| V_1\Jpsi{}_1 \ket$ channel,
 we ensure that the $\rho$ or $\omega$ meson decay occurs 
 just from the color-singlet light quark-antiquark pair
 and that the OZI rule can be applied to the $V\Jpsi$ channel.
 Since the meson sizes are different from each other,
 the 
 $|\Dbar{}_8 D^*{}_8\ket$ and $| V_8\Jpsi{}_8 \ket$
 with orbital excitation can be introduced
 as additional base. 
 We, however, do not take them into account for the sake of simplicity.
 
 The hamiltonian for the two meson becomes
 \begin{align}
 {\mathcal H} &= \bra \psi^{(c)}_i|H_q|\psi^{(c')}_j\ket
 \end{align}
 and one can extract the effective interaction for the $\phi$ base as
 \begin{align}
 {\mathcal V}^\text{eff} &={\mathcal B} {\mathcal H} {}^t{\mathcal B} - 
 \begin{pmatrix}K^{(1)}_\text{mesons} &0\\0 & K^{(2)}_\text{mesons}\\ \end{pmatrix}
 \\
 K^{(c)}_\text{mesons}&= \int 4\pi r^2 \rmd r\phi(\beta_i,r)\left(\sqrt{m_{c}^2+p^2}+\sqrt{M_{c}^2+p^2}\right)\phi(\beta_j,r)~.
 \end{align}
 We derive the strength of the separable potential for the two-meson systems, $v_{ij}$
 in eq.\ (\ref{eq:g}),
 so that their matrix elements  have the same value.
 Namely, we determine the values of  $u$, $v$ and $v'$ in the parameter set QM in Table \ref{tbl:param} from the condition
 \begin{align}
 \int \psi_\alpha V_P \psi_\alpha = \int \psi_\alpha {\mathcal V}^\text{eff} \psi_\alpha 
 ~~~\text{with} ~~~\psi_\alpha(r)\propto {e^{-\alpha r}-e^{-\Lambda r}\over r}~,
 \end{align}
 where $\psi_\alpha$ 
 is the wave function for the separable potential in Eq.\ (\ref{eq:sepawf}) for the state of the binding energy 1 MeV.
 The obtained value for $u$ is used also for the parameter sets A and B.

\section{Results and discussions}
\subsection{The $X$(3872) bound state}

\begin{figure}[tbp]
\begin{center}
\includegraphics[scale=0.45]{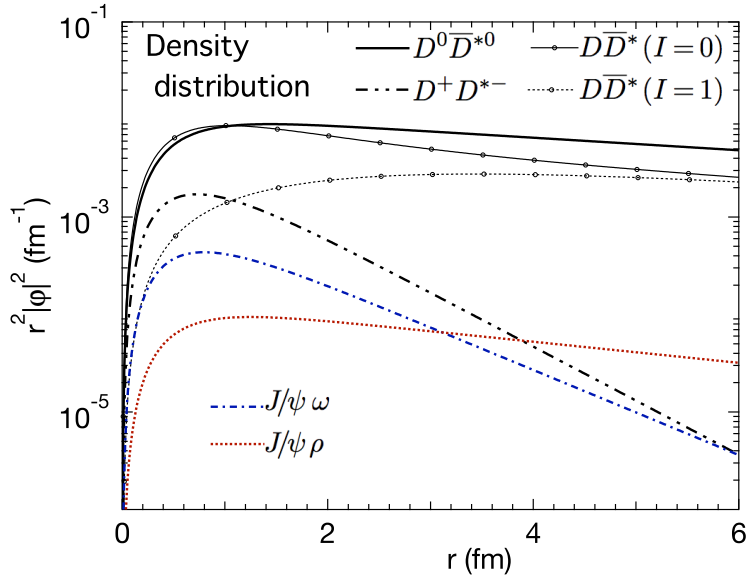}
\caption{The density distribution of the \X\ bound state
against the relative distance of the two mesons, $r$,
for the parameter set A 
calculated without introducing the meson decay width. 
}
\label{fig:dens}
\end{center}
\end{figure}
First we discuss the bound state which corresponds to \X.
In Fig.\ \ref{fig:dens}, the density distribution of the bound state 
for each two-meson channel is plotted against the relative distance of the two mesons, $r$, 
for the parameter set A calculated
without introducing the meson width.
The interaction range of the present model
is $\Lambda^{-1} \sim 0.4$ fm.
The slope of the density distribution outside this range
is essentially determined by the energy difference from each threshold.
The size of the isospin $I=1$ component of $D\overline{D}^*$ is
 small in the short range region.
It, however, becomes the same amount as that of 
the $I=0$ component at the large distance because 
the \DDbarpm\ wave function decreases much faster than that of \DDbarz.
The difference in the slopes of the  $\Jpsi\omega$ and the $\Jpsi\rho$ densities
also comes
from the energy difference of their thresholds.

The largest component of \X\ is \DDbarz\ 
because the lowest threshold is the \DDbarz\ 
and the binding energy is very small, 0.11 MeV.
Though the $\Jpsi\rho$ threshold is similarly low, 
the size of its component
in \X\ is small.
This can be explained because 
 the $\Jpsi\rho$ system has a larger kinetic energy than the \DDbar\ does,
but does not have enough 
attraction to make a state as low as \DDbar\
due to a lack of the coupling to \ccbar.
The size of the $\Jpsi\omega$ component is somewhat
larger than that of the $\Jpsi\rho$ at the short distance
because its isospin is equals to zero.

\begin{table}[tdp]
\caption{Probabilities of each component in the \X\ bound state
calculated by the model without the meson width.}
\begin{center}
\begin{tabular}{lccccccc}\hline
Model & \DDbarz\ & \DDbarpm\ & $\Jpsi\omega$ &  $\Jpsi\rho$ &\ccbar\ \\ \hline
A  & 0.913 & 0.034 & 0.010 & 0.006 & 0.036 \\ 
B  & 0.936 & 0.022 & 0.009 & 0.009 & 0.023 \\ 
C  & 0.916 & 0.020 & 0.019 & 0.022 & 0.023 \\ 
QM & 0.864 & 0.049 & 0.019 & 0.007 & 0.061 \\ 
\hline
\end{tabular}
\end{center}
\label{tbl:BEcompo}
\end{table}%

In Table \ref{tbl:BEcompo}, we show the size of each component 
in the \X\ bound state calculated by the present model
without the meson width. 
The obtained size of the \ccbar\ component  
varies from 0.023 to 0.061 according to the
parameters.
The probability of the \ccbar\ component is 0.036 for the parameter set A, which is 
somewhat smaller, but similar to that
of the $(g/g_0)^2\sim 0.5$ case 
in our previous work \cite{Takizawa:2012hy},
where we investigated the \X\ 
without the $\Jpsi V$ channels.
Including the effective \DDbar\ attraction reduces the \ccbar\ probability 
as seen in Table \ref{tbl:BEcompo} under the entries of the parameter set A-C.

It seems that the $\rho$ and $\omega$ components
of the bound state are
comparable in size.
This does not directly mean that
the $\rho$ and $\omega$ fraction from the \X\ in the $B$ decay 
are comparable.
As we will show in the next subsection,
the $\omega$ fraction in the mass spectrum is enhanced  
because the \X\ forms
from the \ccbar, the isospin-zero state,
and the $\rho$ fraction in turn is enhanced because of its large decay width.

\subsection{The transfer strength from \ccbar\ to the two-meson states}

\begin{figure}[tbp]
\begin{center}
\includegraphics[scale=0.41]{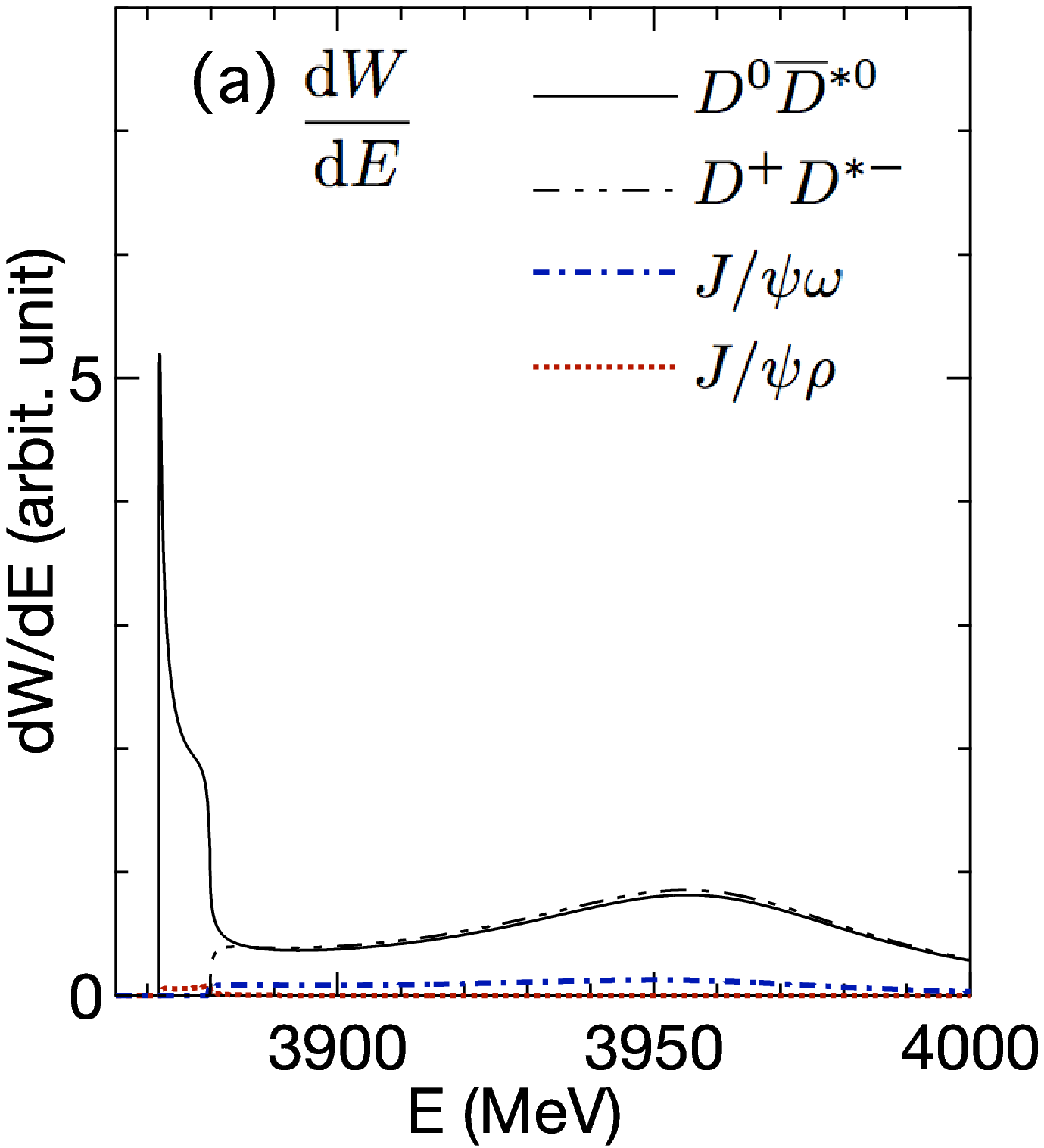}
\includegraphics[scale=0.41]{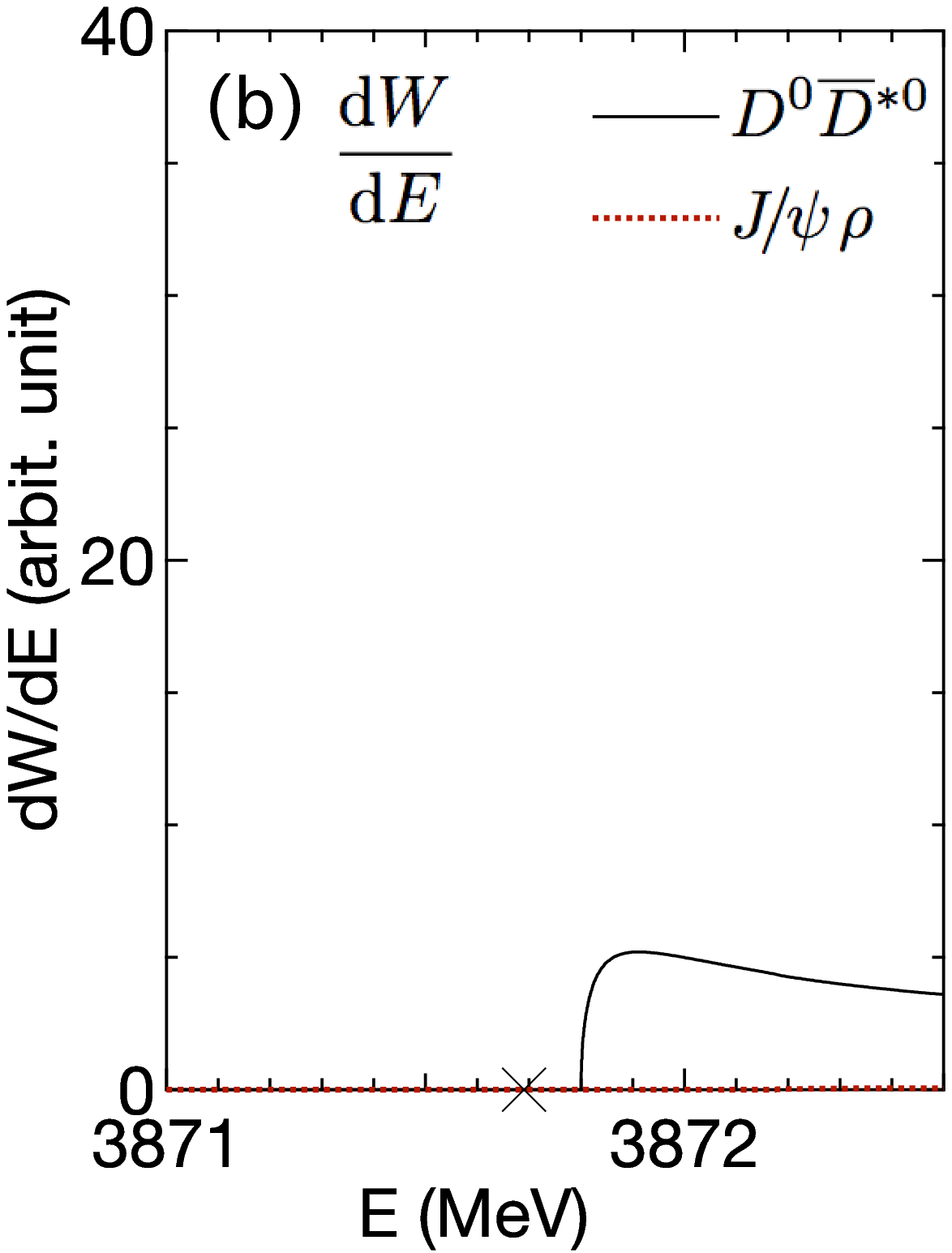}
\caption{The transfer strength from the \ccbar\ quarkonium to the
two-meson states (a) for 3870 MeV $\le E\le$ 4000 MeV and (b) around the \DDbarz\ threshold.
The spectra are for the parameter set A without the meson widths. 
The solid lines are for the transfer strength which goes from \ccbar\ into the \DDbarz\ channel.
Figs.\ (a) and (b) show the same spectra with a different scale both in the vertical and horizontal axes.
In Fig.\ (b), the $\Jpsi\rho$ spectrum is shown but almost invisible in this scale, whereas the $\Jpsi\omega$ and \DDbarpm\ spectra
are not shown because the channels are not open in this energy region.
}
\label{fig:figtra}
\end{center}
\end{figure}

Next we discuss the
transfer strength defined by eq.\ (\ref{eq:17}) from the \ccbar\ quarkonium 
to the final two-meson states, \DDbarz, \DDbarpm, $\Jpsi\rho$ and $\Jpsi\omega$.
In Fig.\ \ref{fig:figtra}, we show them
for the parameter set A without the meson width.
The lines for  \DDbarz, \DDbarpm, and $\Jpsi\rho$ 
correspond to the observed spectrum though the overall 
factor arising from the weak interaction should be multiplied.
In order to obtain the $\Jpsi\pi^3$ spectrum, the fraction 
$\tilde\Gamma_{\omega\rightarrow 3\pi}=\Gamma_{\omega\rightarrow 3\pi}/\Gamma_\omega
= 0.892$ \cite{Agashe:2014kda} should be multiplied furthermore to the $\Jpsi\omega$ spectrum.
The spectra are plotted in Fig.\ (a) for 3870 MeV $\le E\le$ 4000 MeV.
In Fig.\ (b), we plot
the same spectra around the \DDbarz\ threshold in a different scale.
There is a bound \X\ at 3871.69 MeV, which is marked 
by a $\times$ in the Fig.\ \ref{fig:figtra} (b).
The \DDbarpm\ and the $\Jpsi\omega$ spectra are not shown in Fig.\ (b) because they are still closed.
All of the four two-meson channels as well as the \ccbar\ state are included
in the calculations 
throughout the present article.

As seen from Fig.\ \ref{fig:figtra}(a), 
the transfer strength has a peak just above 
the \DDbarz\ threshold.
Such a peak appears because the bound state exists very close to the threshold.
It, however, is probably difficult to distinguish the strength of this peak from that of the bound state by
the experiments of the current resolution. 
Above the \DDbarpm\ threshold, the \DDbarz\ and \DDbarpm\ spectra are 
almost the same. 
The isospin symmetry breaking is restored there,
which can also be seen from the fact that the $\Jpsi\rho$ spectrum is almost invisible there.
The $c\cbar$ quarkonium mass is 3950 MeV when the \ccbar-\DDbar\  
coupling is switched off.
After the coupling is introduced, the pole 
moves from 3950 MeV to  $3959 - {\ii\over 2} 72$ MeV.
All the spectra are found to be rather flat at around 3950 MeV
because the imaginary part of the pole energy is large.
%

\begin{figure}[tbp]
\begin{center}
\includegraphics[scale=0.38]{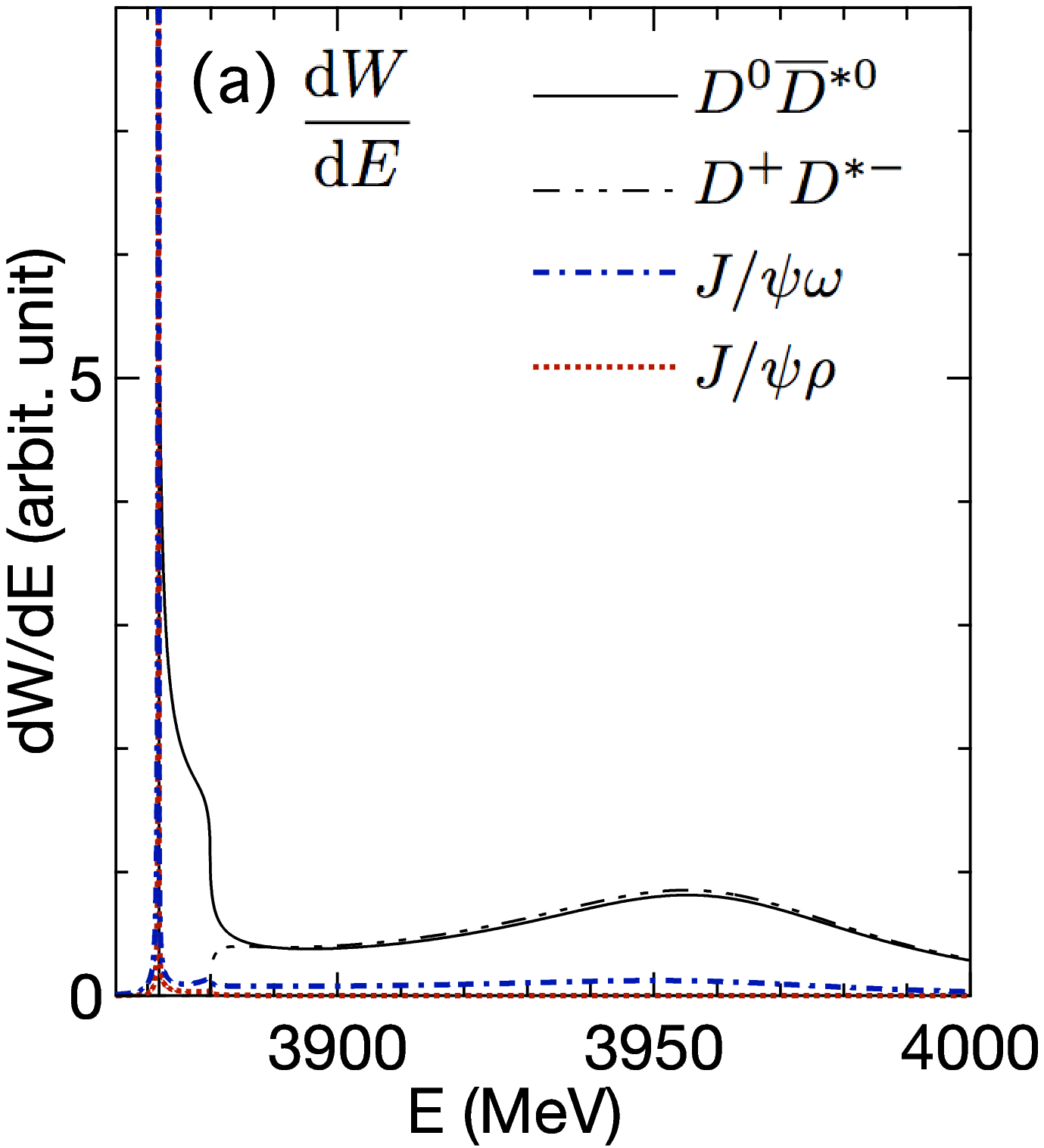}
\includegraphics[scale=0.38]{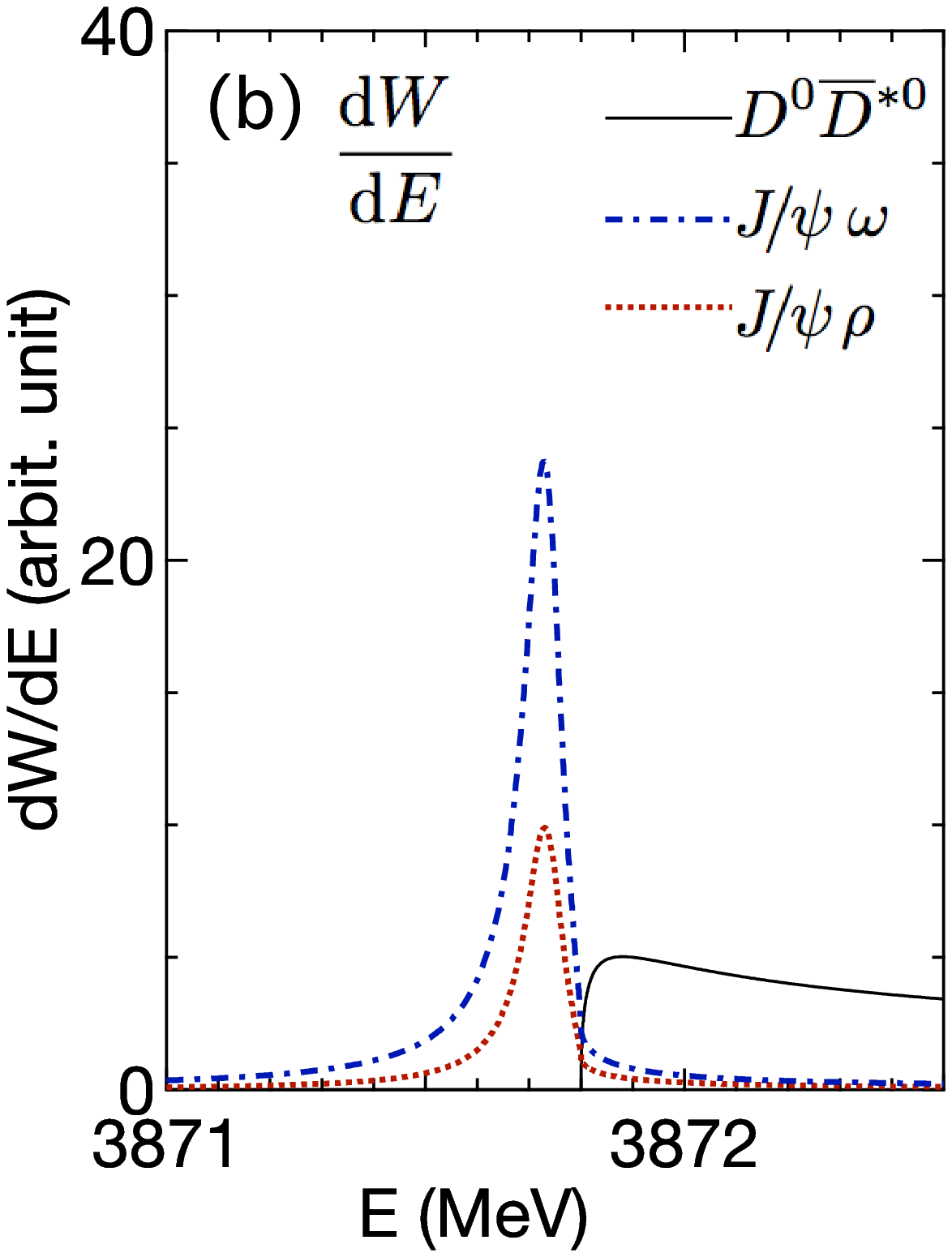}
\includegraphics[scale=0.38]{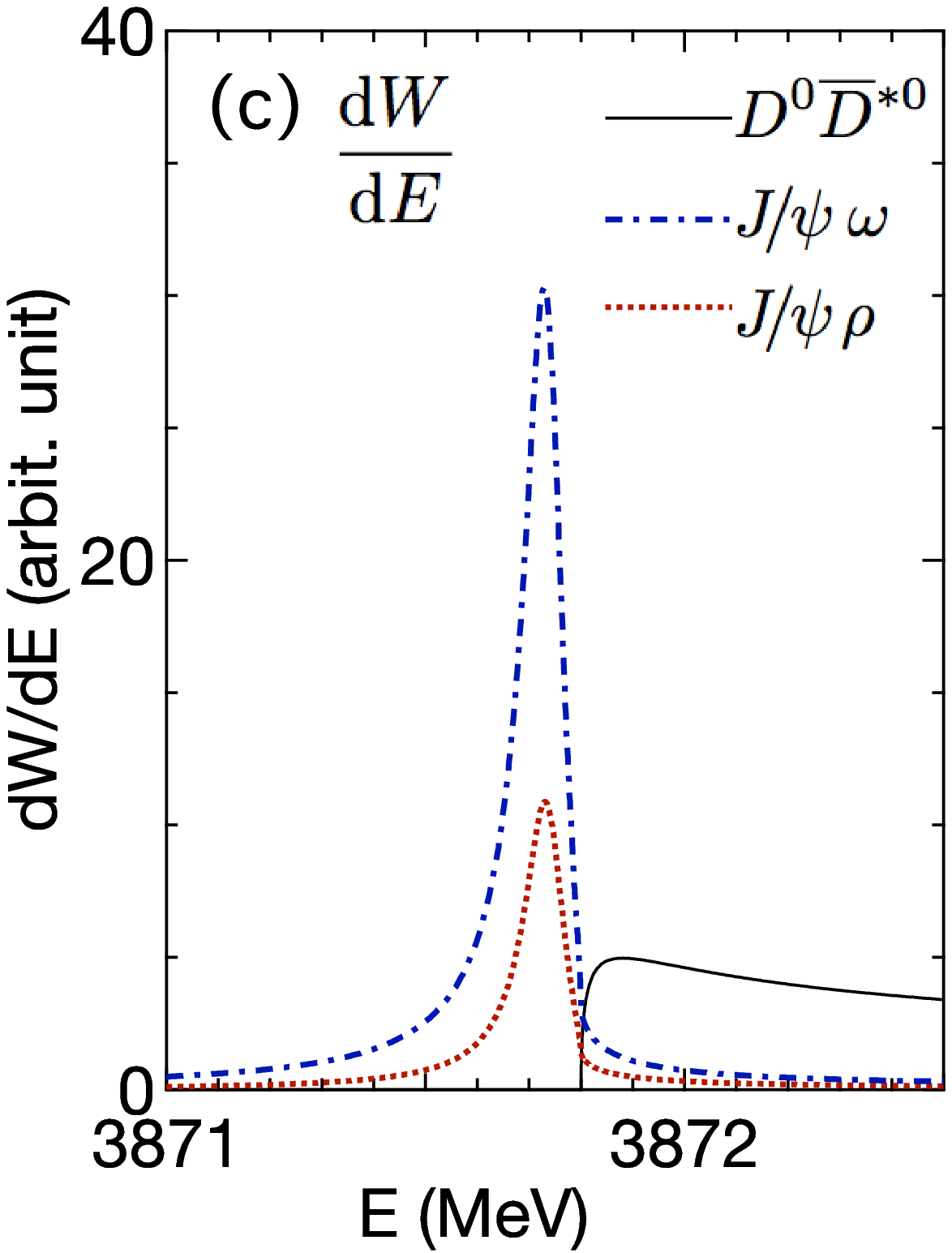}
\caption{The transfer strength from the \ccbar\ quarkonium to the
two-meson states.
 (a) for 3870 MeV $\le E\le$ 4000 MeV and (b) around the \DDbarz\ threshold
by the parameter set A with the $\rho$ and $\omega$ meson widths. 
 The Fig.\ (c) corresponds to those 
by the parameter set A with the energy-independent width.
}
\label{fig:figtraw}
\end{center}
\end{figure}

The transfer strengths calculated with the $\rho$ and $\omega$ meson width 
are shown in Fig.\ \ref{fig:figtraw}, which 
correspond again to the parameter set A.
The overall feature of the \DDbarz\ and \DDbarpm\ spectra do not change much
when the $\rho$ and $\omega$ meson width is introduced.
The \DDbarz\ peak exists naturally above the threshold. 
That means the peak energy is higher 
than that in the $\Jpsi V$ spectrum, which is consistent with the experiment: the \X\ mass from \DDbarz\ mode $= 3872.9^{+0.6}_{-0.4}\,{}^{+0.4}_{-0.5}$ MeV for Belle \cite{Adachi:2008sua}, or
$ 3875.1^{+0.7}_{-0.5}\pm 0.5$ MeV for \BABAR\ \cite{Aubert:2007rva}.
The width of the peak 
from \DDbarz\ mode 
is found to be a few MeV in our calculation,
which is also consistent with the experiments,
$\Gamma_{X\rightarrow \text{\DDbarz}} = 3.9^{+2.8}_{-1.4}\,{}^{+0.2}_{-1.1}$ MeV\cite{Adachi:2008sua},
or
$3.0^{+1.9}_{-1.4}\pm 0.9$ MeV\cite{Aubert:2007rva}.
On the other hand,
the $\Jpsi\rho$ and $\Jpsi\omega$ strength around the \DDbarz\ threshold
change drastically by introducing the width as seen in Fig.\ \ref{fig:figtraw}(b).
They make a very thin peak at the \X\ mass.
Note that the experiments give only an upper limit for the \X\ width, $<$ 1.2 MeV,
in the $\Jpsi \pi^n$ spectrum \cite{Choi:2011fc}.
The widths of the $\Jpsi V$ peaks obtained here are less than 0.2 MeV, 
which are much smaller than the experimental upper limit.
The $\Jpsi\omega$ component appears around the \DDbarz\ threshold
due to the $\omega$ decay width
 though the channel is still closed.
In the Fig.\ \ref{fig:figtraw}(c),
we show the spectrum when the 
meson widths are taken to be energy independent.
The peak reduces when the
 energy dependent widths are introduced.

To look into the isospin symmetry breaking around the \DDbarz\ threshold,
we calculate ratio of the strength integrated over the range of $m_X\pm \epsilon_X$,
where $m_X$ is the average mass of \X, 3871.69 MeV, 
$\epsilon_X$ is the upper limit value of $\Gamma_\X$, 1.2 MeV.
\begin{align}
R_\Gamma&={I_{\Jpsi\omega}(m_X- \epsilon_X,m_X+ \epsilon_X) 
\over I_{\Jpsi\rho}(m_X- \epsilon_X,m_X+ \epsilon_X)}
~{~\tilde\Gamma_{\omega\rightarrow 3\pi}\over \tilde\Gamma_{\rho\rightarrow 2\pi}}
\label{eq:eqR1G}
\\
I_f(E_1,E_2)&=\int_{E_1}^{E_2} \rmd E \;
{\rmd W(c\overline{c}\rightarrow f)\over \rmd E}~.
\end{align}
Here, the factor $\tilde\Gamma_{\omega\rightarrow 3\pi}$ 
is the fraction of $\omega\rightarrow \pi\pi\pi$, $0.892\pm 0.007$, whereas that of $\rho$, $\tilde\Gamma_{\rho\rightarrow 2\pi}$ is $\sim$ 1
\cite{Agashe:2014kda}.
We assumed the value of the ratio of these fractions to be 0.892.
This $R_\Gamma$ defined above should correspond to the experimental ratio, 
eqs.\ (\ref{eq:eq1})
and (\ref{eq:eq2}),
$1.0 \pm 0.4 \pm 0.3$\cite{Abe:2005ix} or $0.8\pm 0.3$\cite{delAmoSanchez:2010jr}.
For the parameter set A, this ratio $R_\Gamma$ is found to be 2.24,
which is somewhat larger than the experiments.
There is an estimate by employing a two-meson model,
where its value is about 2 \cite{Gamermann:2009uq}, 
whereas in the work of the one-boson exchange model,
this value is about 0.3 for a bound state with the binding energy of 0.1 MeV \cite{Li:2012cs}.
The present work, having no isospin breaking term in the interaction,
gives a closer value to the former
 case.

\begin{figure}[tbp]
\begin{center}
\includegraphics[scale=0.38]{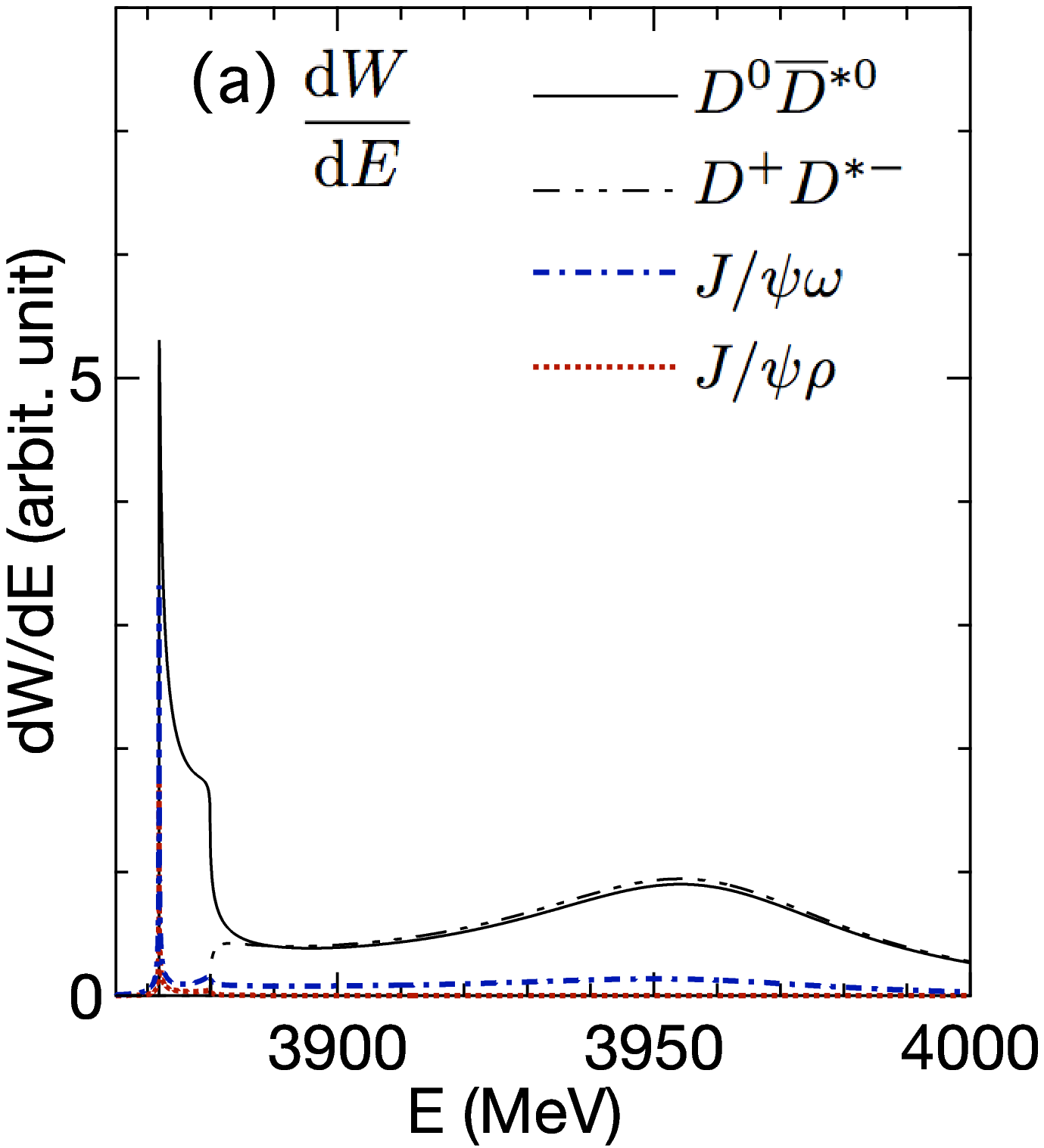}
\includegraphics[scale=0.38]{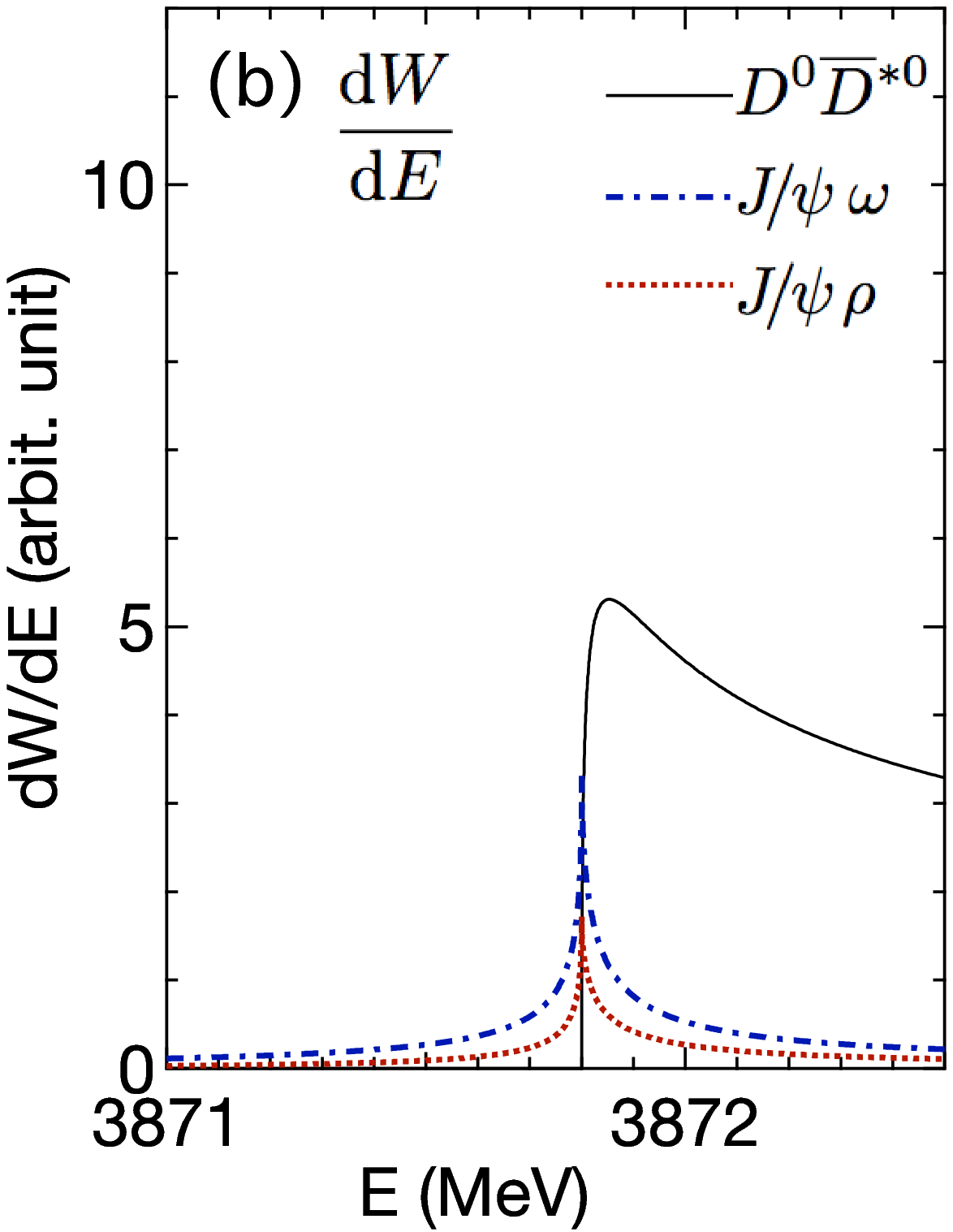}
\includegraphics[scale=0.38]{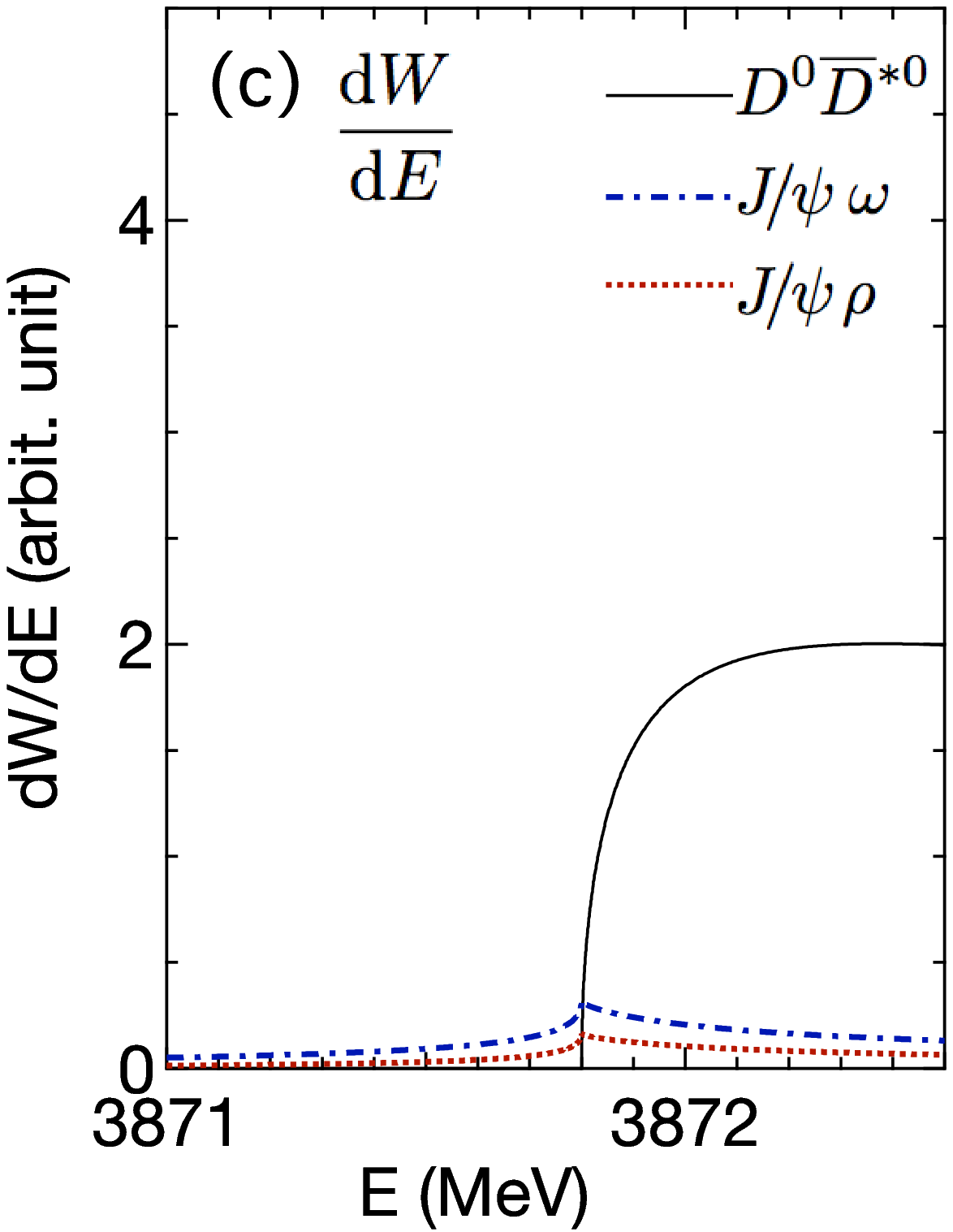}
\caption{The transfer strength from the \ccbar\ quarkonium to the
two-meson states.
Parameter set A with the $\rho$ and $\omega$ meson widths. 
The \ccbar-\DDbar\  coupling $g^2$ is weakened by $0.9g^2$ in Figs.\ (a) and (b),
by $0.8g^2$ in Fig.\ (c).
Note that the scale of the vertical axis of Fig.\ (b) or (c) is different from the
Figs.\ \ref{fig:figtra} or \ref{fig:figtraw}.
}
\label{fig:figtrag9}
\end{center}
\end{figure}

As listed in 
Table \ref{tbl:threshold-mass}, 
the 
peak energy of \X\ corresponds to the threshold energy within the error bars.
There is a possibility that the \X\  is
not a bound state but a peak at the threshold.
In order to see the situation,
we also calculate the spectrum by the parameter set A with weakened
\ccbar-\DDbar\ couplings: the one where the coupling strength $g^2$ is 0.9 times as large as that of the parameter set A 
(denoted by 0.9$g^2$ and shown in Figs.\ \ref{fig:figtrag9}(a) and (b)) 
and that of 0.8 (denoted by 0.8$g^2$ and shown in Fig.\ \ref{fig:figtrag9} (c)).
There is no bound state anymore but a virtual state in both of the cases, but a
peak is still found at the \DDbarz\ threshold for the 0.9$g^2$ case.
The strength of the $\Jpsi V$ channels, however,
becomes considerably smaller.

\begin{figure}[tb]
\begin{center}
\includegraphics[scale=0.45]{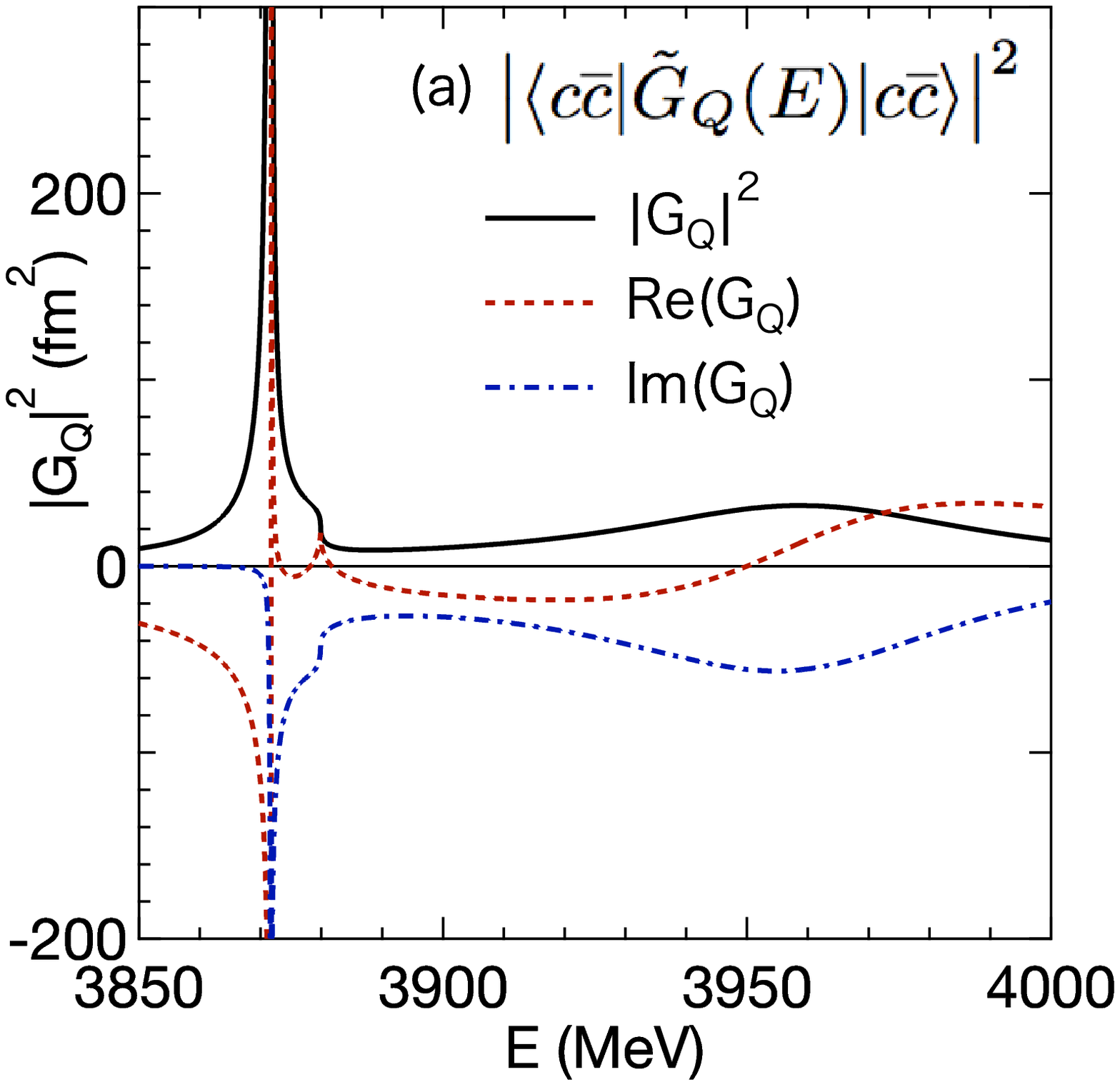}
\includegraphics[scale=0.45]{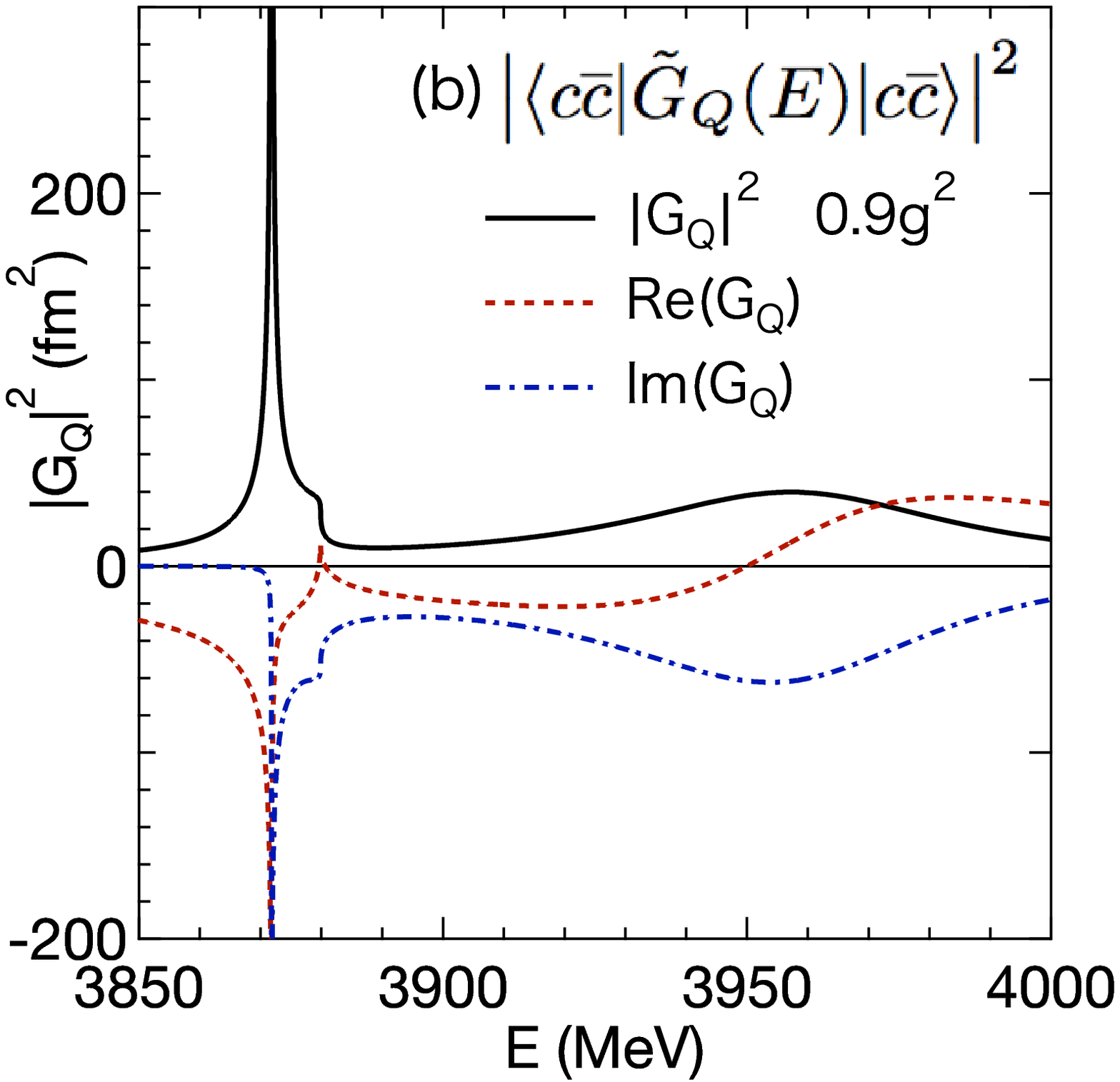}
\includegraphics[scale=0.45]{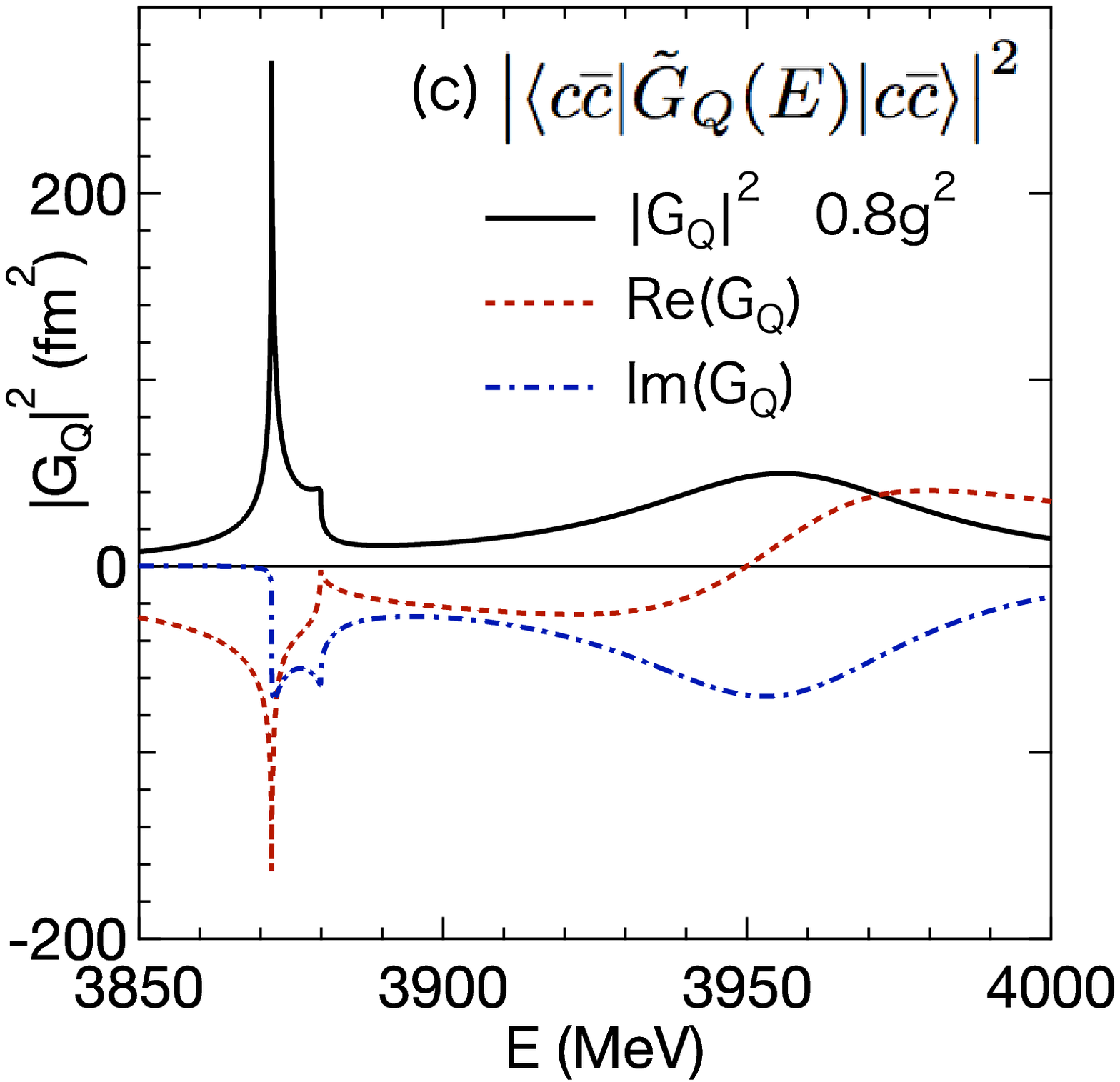}
\caption{Factors of the transfer strength from the \ccbar\ quarkonium to the
two-meson states.
 (a) $|\bra \text{\ccbar}|G_Q|\text{\ccbar}\ket|^2$
in eq.\ (\ref{eq:eq40factors}) for each channel
 around the \DDbarz\ threshold for the parameter set A.
  In Figs.\ (b) and (c), $|\bra \text{\ccbar}|G_Q|\text{\ccbar}\ket|^2$
  for the 0.9$g^2$ and 0.8$g^2$ cases are shown.
 }
\label{fig:compo}
\end{center}
\end{figure}

\begin{figure}[tb]
\begin{center}
\includegraphics[scale=0.43]{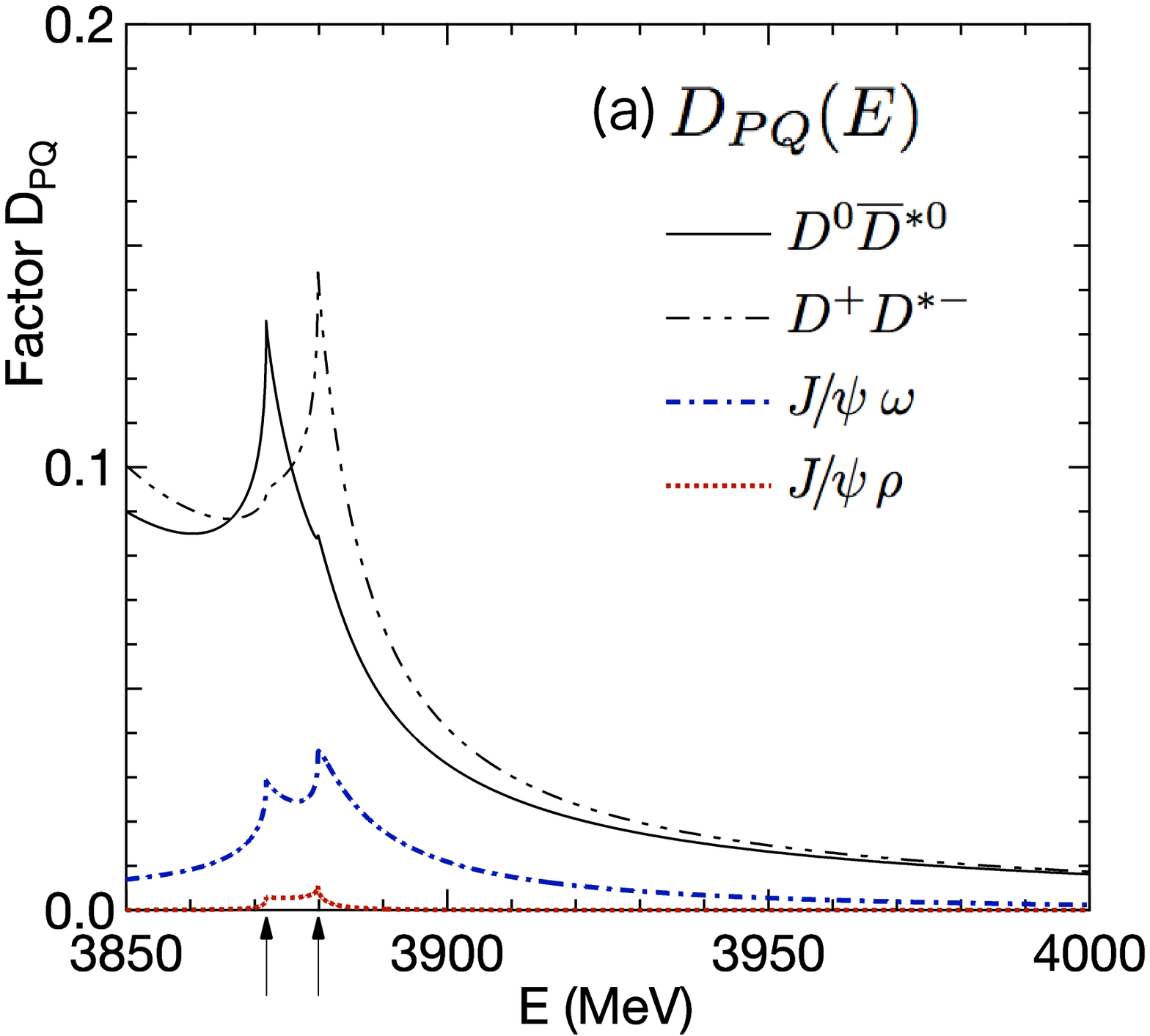}
\includegraphics[scale=0.43]{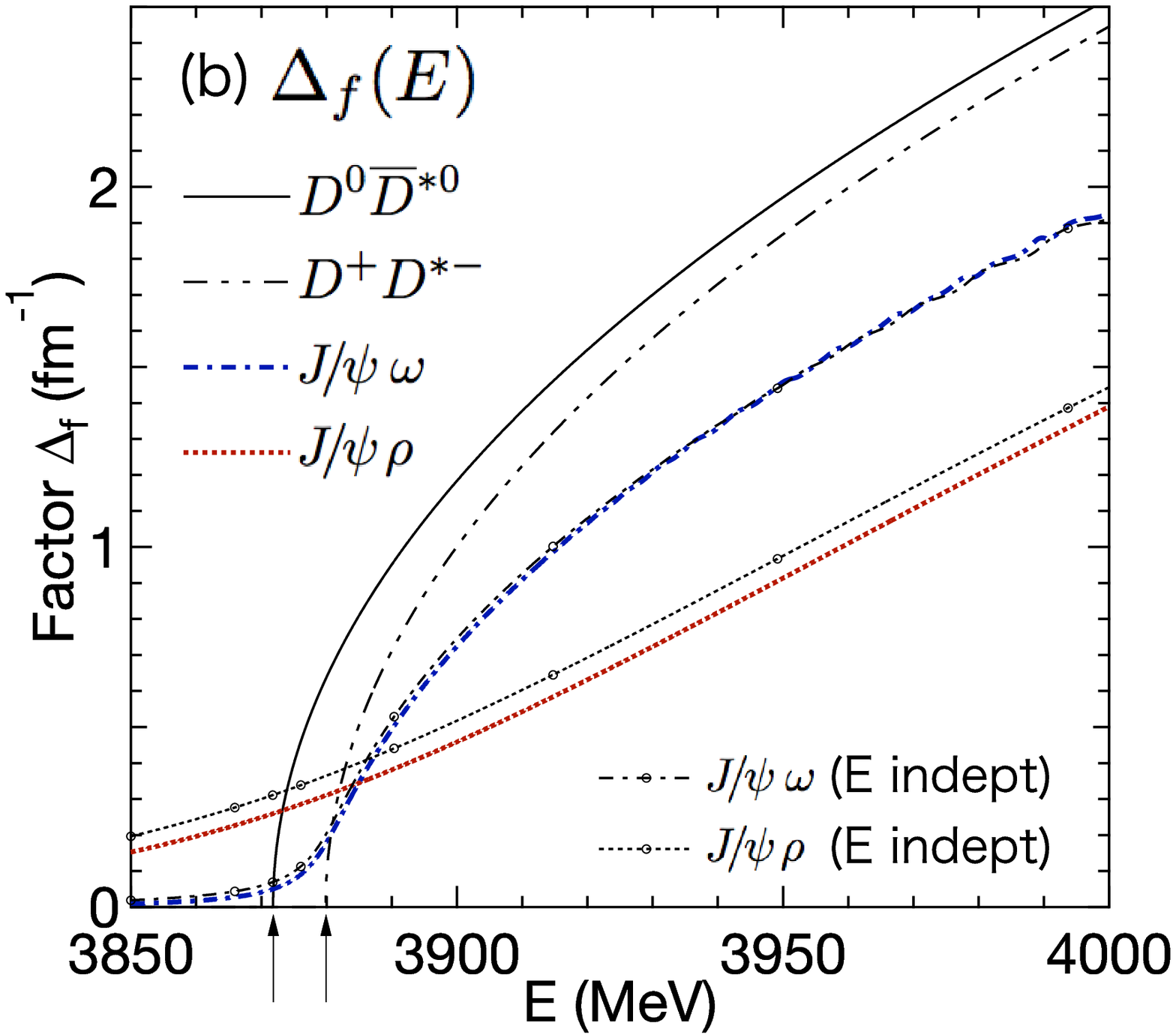}
\caption{Factors of the transfer strength from the \ccbar\ quarkonium to the
two-meson states.
 (a)  the factor $D_{PQ}$ and (b) the factor  $\Delta_f$
 in eq.\ (\ref{eq:eq40factors}) for each channel
 around the \DDbarz\ threshold for the parameter set A.
 The arrows at the horizontal axis correspond to the \DDbarz\ and the \DDbarpm\ threshold energy.
 }
\label{fig:compo2}
\end{center}
\end{figure}

In order to see the mechanism to create a peak at around the threshold
and how the peak of each channel is developed,
we plot each factor defined by eq.\ (\ref{eq:eq40factors})
in Figs.\ \ref{fig:compo} and \ref{fig:compo2}.
From the Fig.\ \ref{fig:compo}, 
one can see that the full propagator of the \ccbar\ space, $G_Q$, is
responsible to make the peak structure.
As $(g/g_0)^2$ is weakened, the bound state becomes a virtual state.
But the $G_Q$ still has a peak at 0.9$g^2$ 
as seen in Fig.\ \ref{fig:compo}(b),
which makes a thin peak in the transfer strength.
The shape of $G_Q$ is essentially determined within 
the \ccbar-\DDbar\ system.
The effect of the $\Jpsi V$ channel is rather small here.

The \ccbar\ state branches out into each two-meson state by the factor $D_{PQ}$.
As seen in Fig.\ \ref{fig:compo2}(a), 
the factor for the $\Jpsi\rho$ component is very small,
while the factors for the \DDbarz\ and \DDbarpm\ are comparable to each other.
All the factors have cusps at both of the two thresholds.

The  $\Delta_f$, which is shown in Fig.\ \ref{fig:compo2}(b), 
is an essentially kinematical factor.
Because of the large $\rho$ meson decay width, 
$\Delta_{\Jpsi\rho}$ is 5.23 times 
larger than $\Delta_{\Jpsi\omega}$
at the \X\ peak energy.
Without this $\Delta_f$ factor, the branching ratio, $R_\Gamma$ defined by eq.\ (\ref{eq:eqR1G}),
is about 11.7, due to the large $D_{PQ}$ for the $\Jpsi\omega$ channel.
Both of $\Delta_{\Jpsi\rho}$ and $\Delta_{\Jpsi\omega}$
become smaller as the energy-dependence of the decay widths are
taken into account:
$\Delta_{\Jpsi\rho}$ at $m_\X$ reduces from 0.311 to 0.259 
while $\Delta_{\Jpsi\omega}$ reduces from 0.069 to 0.050.
This reduction of $\Delta_f$ 
is the reason why the peak with the energy dependent widths
is smaller in Fig.\ \ref{fig:figtraw}.
Our enhancement factor 5.23 is smaller than the value 
given by \cite{Suzuki:2005ha}, 13.3, though
it is probably not excluded by the experimentally required value estimated by
\cite{Braaten:2005ai}, 11.5 $\pm$ 5.7.

\begin{figure}[tb]
\begin{center}
\includegraphics[scale=0.38]{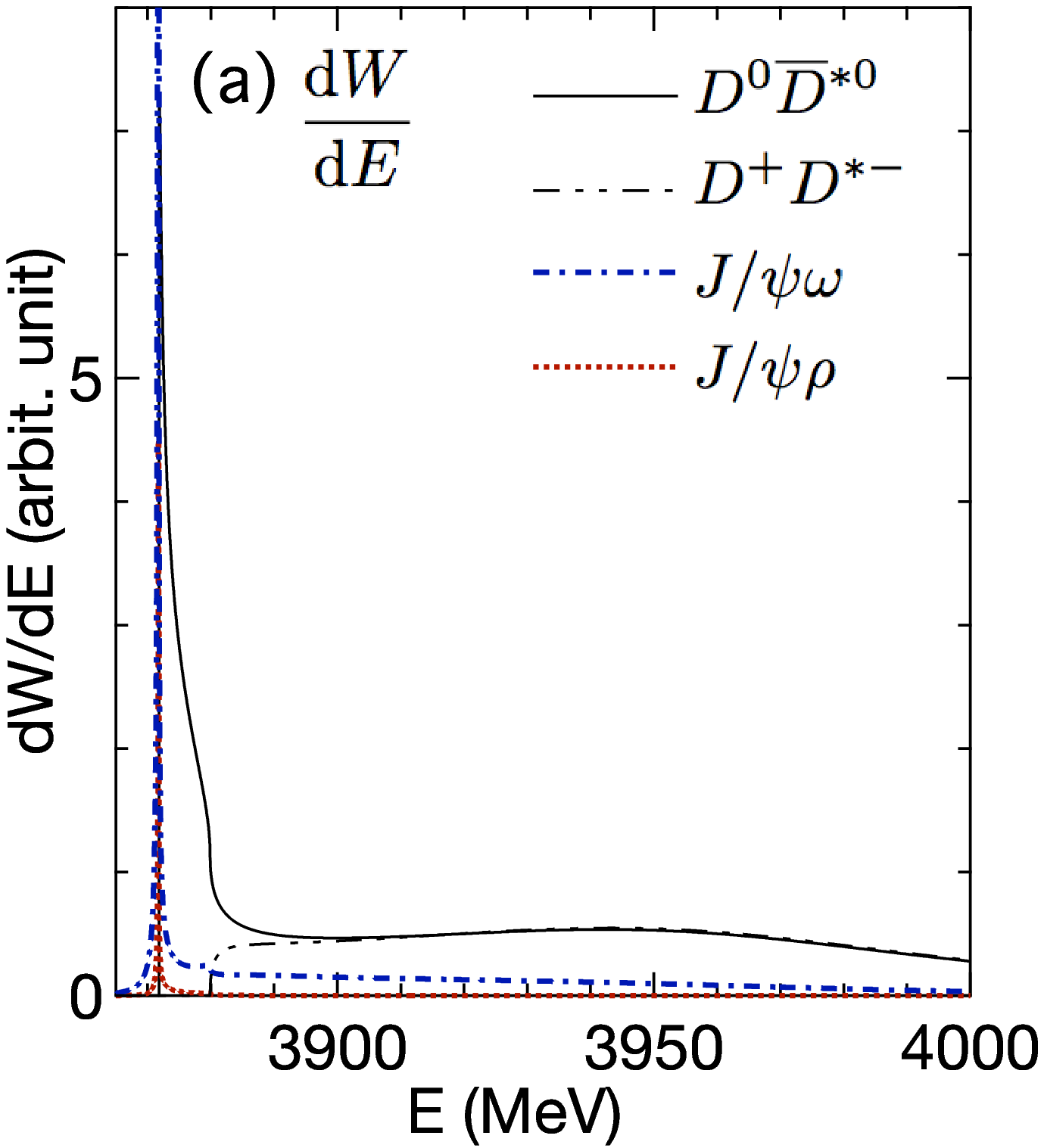}
\includegraphics[scale=0.38]{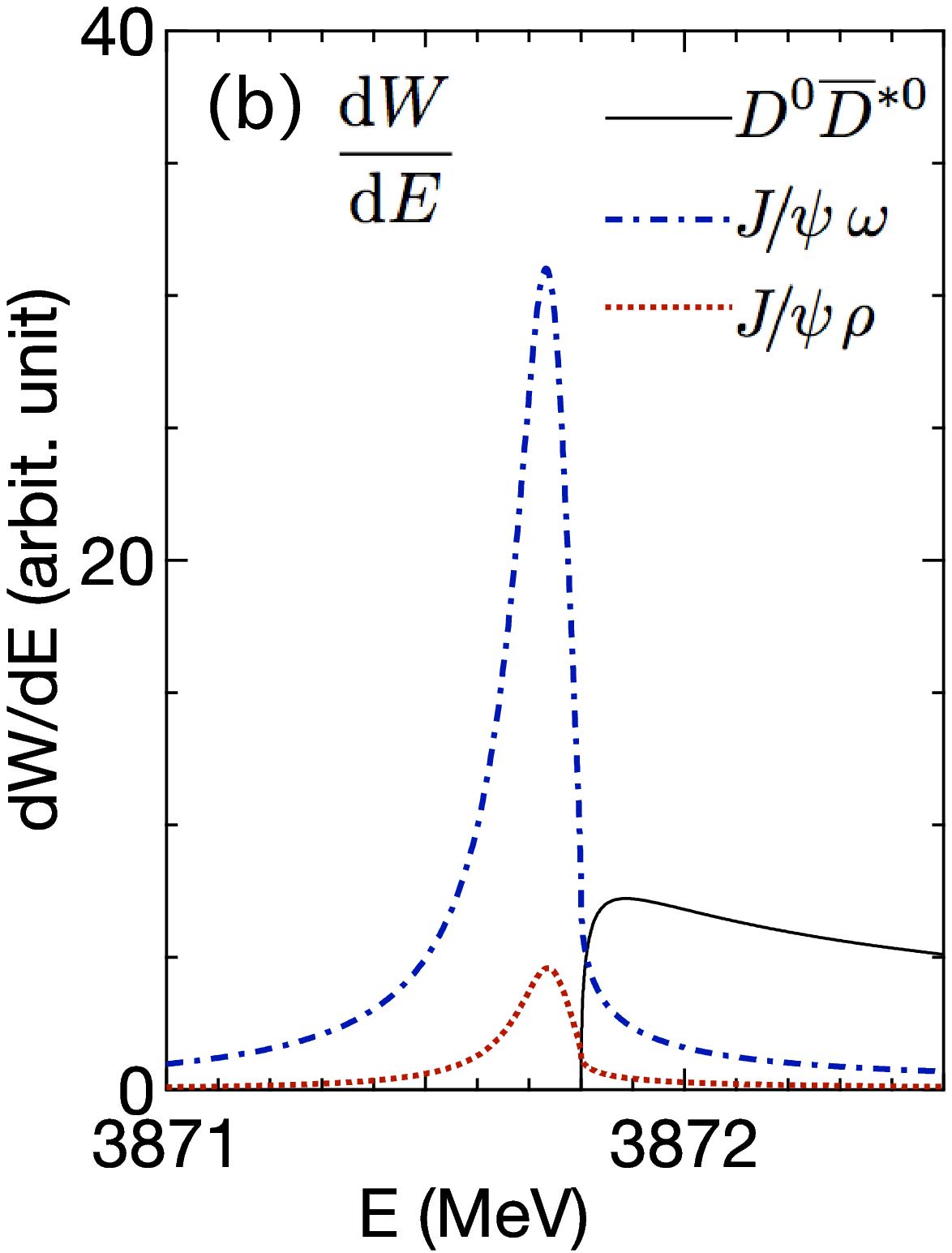}
\includegraphics[scale=0.38]{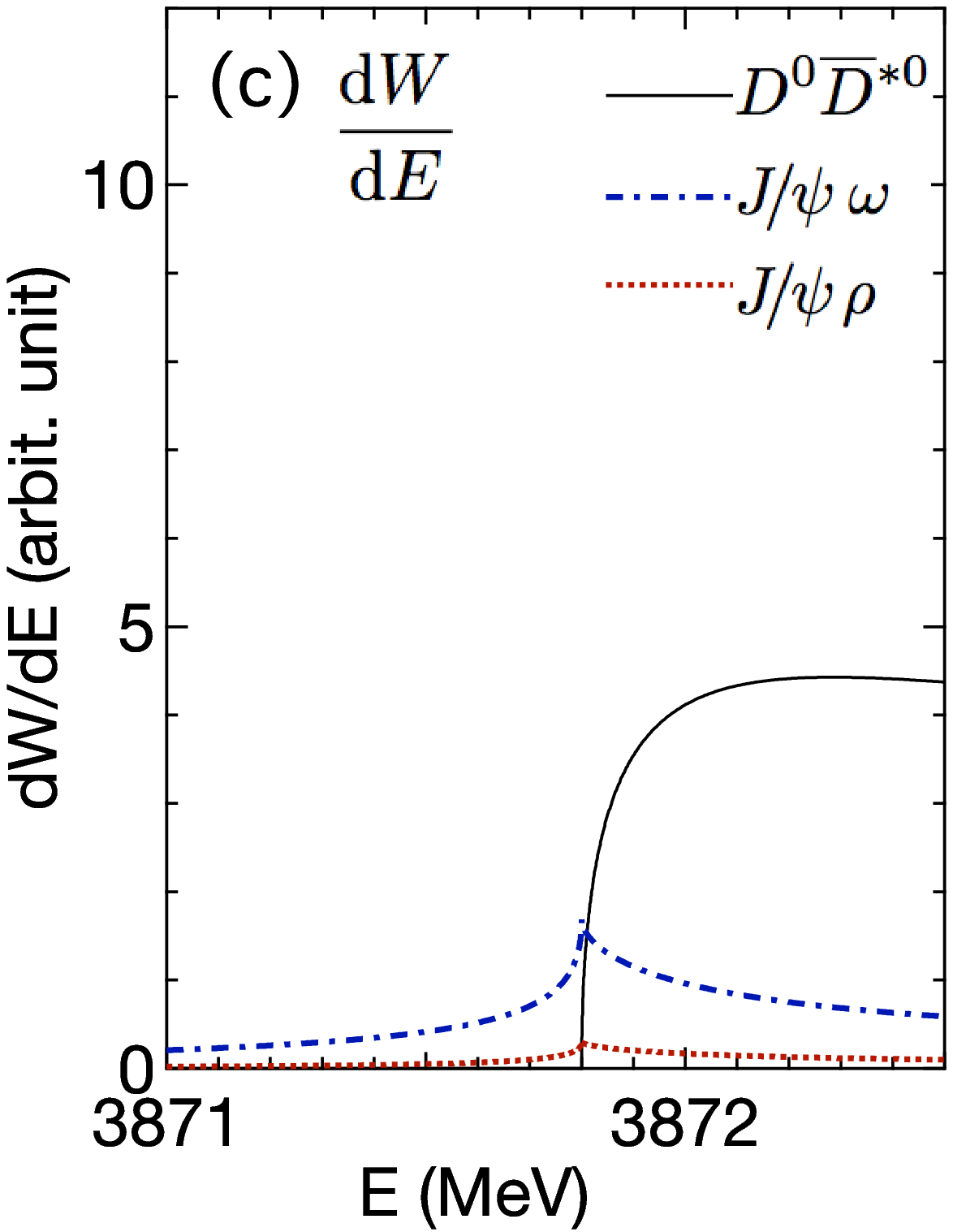}
\caption{The transfer strength from the \ccbar\ quarkonium to the
two-meson states.
 (a) for 3870 MeV $\le E\le$ 4000 MeV and (b) around the \DDbarz\ threshold,
 (c) those with the \ccbar-\DDbar\  coupling weakened by $0.9g^2$.
Parameter set QM. 
}
\label{fig:figtrqw}
\end{center}
\end{figure}
In the parameter set QM,
we use the quark model values
for all the two-meson interactions: the one between the 
$D$ and $\Dbar^{*}$ mesons or the $\Jpsi$ and the light vector mesons,
 as well as the transfer potential between 
\DDbar-$\Jpsi V$ channels.
As seen in Table \ref{tbl:param},
there is no attraction in the \DDbar\ channel,
though there is a considerable attraction appears between $\Jpsi$ and the light vector meson.
This attraction, however, is not large enough to 
make a bound state by itself. 
In this parameter set, most of the attraction to form a bound \X\ comes from the \ccbar-\DDbar\ coupling;
it requires $(g/g_0)^2\sim 1$
to have a bound \X.
As seen from Fig.\ \ref{fig:figtrqw},
the $D\overline{D}^*$ spectrum at around 3950 MeV is very flat,
reflecting the fact that the \ccbar-\DDbar\ coupling is very strong.
There is a large $\Jpsi\omega$ peak at the \DDbarz\ threshold,
while the $\Jpsi\rho$ peak is small.
In the case of the weaker coupling, 0.9$g^2$, (Fig.\ \ref{fig:figtrqw}(c)), there is almost no strength 
in the $\Jpsi\rho$ channel.

\begin{figure}[tb]
\begin{center}
\includegraphics[scale=0.38]{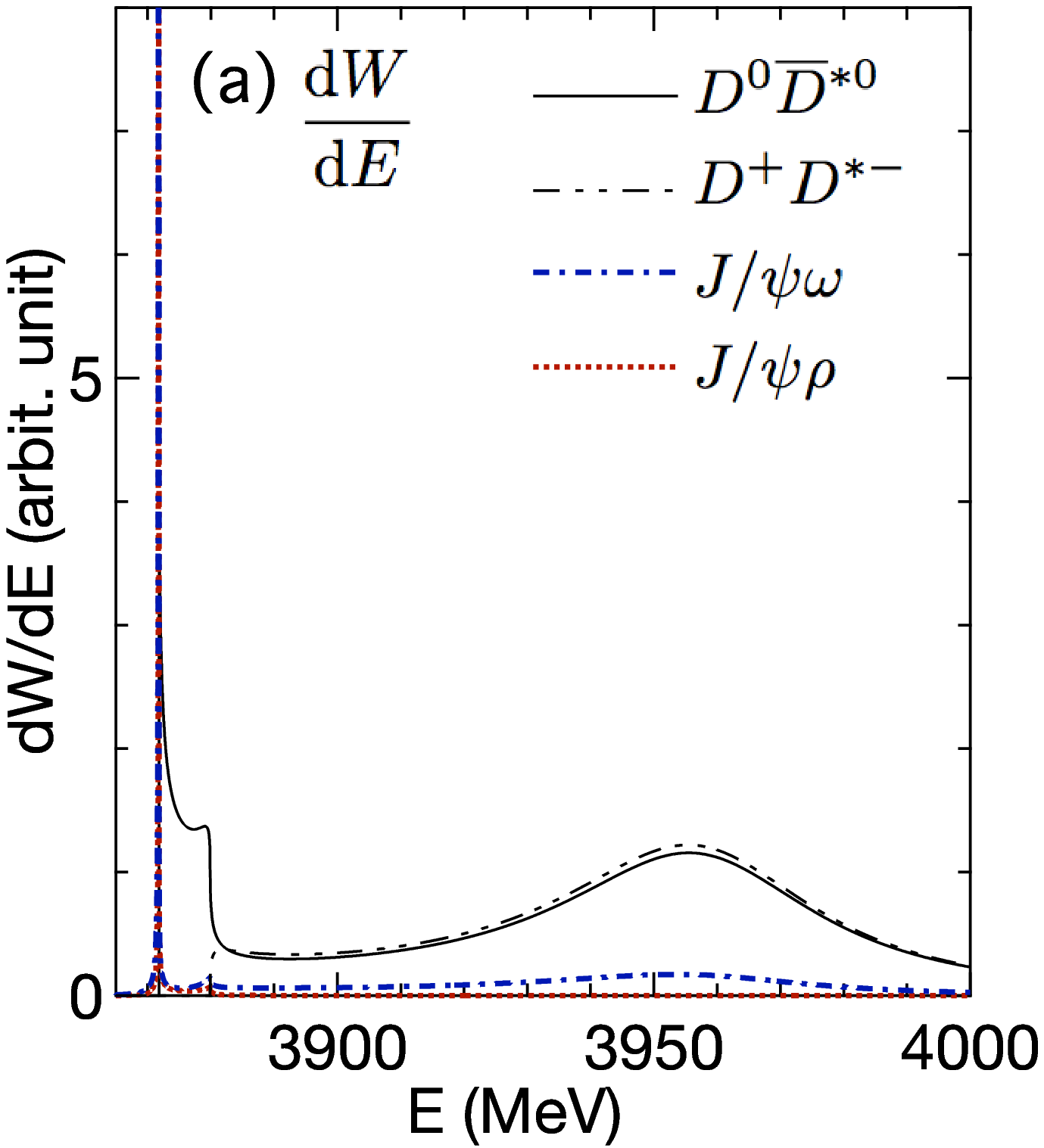}
\includegraphics[scale=0.38]{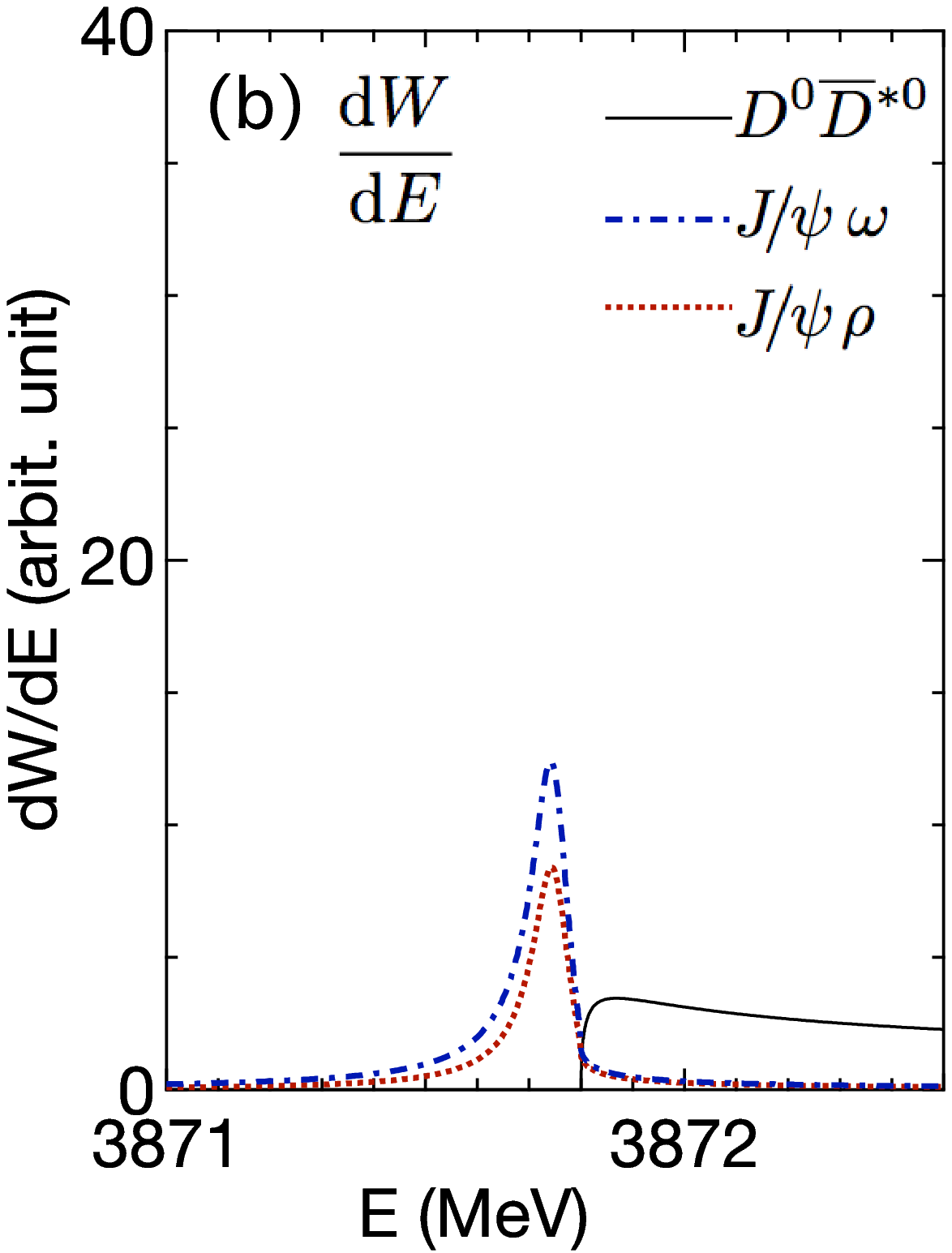}
\includegraphics[scale=0.38]{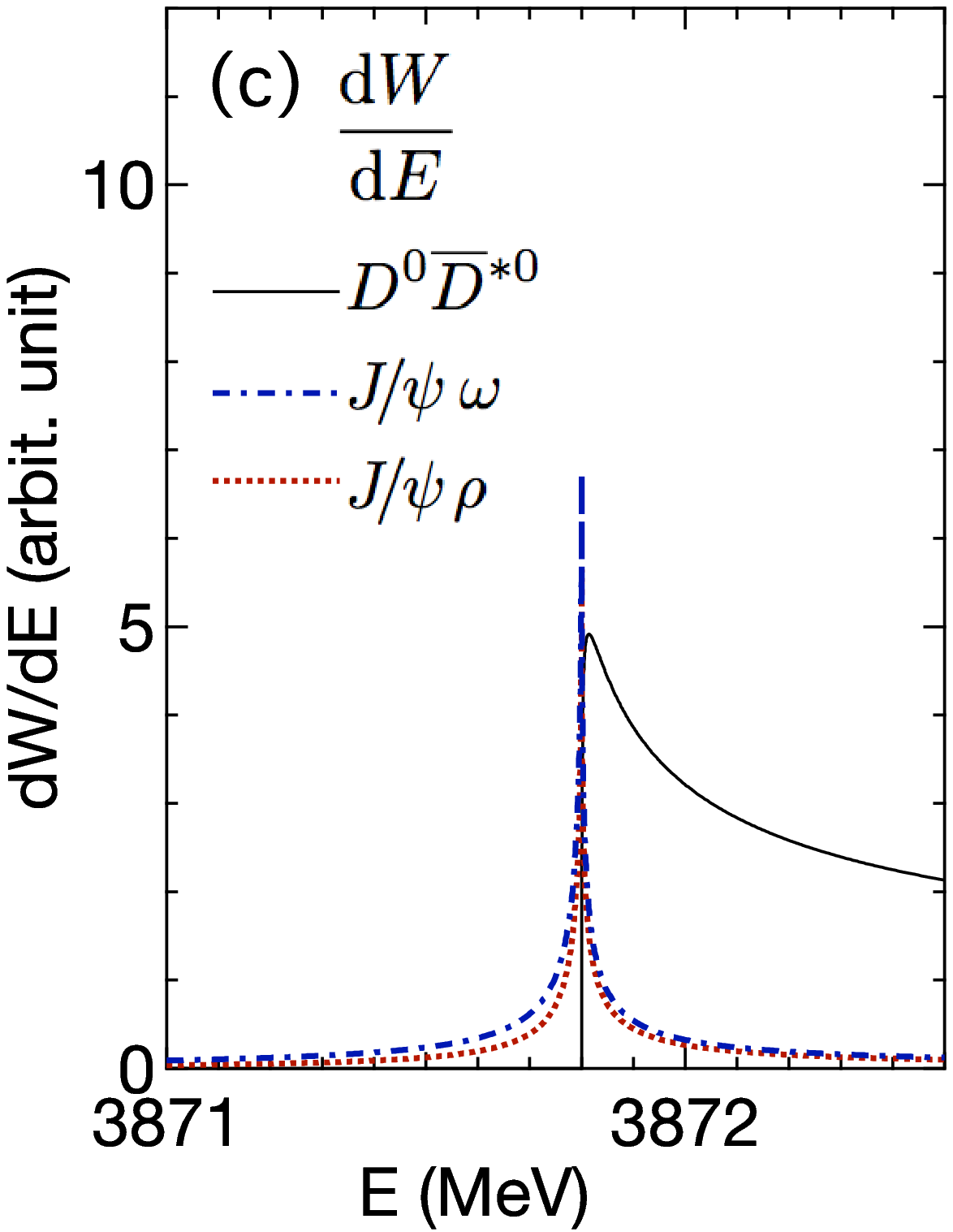}
\caption{The transfer strength from the \ccbar\ quarkonium to the
two-meson states.
 (a) for 3870 MeV $\le E\le$ 4000 MeV and (b) around the \DDbarz\ threshold,
 (c) those with the \ccbar-\DDbar\  coupling weakened by $0.9g^2$.
Parameter set B. 
}
\label{fig:figtrbw}
\end{center}
\end{figure}
\begin{figure}[tb]
\begin{center}
\includegraphics[scale=0.38]{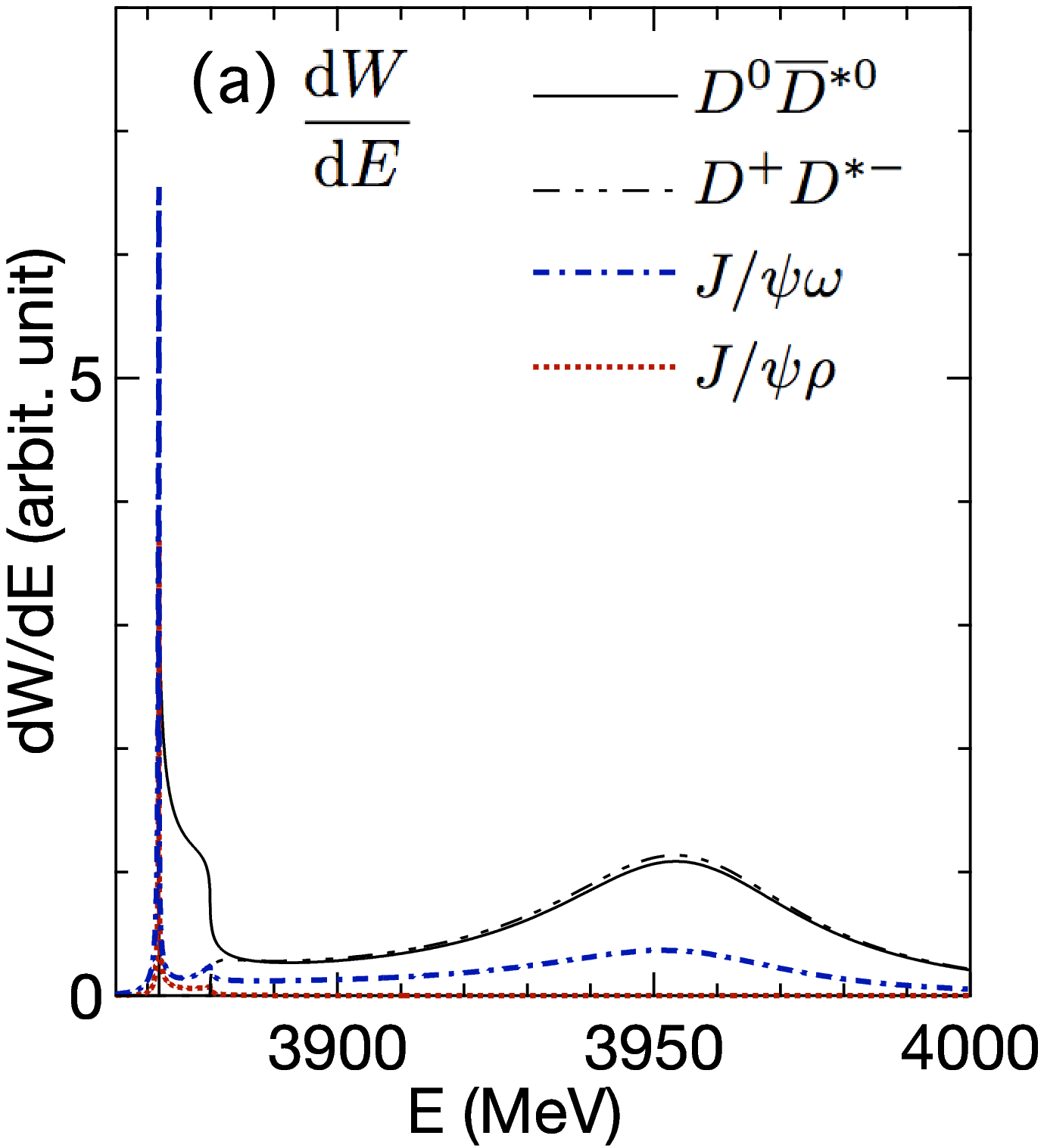}
\includegraphics[scale=0.38]{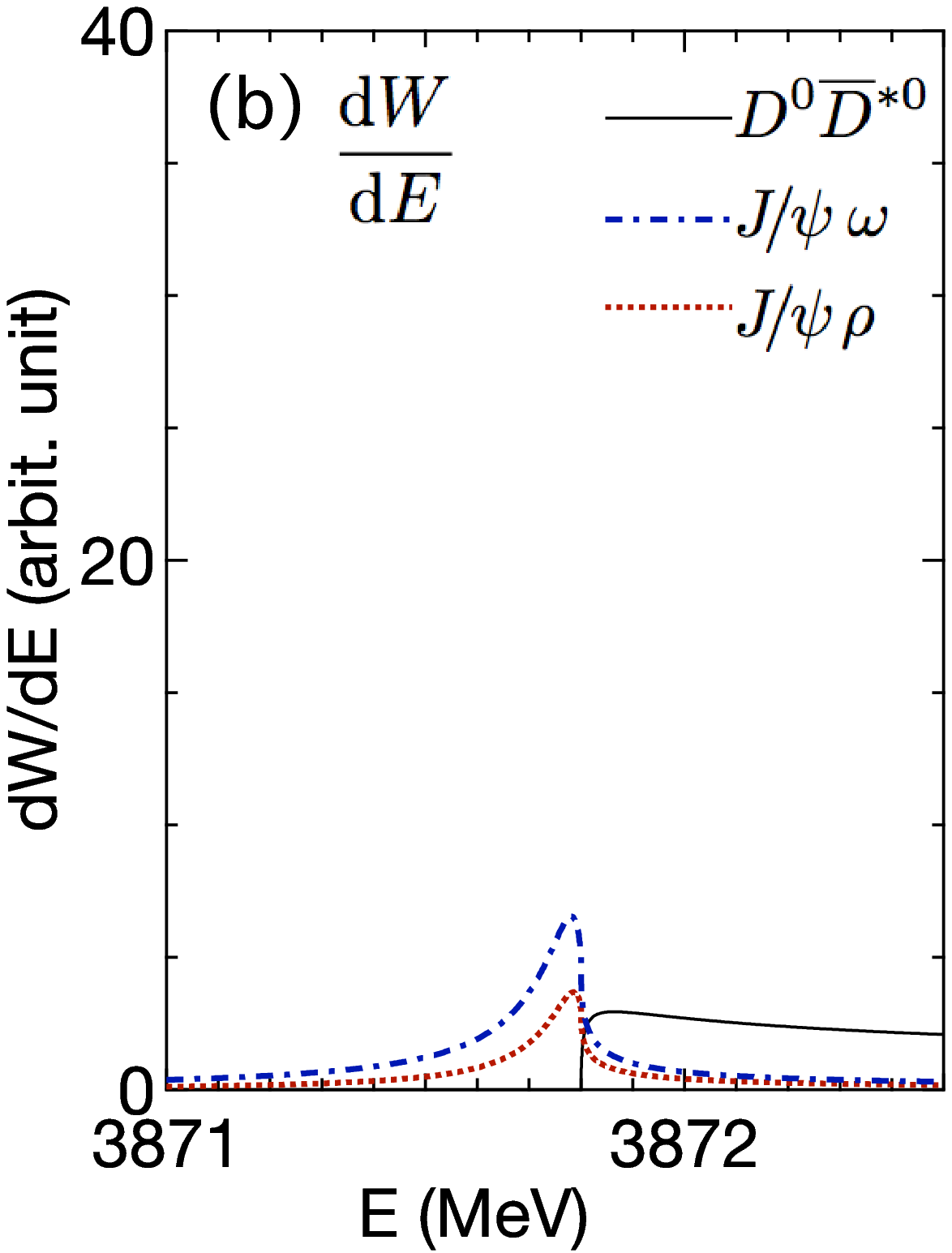}
\includegraphics[scale=0.38]{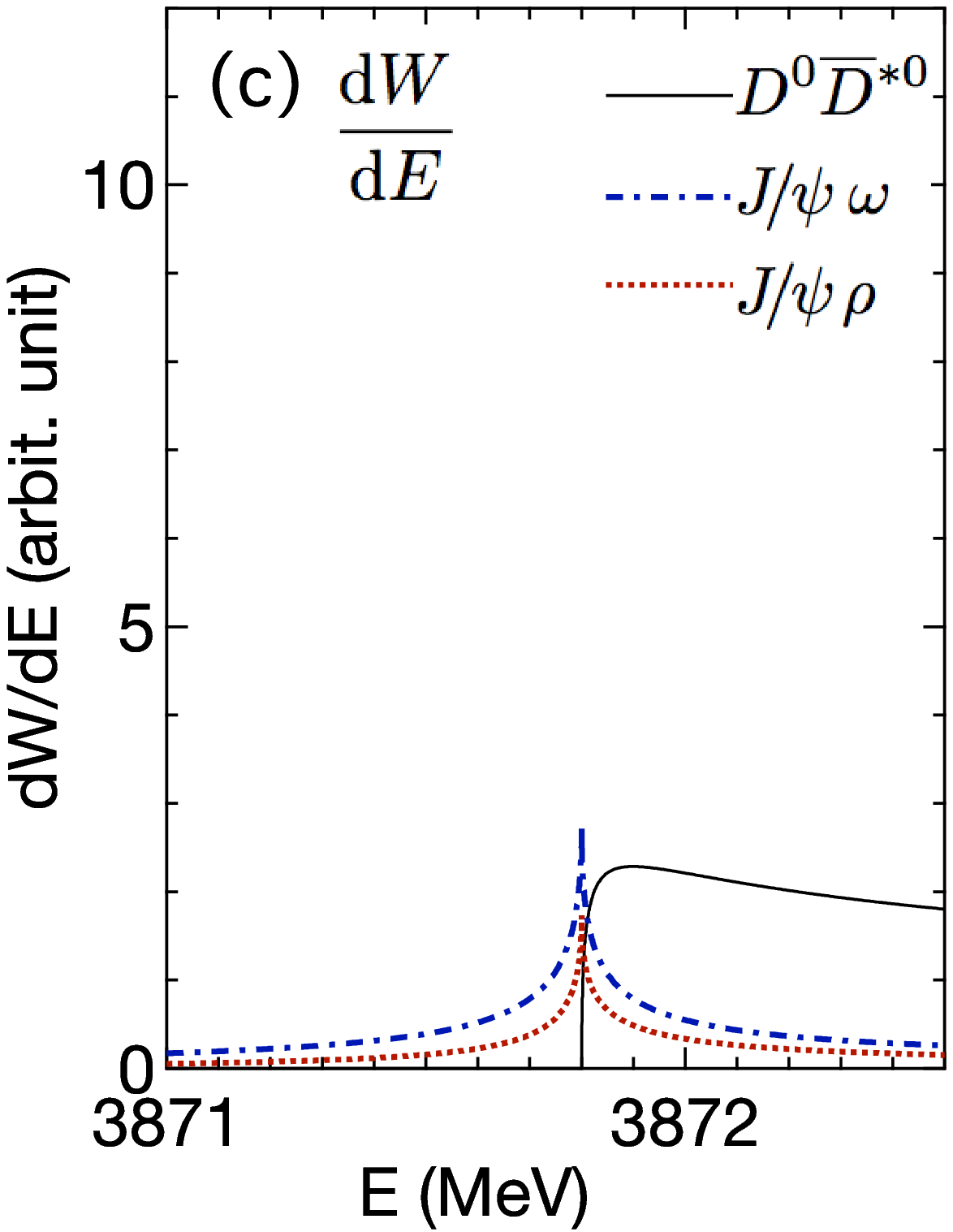}
\caption{The transfer strength from the \ccbar\ quarkonium to the
two-meson states.
 (a) for 3870 MeV $\le E\le$ 4000 MeV and (b) around the \DDbarz\ threshold,
 (c) those with the \ccbar-\DDbar\  coupling weakened by $0.9g^2$.
Parameter set C. 
}
\label{fig:figtrcw}
\end{center}
\end{figure}

To see the parameter dependence, we use a different size of $v$ or $u$ in the parameter set B or C,
respectively.
Their transfer strengths are shown in Figs.\ \ref{fig:figtrbw} and \ref{fig:figtrcw}.
The overall feature of the \DDbar\ channels is similar to that of the parameter set A.
As for the parameter set B, 
the bump at $E=3950$ MeV is enhanced slightly.
Also, a small shoulder appears at the \DDbarpm\ threshold.
This shoulder develops to an actual peak as the attraction $v$ becomes stronger. 
When $g\sim 0$, there will be three peaks:
if $v$ is strong enough to have a bound state in \DDbarz, 
then there will also be a bound state in \DDbarpm\
provided that the mixing between the \DDbarz\ and \DDbarpm\ is small.
Moreover, there should be a peak of \ccbar, which couples to 
\DDbar\ only weakly.

\subsection{Various ratios of the transfer strength}

\begin{table}[tdp]
\caption{Various ratios of the transfer strength 
for the original parameter set A, B, C, and QM, 
and those with the weakened \ccbar-\DDbar\ coupling, which are denoted by 0.9$g^2$.
A$_0$ is the parameter set A with the energy-independent meson width. 
As for the definition of the ratios, see text.
We put B/V in the last column depending on whether the calculated \X\ is a bound or virtual state.
The Belle experiment of $R_\Gamma$ is taken from ref.\ \cite{Abe:2005ix},
while that of \BABAR\ is taken from ref.\ \cite{delAmoSanchez:2010jr}.
As for the $r_\text{\DDbarz}$, the Belle value is taken from refs.\ \cite{Choi:2011fc,Adachi:2008sua} while that of \BABAR\ is taken from refs.\ \cite{Aubert:2007rva,Aubert:2008gu}.
}
\begin{center}
\begin{tabular}{lcccccccccc}\hline
Model~~~&$(g/g_0)^2$
&$R_\Gamma$ & $r_\text{\DDbarz}$(4MeV)&$r_\text{\DDbarz}$(8MeV) &$D_{I=1/0}$ &Bound
\\ \hline
A                & 0.655 & 2.24 & ~6.52 & ~9.91 & 0.0867 & B \\
A($0.9g^2$)      & 0.589 & 2.02 & 21.92 & 28.52 & 0.0850 & V \\
A$_0$            & 0.655 & 2.60 & ~5.63 & ~8.55 & 0.0861 & B \\
A$_0$($0.9g^2$)  & 0.589 & 2.34 & 18.37 & 23.98 & 0.0843 & V\\
B                & 0.472 & 1.39 & ~5.12 & ~7.96 & 0.0467 & B\\
B($0.9g^2$)      & 0.425 & 1.27 & 12.05 & 16.22 & 0.0423 & V\\
C                & 0.491 & 1.83 & ~5.40 & ~7.61 & 0.0528 & B\\
C($0.9g^2$)      & 0.442 & 1.74 & ~8.59 & 11.18 & 0.0463 & V\\
QM               & 1.003 & 6.34 & 12.55 & 17.61 & 0.1483 & B\\
QM($0.9g^2$)     & 0.903 & 5.79 & 42.14 & 53.07 & 0.1497 & V\\
\hline
Belle && $1.0 \pm 0.4 \pm 0.3$ 
&
\multicolumn{2}{c}{8.92 $\pm$ 2.42  }
\\
\BABAR\ && $0.8\pm 0.3$ &
\multicolumn{2}{c}{ 19.9 $\pm$ 8.05  }
\\
\hline
\end{tabular}
\label{tbl:compo}
\end{center}
\end{table}%

In the previous subsection, we show that 
all of the present parameter sets 
produce a thin $\Jpsi\pi^n$ peak at around the \DDbarz\ threshold.
The mechanism to form \X, however, can be different from each other.
To look into what kinds of observables
can be used to distinguish the models,
we listed the values of various ratios of the transfer strength in Table \ref{tbl:compo}.

\begin{figure}[tb]
\begin{center}
\includegraphics[scale=0.52]{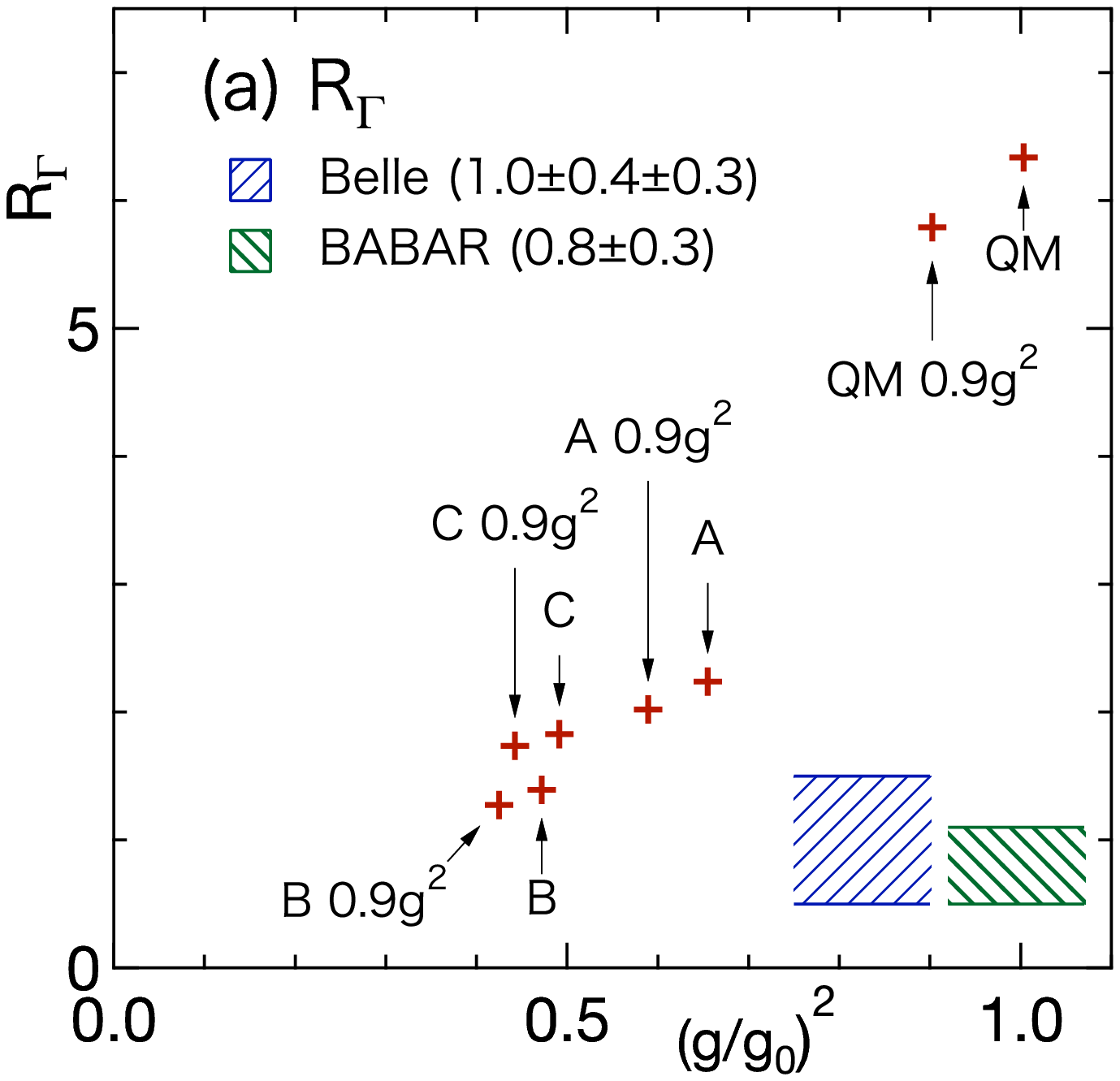}\hfill
\includegraphics[scale=0.52]{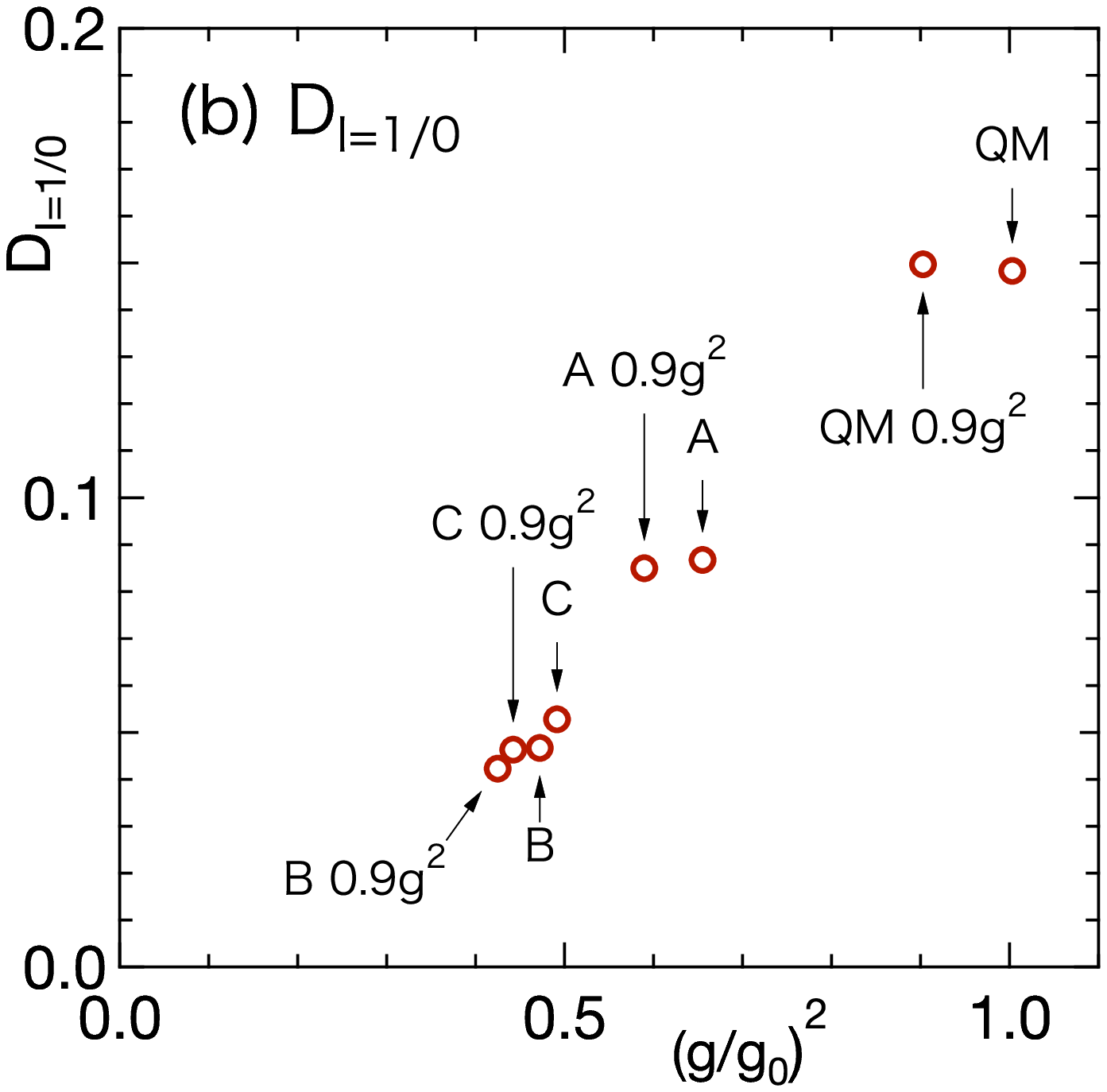}
\caption{The $\Jpsi\pi^3$-$\Jpsi\pi^2$ ratio at the \X\ peak, $R_\Gamma$,
and the \DDbarz-\DDbarpm ratio integrated over the scattering state, $D_{I=1/0}$.
In the Fig.\ (a), the $R_\Gamma$ is plotted
 against $(g/g_0)^2$, while $D_{I=1/0}$ is plotted in Fig.\ (b).
The experimental results for $R_\Gamma$ \cite{Abe:2005ix,delAmoSanchez:2010jr},
which do not depend on the $(g/g_0)^2$, are shown in Fig.\ (a) by the hatched areas.
}
\label{fig:gR}
\end{center}
\end{figure}

First let us discuss 
the ratio $R_\Gamma$ defined by Eq.\ (\ref{eq:eqR1G}).
This $R_\Gamma$ is defined by integrating the strength over 
$m_X\pm$ 1.2 MeV.
The values of $R_\Gamma$ do not change much if we integrate the strength 
over $m_X\pm$ 2.4  MeV; the largest deviation is about 3\%
of the listed value.
The ratio $R_\Gamma$ varies rather widely according to the parameters $(g/g_0)^2$.
As the $(g/g_0)^2$ becomes smaller, the ratio $R_\Gamma$
becomes smaller, and the degree of the isospin symmetry breaking becomes larger. 
On the other hand, the $R_\Gamma$ does not change much if the bound state exists.
%
The situation is illustrated in Fig.\ \ref{fig:gR} (a).
The parameter QM, where $(g/g_0)^2$ is about 1, the ratio $R_\Gamma$ is 6.34.
For the parameter set A, where $(g/g_0)^2=0.655$, the value is 2.24.
For the parameter sets B or C, the value becomes around 1.27--1.83.
Though the values we obtained here are still larger than the observed ones,
they agree with the experiment qualitatively. 
The experimental results
suggest that $(g/g_0)^2\sim 0.3$--0.5.
 The relative importance of the \ccbar-\DDbar\ coupling,  $(g/g_0)^2$,
together with the kinematical enhancement $\Delta_{f}$, surely play important roles 
in the mechanism to have the isospin symmetry breaking of this size.
Oppositely, one can 
estimate the sizes of the \ccbar-\DDbar\ coupling as well as the attraction 
between $D$ and $\overline{D}{}^*$ from the observed size of the isospin symmetry breaking.

Next we discuss
the ratio between the \DDbarz\ and \DDbarpm\ strengths:
\begin{align}
D_{I=1/0}
&=
{
I_\text{\DDbarz}(m_{D^0}+m_{\overline{D}{}^{*0}},\infty)-I_\text{\DDbarpm}(m_{D^+}+m_{{D}^{*-}},\infty)
\over 
I_\text{\DDbarz}(m_{D^0}+m_{\overline{D}{}^{*0}},\infty)+I_\text{\DDbarpm}(m_{D^+}+m_{{D}^{*-}},\infty)
}~,
\end{align}
which is shown in Fig.\ \ref{fig:gR} (b) and listed in Table \ref{tbl:compo}.
This $D_{I=1/0}$ essentially describes the 
ratio of the \DDbar\ strength below and above the \DDbarpm\ threshold,
which is
found to be again 
governed by the relative importance of the \ccbar-\DDbar\ coupling
against the $D$-$\overline{D}{}^*$ attraction,
$(g/g_0)^2$.
No experimental result has been reported  for this value, 
but with this and the size of the isospin symmetry breaking,
the information on the \X\ structure, or on the size of the 
the \ccbar-\DDbar\ coupling 
or the heavy meson interaction will become much clearer.

Lastly, we discuss the ratio $r_\text{\DDbarz}$, which is defined as
\begin{align}
r_\text{\DDbarz}&={
I_\text{\DDbarz}(m_\X- \epsilon,m_\X+ \epsilon)
\over 
I_{\Jpsi\rho}(m_\X- \epsilon,m_\X+ \epsilon)}~.
\label{eq:rDD}
\end{align}
We listed $r_\text{\DDbarz}$ for $\epsilon=4$ MeV and 8 MeV 
in Table \ref{tbl:compo}.
It is found that for the parameter sets which are $(g/g_0)^2\sim 0.5$,
this $r_\text{\DDbarz}$ is about 5.12--9.91 if the \X\ is a bound state,
while the value is more than 8.59 if there is no bound state, which is denoted by V. 
The results suggest that 
one can judge whether the \X\ is a bound state
by looking into the ratio of the partial decay width of \X\ in the \DDbarz\ channel
to that in the $\Jpsi\rho$ channel.
As we mentioned in the introduction, the experiments for this ratio 
is still controversial.
More precise measurements will help to determine whether 
the \X\ is a bound state or not.

\subsection{Model features}

In this subsection, we discuss the model features of the present work.

We have argued that the \X\ is a hybrid state of the \ccbar\ and the two-meson molecule:
a superposition of the \DDbarz,  \DDbarpm, $\Jpsi\rho$ and $\Jpsi\omega$ states
and the $c\bar c(2P)$ quarkonium.
The scattering states of the $J^{PC}=1^{++}$ channel
also consist of the above degrees of freedom.
Our approach is unique in the sense that (1) we simultaneously calculated the mass spectrum 
from the \DDbarz\ threshold up to 4 GeV in addition to the bound state,
(2) all the two-meson states are treated dynamically,
(3)  each of the final branching fractions of the \X\ peak is investigated 
separately,
(4) the energy dependent $\rho$ and $\omega$  meson widths are introduced,
(5) the interaction between the \DDbar\ and the $\Jpsi V$ channels is 
derived from the quark model. 
%

We assume that there is an attraction between the two mesons in the \DDbar\ channels.
The size of the attraction is consistent 
with the fact that no $B\overline{B}{}^*$ bound state has been found yet.
The parameter set C, where the \DDbar\ attraction is set to be stronger,
is an exception and give a bound state with the binding energy 2.4 MeV
if the interaction is applied to the $B\overline{B}{}^*$ system as it is.
The \DDbar\ system, however,
 does not have a bound state for all the parameter sets
if only the attraction in the \DDbar\ channel is taken into account
because the system has a smaller reduced mass.
The \ccbar-\DDbar\ coupling 
gives the required extra attraction to make the \X\ peak.

We take only the \ccbar$(2P)$, and not \ccbar$(1P)$ for example, 
as the source or the component of \X\
because this \ccbar$(2P)$ state has the closest   
mass to the \X\ among the $J^{PC}=1^{++}$ \ccbar\ series calculated by the quark model.
Including \ccbar$(1P)$ in addition to $\chi_{c1}(2P)$ does not change the mass spectrum or the 
\X\ state much
\cite{Takizawa:2012hy}.
Since there is no \ccbar$(2P)$ peak is observed experimentally,
one may only include the $\chi_{c1}(1P)$, whose existence is confirmed experimentally,
 in the two-meson system
as the source of the \X.
In such a case, however, 
the \ccbar-\DDbar\ coupling gives only a repulsion 
to the two-meson channels;
the required attraction to make the \X\ peak 
must come from the
interaction between the $D^{(*)}$ and $\overline{D}^{(*)}$ mesons.
Considering the heavy quark symmetry, 
this will most probably cause a bound state in the $B\overline{B}{}^*$ systems,
which has not been found experimentally.
Considering also that radiative decay of $X(3872)\rightarrow \psi(2S)\gamma$ 
is large \cite{Aubert:2008ae,Bhardwaj:2011dj,Aaij:2014ala}, 
we argue that there is a \ccbar$(2P)$ state though it is not seen directly
in the $\Jpsi \pi^n$ spectrum.
When one investigates the radiative decay,
the other \ccbar$(nP)$ states may become important
because each \ccbar\ core decays differently to the final
$\Jpsi\gamma$ or $\psi(2S)\gamma$ states \cite{Takizawa:2014nma}.
It is interesting but we discuss the problem elsewhere.

In the present calculation, the potential range is taken to be $\Lambda=500$ MeV,
which is a typical hadron size.
The attraction between the $D$ and $\overline{D}{}^*$ mesons,
however, is considered to come from the $\pi$- and $\sigma$-meson exchange, 
which has much longer range than that of the \DDbar-$\Jpsi V$ 
 or \ccbar-\DDbar\  coupling.
The $\Lambda$ dependence of the mass spectrum is investigated in \cite{Takizawa:2012hy};
when we take $\Lambda$ = 300 MeV, the enhancement of the \DDbar\ mass spectrum
at around 3950 MeV becomes larger.
The present results may change quantitatively if one introduces more realistic interaction.
We expect, however, that
the mechanism to have a thin peak or to have a large $I=1$ component 
will not change.

The production rates of \X\ are one of the important observables and have been discussed 
in \cite{Braaten:2004rw,Eichten:2004uh,Braaten:2004fk,Braaten:2005ai,Braaten:2005jj,Braaten:2004jg,
Bignamini:2009sk,Artoisenet:2009wk,Zanetti:2011ju}.
Experimentally, it is reported that the \X\ production rate in the $p\bar p$ collision
 is more than 0.046 times that of $\psi(2S)$ \cite{Bignamini:2009sk}.
Since the production rate of \ccbar\ of the opposite parity 
in the  $p\bar p$ collision
is probably not the same as that of the \ccbar\ which couples to the \X,
they cannot be compared each other directly.
The fact that there is a non-negligible component of the \ccbar\ in the \X\ wave function,
however, supports qualitatively that the \X\ production rate is larger
than expected from the meson-molecule picture.
Ortega {\it et.\ al.}\ solved the four quark system for the \ccbar\ and \DDbar\ systems,
and extract the \ccbar-\DDbar\ coupling.
The \DDbar\ system is solved as a hadron model with this coupling.
The parameter QM in the present work is similar to the model C
in ref.\ \cite{Ortega:2010qq}, where the \ccbar($2P$) is found to be 7\%.
Our result, 6.1\%, is consistent with their result.
%
In the present work, the \ccbar\ component in the \X\ is 0.023--0.061.
The \ccbar\ mixing of this size seems common to the hybrid picture.
Whether this mixing can explain the observed formation rate of \X\
quantitatively is still a open problem.

The peak shape of the \X\ in the $J^{PC}=1^{++}$ is discussed in refs.\  \cite{Kalashnikova:2005ui,Coito:2010if,Ortega:2010qq}.
The shape of the \DDbarz\ spectrum 
around the threshold in these works including ours is essentially the same;
The spectrum rises sharply at the threshold, and decreases slowly as the energy increases.
The pole position in the \DDbarz\ channel with the \ccbar\ state is investigated extensively in ref.\ \cite{Coito:2012vf},
which is also similar to the present work.
 It seems that the size of the $\Jpsi V$ component in the \X\  is not large.
Its effect on the transfer spectrum in the higher energy region is
 not large, either.
 Introducing the $\Jpsi V$ channels, however, 
 changes the phenomena at the \DDbarz\ threshold drastically.
 As for the $\Jpsi \rho$ or $\Jpsi \omega$ spectrum,
Coito {\it et.\ al.}\ has shown that a thin peak can be reproduced 
by employing the resonance spectrum expansion \cite{Coito:2010if}.
They assume that there is a direct coupling between the \ccbar\ and $\Jpsi V$ channels
in addition to the \ccbar-\DDbar\ and -$D_s\overline{D}{}_s^*$ couplings.
The model by Ortega {\it et.\ al.}, where 
the $\Jpsi V$ decay channel is added perturbatively via the quark rearrangement,
 also give a thin peak\cite{Ortega:2010qq}.
The present model, where the $\Jpsi V$ couples to \ccbar\ only via \DDbar\ channels,
again gives the thin peak.
As seen in the previous section, the mechanism to have a thin peak is a robust one.

%
%


The isospin symmetry breaking found in the \X\ decay
has been discussed in various ways.
%
For example,
the kinetic factor which enhances the isospin $I=1$ component is discussed in ref.\ 
\cite{Suzuki:2005ha},
the contribution from the $\rho^0$-$\omega$ mixing is pointed out in ref.\ \cite{Terasaki:2009in}, 
an estimate by a two-meson model with the realistic meson masses and the widths
is reported  in ref.\ 
\cite{Gamermann:2009fv}, the isospin breaking in the one-boson exchange interaction
is investigated in ref.\ \cite{Li:2012cs}.
Let us note that our results do not exclude that the existence of other sources of 
the isospin symmetry breaking, which contribute 
to reduce the ratio $R_\Gamma$.
It will be interesting to see how the combined effects
change the ratio.

We look into the 
parameter dependence of various ratios
of the decay fractions.
There we found that
the  ratio $R_\Gamma$ become smaller as the size of the \ccbar-\DDbar\ coupling becomes smaller.
The present experiments on this ratio
suggest that the about one-third of the attraction in \X\ comes from this coupling.
The relative strength of \DDbarpm\ to \DDbarz\ is also closely related to the size of the coupling.
With these two observables combined,
one may extract the condition over the size of interaction among the heavy quark systems.
We also found that the ratio $r_\text{\DDbarz}$ reflects the binding energy of the \X\ rather strongly.
It will help to understand the \X\ state if this value is determined experimentally.

In our model, the energy sum rule eq.\ (\ref{eq:energy_sum_rule}) 
holds approximately even after the inclusion of the 
energy dependent widths
because  all the two-meson channels as well as the meson decay width are
properly introduced.
If we introduces the width only $\Delta_{\Jpsi V}$ in eq.\ (\ref{eq:eq40delta}),
for example,
the energy sum deviates largely and 
 an artifactual enhancement of the strength occurs at around the threshold. 
Our treatment enables us to compare the strength of different energy regions
and to discuss the relation between the ratios and the size of the coupling
or the binding energy.


%

%

Recently, the $Z_c(3900)^\pm$ resonance
 has been found 
in the $J/\psi \pi^\pm$ mass spectrum \cite{Ablikim:2013mio,Liu:2013dau}.
It is a charged charmonium-like state,
a genuine exotic state whose
minimal quark component is $c \bar c q \bar q$.
Since there is no `\ccbar\ core' for this state, the present picture of the \X\
cannot be applied directly to the $Z_c(3900)^\pm$ resonance;
it is considered that the $Z_c(3900)$ is not a simple $I=1$ counter part of \X.
There is a report that the peak
may not be a resonance but a threshold effect \cite{Chen:2013coa}.
Further works will be necessary to understand this resonance.
As for the charged bottomonium-like resonances,  $Z_b(10610)^{\pm}$ and $Z_b(10650)^{\pm}$ \cite{Belle:2011aa},
the present model cannot be applied directly, either,
because these states again have a nonzero charge and do not couple to the bottomonium states. 
The $Z_b(10610)^0$ resonance 
\cite{Krokovny:2013mgx} is probably be the 
neutral partner of $Z_b(10610)^{\pm}$, the $I=1$ state. 
Since the masses of the charged and the neutral $Z_b(10610)$'s
are essentially the same, the isospin symmetry breaking of this system 
 must be small;
the mixing of the $b\overline{b}$ state is probably negligible.
%
%
There, the most important interaction in the $B\overline{B}{}^*$ system
will be the interaction between the $B$ and $\overline{B}{}^*$ mesons
unlike the \X\ case.
The interaction between the two heavy mesons,
empirically obtained here, may be tested in such systems.
Our results of $R_\Gamma$ suggests
that a larger attraction in the \DDbar\ channel 
than the parameter set A is favored.
There will be a bound state in the $B\overline{B}{}^*$ system 
if such a larger attraction
is applied to  the two heavy mesons as it is.
It will be very interesting and contributing to understand the heavy quark physics 
if one finds out whether such a bound state exists in the $B\overline{B}{}^*$ systems.

\section{Summary}

The $X$(3872) 
and the two-meson spectrum from the $B$-decay
are investigated by a $c\bar c$-two-meson hybrid model
for the energy from around the \DDbarz\ threshold up to 4 GeV.
The two-meson state consists of the $D^0\bar D{}^{*0}$, 
$D^+D^{*-}$, $J/\psi\rho$, and $J/\psi\omega$.
The final states are investigated separately for each channel.
The energy dependent decay widths of the $\rho$ and $\omega$ mesons are taken into account.
The strength of the coupling between the \DDbar\ and the $\Jpsi V$ channels is 
determined from the quark model.
The attraction between the $D$ and $\overline{D}{}^{*}$
is determined so that it produces a zero-energy resonance 
but no bound state if the attraction of the same size is
introduced in the $B\overline{B}{}^*$ system. 
The strength of the \ccbar-\DDbar\ coupling is taken to be a free parameter to
give the correct \X\ mass.

We have
found that the $X$(3872) can be a shallowly bound state or a $S$-wave virtual state.
For both of the cases, 
the following notable features are found: 
(1) both of the \ccbar\ $\to\Jpsi\rho$ and \ccbar\ $\to\Jpsi\omega$ mass spectra have 
a very narrow peak below or 
on the \DDbarz\ threshold, 
(2) the peak of \DDbarz\ spectrum has the width of a few MeV,
(3) there is no sharp peak at around 3950 MeV, which is the $\chi_{c1}(2P)$ mass predicted
by the quark model,
(4) the strength of the $\Jpsi\pi^2$ peak is comparable to that of the $\Jpsi\pi^3$ peak,
and 
(5) the ratios of some transfer strength give us information on the position of the \X\ pole,
the size of the \ccbar-\DDbar\ coupling, and the size of the $D$ and $\Dbar^*$ interaction.

The feature (1) implies that the observed peak found in the $\Jpsi \pi^n$ spectrum may not 
directly correspond to the pole energy of the \X.
It may be a peak at the threshold caused by a virtual state.
If the \X\ is a bound state, then the peak which corresponds to the pole energy appears below the \DDbarz\ threshold. The current experiments cannot distinguish these two cases.
The features (2) and (3) correspond to the experimental facts that 
the \X\ width from \DDbarz\ mode is about 3 or 4 MeV and that no $\chi_{c1}(2P)$
peak has been found, respectively. 
As for the feature (4), 
the present work shows that the isospin symmetry breaking caused by the 
neutral and charged $D$ and $D^{*}$ meson mass difference
seems to be large enough to explain the experiments
owing to the enhancement by the large $\rho$ meson width.
When the peak strength is integrated over the interval $E=m_{X(3872)}\pm 1.2$ MeV, 
the decay ratio, $R_\Gamma$, becomes 1.27--2.24.
Though this is still larger than the observed values,
$1.0 \pm 0.4 \pm 0.3$ or $0.8\pm 0.3$,
the obtained values agree with the experiment qualitatively. 

The size of the isospin symmetry breaking in the transfer strength 
becomes larger as the $c\bar c$-$D\bar D{}^*$ coupling becomes weaker.
The relative strength of the $D^0\bar D{}^{*0}$ below the $D^+D^{*-}$ threshold also varies largely according to the size of this coupling.
We would like to point out,
as we mentioned above as the feature (5), 
that from these two observables combined, the information on the size of the 
the \ccbar-\DDbar\ coupling 
or the heavy meson interaction can be obtained more clearly.
It is also found that the branching ratio of the $D^0\bar D{}^{*0}$ to the $J/\psi\rho$, which is still controversial experimentally,
is a good indicator of evaluating whether the $X$(3872) peak is a bound state or a virtual state.
Investigating the \X\ properties really gives us rich information 
on the heavy quark physics.


\appendix

\section{Appendix: Width of the $\rho$ and $\omega$ mesons}

\subsection{Kinematics}
The B$^+$ meson at rest has the mass
$m_B= 5279.26 \pm 0.17$ MeV \cite{Agashe:2014kda}.
It can decays into K$^+$ and a \ccbar\ pair by the weak interaction.
When this K meson has the momentum $\vecp_K$, then the \X, which is generated from 
the \ccbar\ pair,
has the energy $E_X$ as
\begin{equation}
E_X=m_B-
\sqrt{m_K^2+p_K^2}
\label{eq:EX}
\end{equation}
with the momentum
 $\vecp_X=-\vecp_K$.

Suppose the \X\ is a bound state and 
does not decay, 
it has the center of mass momentum $\vecp_X$
and the energy $E_X=\sqrt{m_{\X}^2+p_X^2}$. 
Thus the size of $\vecp_K$ is uniquely determined once 
$m_\X$ is given: {\it e.g.}\
when $m_\X=3871.68$ MeV, $p_K$ 
= 5.78 fm$^{-1}$.

On the other hand, suppose the \X\ is a resonance and the final states are the 
scattering two mesons,
 the phase space of the kaon momentum $\vecp_K$ becomes a continuum.
The energy of the two mesons in the $f$-th channel, whose center of mass momentum is $\vecp_X = -\vecp_K$,
can be written as
 \begin{eqnarray}
E_X &=&
\sqrt{(M_f+m_f)^2 + p_X^2}+{k_f^2\over 2\mu_f}~,
\label{eq:34}
\end{eqnarray}
where  $m_f$ and $M_f$ are each of the masses of the final two mesons,
the $\mu_f$ their reduced mass, 
and $\veck_f$
 the relative momentum of the two mesons.
Here we extract the relative motion in a nonrelativistic way.
Since we investigate the reaction only slightly above the threshold, 
$k_f$ is considered to be small comparing to the meson masses.

The energy of the two-meson system at rest, $E_f$, can be defined as
 \begin{eqnarray}
E_f &=& 
M_f+m_f+ {k_f^2\over 2\mu_f}
\label{eq:35}
\\
&=&M_f+m_f+E_X-\sqrt{(M_f+m_f)^2 + p_X^2}~.
\end{eqnarray}
The figures in this paper are plotted against this energy $E_f$ for the 
\DDbarz\ channel, $E_{\text{\DDbarz}}$.

When the final two mesons are $\Jpsi$ and $\rho$, for example,
 the above $E_f$ becomes
\begin{eqnarray}
E_{\Jpsi\rho} &=& 
m_\Jpsi+m_\rho+ {k_{\Jpsi\rho}^2\over 2\,\mu_{\Jpsi\rho}}~,
\label{eq:42}
\end{eqnarray}
where $k_{\Jpsi\rho}$ is the relative momentum of $\Jpsi$ and $\rho$
when the $\Jpsi\rho$ system is at rest.

For a given $|\vecp_X|(=|\vecp_K|)$, $E_X$ is determined by
Eq.\ (\ref{eq:EX}).
Then the momentum $k_{\Jpsi\rho}$ is obtained by
Eq.\ (\ref{eq:34}), and $E_{\Jpsi\rho}$ by Eq.\  (\ref{eq:35}).

When the $\rho$ meson decays into the two-pion state, 
that $E_{\Jpsi\rho}$ can be expressed also by
\begin{eqnarray}
E_{\Jpsi\rho} &=&
\sqrt{m_\Jpsi^2+k^2}
+\sqrt{(2m_\pi)^2+k^2}
-2m_\pi+E_{2\pi}
\\
E_{2\pi}&=&2\sqrt{m_\pi^2+q^2}
~~~~\text{or}~~~
q^2~=~{1\over 4}E_{2\pi}^2 -m_\pi^2~.
\label{eq:qrho}
\end{eqnarray}
Here,
$k$ is the relative momentum between $\Jpsi$ and the center of mass motion of the 
two pions.
The relative momentum between the two pions is denoted as $q$,
and $E_{2\pi}$ is the energy of the two pions whose center of mass motion is zero.
The energy $E_{2\pi}$
becomes a function of $k$ and
$k_{\Jpsi\rho}$, 
 $E_{2\pi}(k,k_{\Jpsi\rho})$.
Note that $k$ can be different from $k_{\Jpsi\rho}$;
$k_{\Jpsi\rho}$ and $k$ correspond to $k_f$ and $k$ in Eq.\ (\ref{eq:eq32}) respectively.

When the final two mesons are $\Jpsi$ and $\omega$, which decays 
into the three-pion state,
the center of mass energy of the $\Jpsi$ and $\omega$ system,
$E_{\Jpsi\omega}$ can be rewritten similarly by
\begin{eqnarray}
E_{\Jpsi\omega}&=&
\sqrt{m_\Jpsi^2+k^2}+\sqrt{(3m_\pi)^2+k^2}
-3m_\pi+E_{3\pi}~,
\end{eqnarray}
where $E_{3\pi}$ is the energy of the three pions whose center of mass momentum equals to zero.
Again, the energy $E_{3\pi}$
becomes a function of $k$ and
$k_{\Jpsi\omega}$, $E_{3\pi}(k,k_{\Jpsi\omega})$.
For the later convenience, we define the `average' momentum, $\overline{q}$, as
\begin{eqnarray}
\overline{q}^2&=&{1\over 9}E_{3\pi}^2 -m_\pi^2~.
\label{eq:qomega}
\end{eqnarray}

\subsection{The $\rho$ and $\omega$ meson width}

In this appendix,
we show how we obtain the energy dependence of the $\rho$ and $\omega$ meson width.
Since our main interest is on the X(3872), we only consider the major decay mode 
for both of the $\rho$ and $\omega$ mesons \cite{Agashe:2014kda}.
%
%
By assuming that the non-resonant term is small, 
the cross section, $\sigma$,  of the mesons can be written as
\begin{eqnarray}
\sigma(E_{n\pi}) &\propto&  {12\pi\over q^2}  
{{1\over 4}\Gamma_V(E_{n\pi})^2 \over (E_{n\pi}-\tilde m_V)^2+{1\over 4}\Gamma_V(E_{n\pi})^2}~,
\end{eqnarray}
Here $\tilde m_V$ and $\Gamma_V(E_{n\pi})$ are the mass and the width of the $\rho$ and $\omega$ mesons, 
respectively,
and $q$ stands for the relative momentum of the two pions which decay from the $\rho$ meson, Eq.\  (\ref{eq:qrho}),
or for the average momentum of three pions from the $\omega$ meson, Eq.\ (\ref{eq:qomega}).


The major decay mode of the $\rho$ meson is $\rho\rightarrow \pi\pi$ ($P$-wave).
The width has a large energy dependence.
We rewrite  the width  as:
\begin{eqnarray}
\Gamma_\rho (E_{2\pi}) &=& \Gamma^{(0)}_\rho {F_\rho(E_{2\pi})\over F_\rho(\tilde m_\rho)}~.
\label{eq:Frho}
\end{eqnarray}
Here $\Gamma^{(0)}_\rho$ is a constant 
and corresponds to the $\rho$ meson width at $E=\tilde m_\rho$,
for which we use the observed value.
We assume the following function form
for 
$F_\rho(E_{2\pi})$.
\begin{eqnarray}
F_\rho(E_{2\pi}) &=& q^3\left({\Lambda_V^2\over\Lambda_V^2+q^2}\right)^2~,
\label{eq:Frho1}
\end{eqnarray}
where $q^2={1\over 4}E_{2\pi}^2 -m_\pi^2$ is  the relative momentum of the pions
and 
$\Lambda_V$ is a momentum cutoff.
This corresponds to the one with the  monopole form factor
 for relative $P$-wave pions.

\begin{figure}[t]
\includegraphics[scale=0.45]{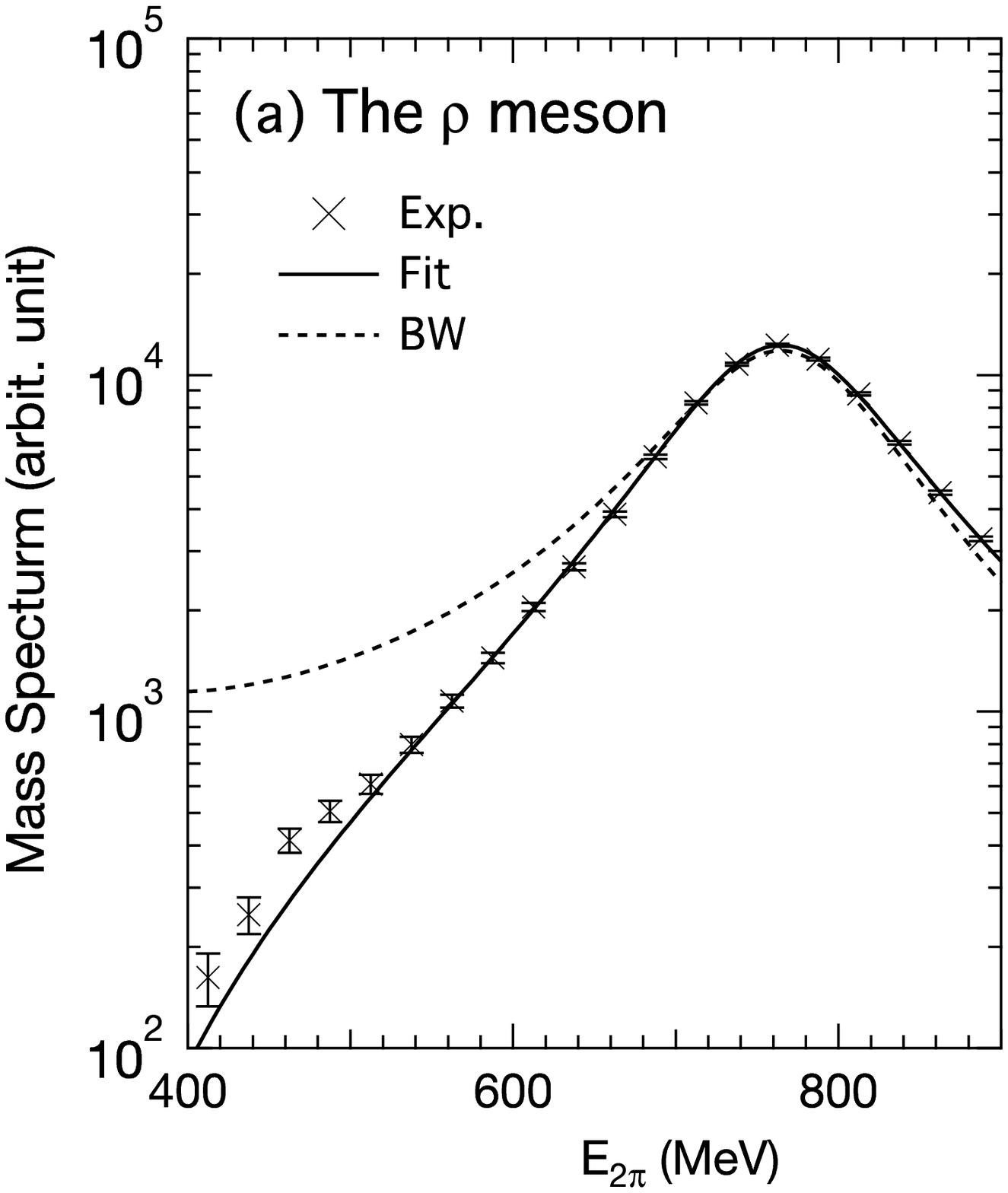}
\includegraphics[scale=0.45]{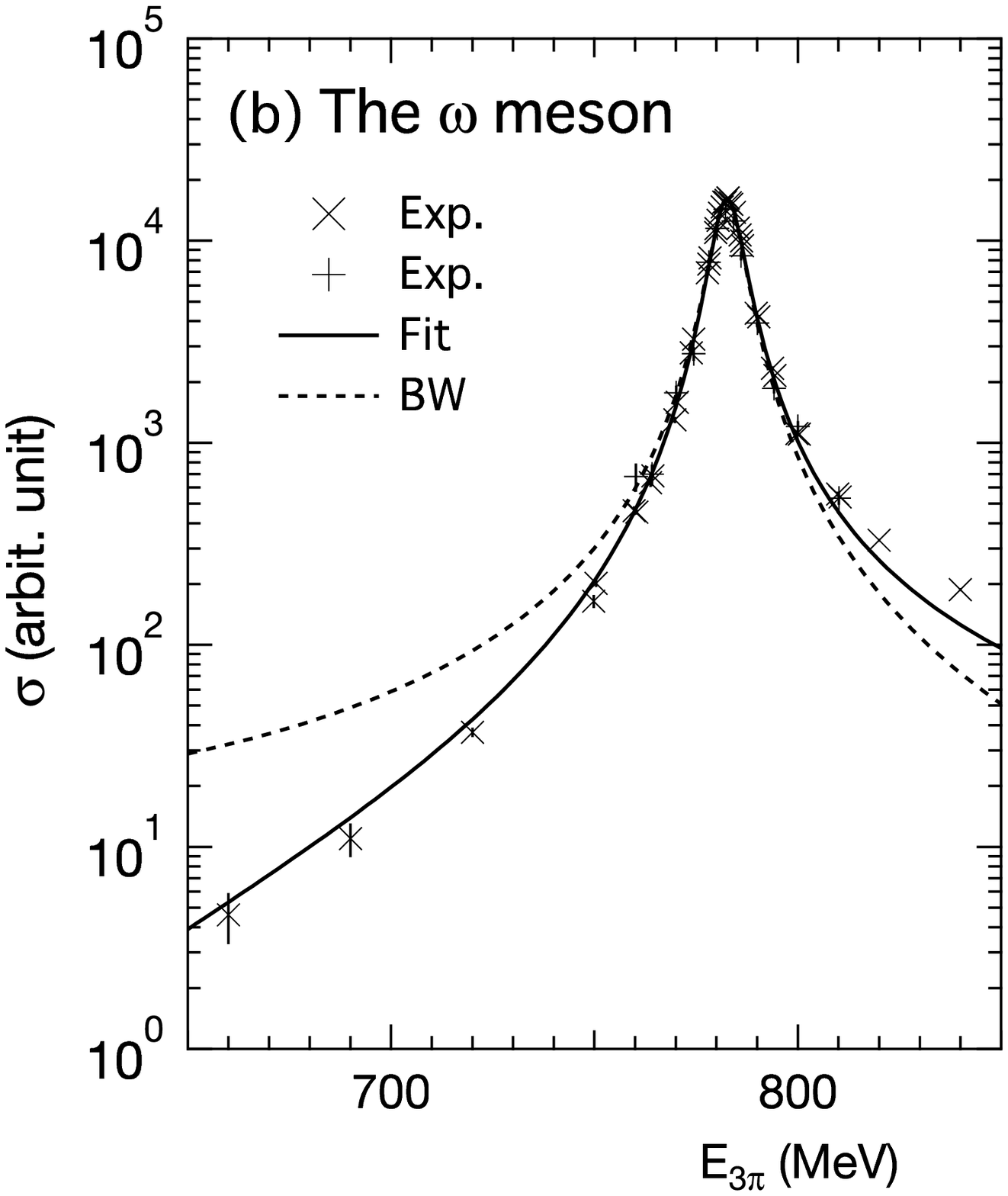}
\caption{%
The $\rho$ and $\omega$ meson decay:
(a) the mass spectrum $\tau^-\rightarrow \pi^-\pi^0\nu_\tau$ decay, where the data are taken from ref.\ \cite{Anderson:1999ui} and (b) the 
$e^+e^-\rightarrow \pi^+\pi^-\pi^0$ cross section  where data are taken from ref.\ \cite{Achasov:2003ir,Akhmetshin:2003zn}.
The solid lines are fitted results by employing the energy-dependent width,  $\Gamma_V(E)$, while dotted lines (BW) are obtained with a energy-independent width,
$\Gamma^{(0)}_V$.
}
\label{fig:vectormesonwidth}
\end{figure}

In Fig.\ \ref{fig:vectormesonwidth}(a), the mass spectrum of the $\rho$ meson,
${\sigma  q}$, 
is plotted  against $E_{2\pi}$.
The experimental data  taken from ref.\ \cite{Anderson:1999ui}
are shown with the error bars.
The solid line is the one we calculated with the energy dependent width,
where we use the values of $\tilde m_\rho$ and $\Lambda_V$ 
as well as the absolute size of the spectrum
as fitting parameters.
They are shown in Table \ref{tbl:paramwidth} with the observed width $\Gamma^{(0)}_\rho$.
The dotted line 
corresponds to the one without energy dependence, $\Gamma_\rho= \Gamma^{(0)}_\rho$.

When we apply the width to the \X, 
the factor $\Delta_f(E)$
appears as 
seen in Eq.\ (\ref{eq:eq40delta}).
For the energy around the \DDbarz\ threshold,
this factor for the $\Jpsi$-$\rho$ channel
is sizable only at around $0<k\lesssim 3$ fm$^{-1}$, 
and takes a maximum value
at $k\sim 1.26$ fm$^{-1}$.
This corresponds to 
$E_{2\pi}=340\sim 775$ MeV with the maximum at around 
670 MeV.
Thus we fit rather lower energy region of the $\rho$ meson peak, 400-900 MeV,
to obtain the energy dependent $\rho$-meson width.

\begin{table}[tdp]
\caption{Parameters for the $\rho$ and $\omega$ meson width.
The values for $\Gamma^{(0)}_\rho$, 
$m_\omega$, and $\Gamma^{(0)}_\omega$ are the observed ones.
All entries are in MeV.}
\begin{center}
\begin{tabular}{cccccccc}\hline
$\tilde m_\rho$ & $\Gamma^{(0)}_\rho$ & 
$m_\omega$ & $\Gamma^{(0)}_\omega$ & 
$\Lambda_V$ &\\ \hline
 768.87  & 149.1  
 & 782.65  &8.49 & 291.05& &\\ \hline
\end{tabular}
\label{tbl:paramwidth}
\end{center}
\end{table}

The $\omega$ meson decays occurs 
 mainly via a $\rho\pi$ state (Gell-Mann Sharp Wagner (GSW) mode\cite{GellMann:1962jt})
at around the peak energy.
Also for the $\omega$ meson we rewrite the width as
\begin{eqnarray}
\Gamma_\omega (E) &=& 
\Gamma^{(0)}_\omega {F_\omega(E)\over F_\omega(m_\omega)}~,
\end{eqnarray}
where $\Gamma^{(0)}_\omega$ is a constant 
and corresponds to the  $\omega$ meson width at $E= m_\omega$,
which we use again the observed  total decay width of $\omega$.
We use a simple form for the energy dependence  also for the $\omega$ meson,
\begin{eqnarray}
F_\omega(E) &=& \overline{q}^6\left({\Lambda_V^2\over\Lambda_V^2+\overline{q}^2}\right)^4~.
\end{eqnarray}
where $\overline{q}^2={1\over 9}E^2 -m_\pi^2$. 
Here we use the same value for the momentum cut-off $\Lambda_V$ as that we obtained for the $\rho$  meson.
This shape of the energy dependence can be derived by assuming the $\rho\pi$ $P$-wave decay 
has also the monopole form factor, and the energy dependence 
of the imaginary part of the $\rho$ meson propagator is governed by that of the $\rho$ meson form factor.
Here we do not discuss whether this assumption is  appropriate.
We employ the above function form 
because it is simple and the fitting is good enough to perform our \X\ calculation.

In Fig.\ \ref{fig:vectormesonwidth}(b), 
the cross sections of $\omega$ meson are shown.
The data are taken from \cite{Achasov:2003ir,Akhmetshin:2003zn}.
The solid line stands for the one with the energy dependent width, and the
dotted one is the one without the energy dependence.
For the $\omega$ meson, the factor $\Delta_f(E)$ in Eq.\ (\ref{eq:eq40delta}) 
has a sizable value
at around $0<k\lesssim 2$ fm$^{-1}$, 
and takes a maximum value
at $k\sim 0.5$ fm$^{-1}$.
This corresponds to 
$E_{3\pi}=600\sim 775$ MeV with the maximum at 
762 MeV.
We fit the data in the energy region 660-786 MeV for the $\omega$ meson peak.

For both of the $\rho$ and $\omega$ decay, 
we can fit the data 
with an enough accuracy for the current purpose.
The values of parameters are summarized in Table \ref{tbl:paramwidth}.
We use only the fitting parameter $\Lambda_V$ (and function forms of the energy dependence,
$F_\rho$ and $F_\omega$) for the \X\ calculation.

\section{Appendix: Meson Interaction obtained from a quark model}

\subsection{Base of the two-meson wave functions}

The color-spin-flavor part of the wave function for the $J^{PC}=1^{++}$ $q\qbar c\cbar$ state
has two components, which may be written 
by the color singlet and octet $\Jpsi$ with the light vector meson: 
\begin{eqnarray}
| V_1\Jpsi{}_1\ket &=&\Big[ |q\qbar ~S=1,~\text{color 1}\ket 
\otimes  |c\cbar ~S=1, ~\text{color 1}\ket \Big]_\text{color 1}
\\
| V_8\Jpsi{}_8\ket &=&\Big[ |q\qbar ~S=1, ~\text{color 8}\ket  
\otimes  |c\cbar ~S=1,~\text{color 8}\ket \Big]_\text{color 1}~,
\end{eqnarray}
where $V=\omega$ or $\rho$, $q$ stands for one of the light quarks, $u$ and $d$,
$S$ is the spin of the two quarks or the two antiquarks, 
and color 1 [color 8] stands for the color singlet [octet] state.
These components can be expressed by rearranged ones, such as
\begin{align}
| \Dbar{}_1 D^*{}_1\ket ={1\over \sqrt{2}}&\Big(\Big[ |q\cbar ~S=0,~\text{color 1}\ket 
\otimes  |c\qbar ~S=1, ~\text{color 1}\ket \Big]_\text{color 1}\nonumber\\&
-\Big[ |q\cbar ~S=1,~\text{color 1}\ket 
\otimes  |c\qbar ~S=0, ~\text{color 1}\ket \Big]_\text{color 1}\Big)
\\
| \Dbar{}_8 D^*{}_8\ket ={1\over \sqrt{2}}&\Big(\Big[ |c\cbar ~S=0, ~\text{color 8}\ket  
\otimes  |q\qbar ~S=1,~\text{color 8}\ket \Big]_\text{color 1}\nonumber\\&
-\Big[ |c\cbar ~S=1, ~\text{color 8}\ket  
\otimes  |q\qbar ~S=0,~\text{color 8}\ket \Big]_\text{color 1}\Big)~.
\end{align}
These two color-spin-flavor base functions can be transferred from each other as:
\begin{align}
\begin{pmatrix}
|\Dbar{}_1 D^*{}_1\ket\\
|\Dbar{}_8 D^*{}_8\ket\\
\end{pmatrix}
&=
\begin{pmatrix}
\sqrt{1\over9}&
\sqrt{8\over9}\\
\sqrt{8\over9}&
-\sqrt{1\over9}\\
\end{pmatrix}
\begin{pmatrix}
| V_1\Jpsi{}_1 \ket\\
| V_8\Jpsi{}_8 \ket\\
\end{pmatrix}~.
\end{align}
When one considers the hadronic system, the color-spin-flavor base will be
$|\Dbar{}_1 D^*{}_1\ket$
and 
$| V_1\Jpsi{}_1\ket$, which are not orthogonal to each other from the quark model viewpoint,
especially at the short distance.
The normalization in the color-spin-flavor space becomes
\begin{align}
N&=
\begin{pmatrix}1 &\frac{1}{3}\\ \frac{1}{3}&1\\ \end{pmatrix}.
\end{align}

\section*{Acknowledgement}

This work is partly supported by Grants-in-Aid
for scientific research  (20540281 and 21105006).


\begin{thebibliography}{99}

\bibitem{Choi:2003ue} 
  S.~K.~Choi {\it et al.}  [Belle Collaboration],
  Phys.\ Rev.\ Lett.\  {\bf 91}, 262001 (2003).
  
\bibitem{Acosta:2003zx} 
  D.~Acosta {\it et al.}  [CDF Collaboration],
  Phys.\ Rev.\ Lett.\  {\bf 93}, 072001 (2004).

\bibitem{Abazov:2004kp} 
  V.~M.~Abazov {\it et al.}  [D0 Collaboration],
  Phys.\ Rev.\ Lett.\  {\bf 93}, 162002 (2004).

\bibitem{Aubert:2004ns} 
  B.~Aubert {\it et al.}  [BaBar Collaboration],
  Phys.\ Rev.\ D {\bf 71}, 071103 (2005).

\bibitem{Aaij:2011sn} 
  R.~Aaij {\it et al.}  [LHCb Collaboration],
  Eur.\ Phys.\ J.\ C {\bf 72}, 1972 (2012).


\bibitem{Agashe:2014kda} 
  K.~A.~Olive {\it et al.}  [Particle Data Group Collaboration],
  Chin.\ Phys.\ C {\bf 38}, 090001 (2014).
  
\bibitem{Choi:2011fc} 
  S.~-K.~Choi, S.~L.~Olsen, K.~Trabelsi, I.~Adachi, H.~Aihara, K.~Arinstein, D.~M.~Asner and T.~Aushev {\it et al.},
  Phys.\ Rev.\ D {\bf 84}, 052004 (2011).
  
\bibitem{Abulencia:2006ma}
  A.~Abulencia {\it et al.}  [CDF Collaboration],
  Phys.\ Rev.\ Lett.\  {\bf 98}, 132002 (2007).

\bibitem{Aaij:2013zoa} 
  R.~Aaij {\it et al.}  [LHCb Collaboration],
  Phys.\ Rev.\ Lett.\  {\bf 110}, no. 22, 222001 (2013).


\bibitem{Adachi:2008sua} 
  T.~Aushev {\it et al.}  [Belle Collaboration],
  Phys.\ Rev.\ D {\bf 81}, 031103 (2010).

\bibitem{Aubert:2007rva} 
  B.~Aubert {\it et al.}  [BaBar Collaboration],
  Phys.\ Rev.\ D {\bf 77}, 011102 (2008).
  
\bibitem{Aubert:2008gu} 
  B.~Aubert {\it et al.}  [BaBar Collaboration],
  Phys.\ Rev.\ D {\bf 77}, 111101 (2008).


\bibitem{Abe:2005ix}
  K.~Abe {\it et al.}  [Belle Collaboration],
  arXiv:hep-ex/0505037.

\bibitem{delAmoSanchez:2010jr} 
  P.~del Amo Sanchez {\it et al.}  [BaBar Collaboration],
  Phys.\ Rev.\ D {\bf 82}, 011101 (2010).


\bibitem{Chiu:2006hd} 
  T.~-W.~Chiu {\it et al.}  [TWQCD Collaboration],
  Phys.\ Lett.\ B {\bf 646}, 95 (2007).

\bibitem{Prelovsek:2013cra} 
  S.~Prelovsek and L.~Leskovec,
  Phys.\ Rev.\ Lett.\  {\bf 111}, 192001 (2013).


\bibitem{Brambilla:2010cs}
  N.~Brambilla, S.~Eidelman, B.~K.~Heltsley, R.~Vogt, G.~T.~Bodwin, E.~Eichten, A.~D.~Frawley, A.~B.~Meyer {\it et al.},
  Eur.\ Phys.\ J.\  {\bf C71}, 1534 (2011).

\bibitem{Swanson:2006st} 
  E.~S.~Swanson,
  Phys.\ Rept.\  {\bf 429}, 243 (2006).
  
\bibitem{Godfrey:2008nc} 
  S.~Godfrey and S.~L.~Olsen,
  Ann.\ Rev.\ Nucl.\ Part.\ Sci.\  {\bf 58}, 51 (2008).


\bibitem{Godfrey:1985xj} 
  S.~Godfrey and N.~Isgur,
  Phys.\ Rev.\ D {\bf 32}, 189 (1985).


\bibitem{Li:2009zu} 
  B.~Q.~Li and K.~T.~Chao,
  Phys.\ Rev.\ D {\bf 79}, 094004 (2009)
  [arXiv:0903.5506 [hep-ph]].

  
\bibitem{Barnes:2003vb} 
  T.~Barnes and S.~Godfrey,
  Phys.\ Rev.\ D {\bf 69}, 054008 (2004).
  
\bibitem{Barnes:2005pb} 
  T.~Barnes, S.~Godfrey and E.~S.~Swanson,
  Phys.\ Rev.\ D {\bf 72}, 054026 (2005).
  
\bibitem{Butenschoen:2013pxa} 
  M.~Butenschoen, Z.~-G.~He and B.~A.~Kniehl,
  Phys.\ Rev.\ D {\bf 88}, 011501 (2013).

\bibitem{Aubert:2008ae} 
  B.~Aubert {\it et al.}  [BaBar Collaboration],
  Phys.\ Rev.\ Lett.\  {\bf 102}, 132001 (2009).

\bibitem{Bhardwaj:2011dj} 
  V.~Bhardwaj {\it et al.}  [Belle Collaboration],
  Phys.\ Rev.\ Lett.\  {\bf 107}, 091803 (2011).

\bibitem{Aaij:2014ala} 
  R.~Aaij {\it et al.}  [LHCb Collaboration],
  arXiv:1404.0275 [hep-ex].

\bibitem{Swanson:2004pp} 
  E.~S.~Swanson,
  Phys.\ Lett.\ B {\bf 598}, 197 (2004).


\bibitem{Maiani:2004vq} 
  L.~Maiani, F.~Piccinini, A.~D.~Polosa and V.~Riquer,
  Phys.\ Rev.\ D {\bf 71}, 014028 (2005).

\bibitem{Matheus:2006xi} 
  R.~D'E.~Matheus, S.~Narison, M.~Nielsen and J.~M.~Richard,
  Phys.\ Rev.\ D {\bf 75}, 014005 (2007).

\bibitem{Maiani:2007vr} 
  L.~Maiani, A.~D.~Polosa and V.~Riquer,
  Phys.\ Rev.\ Lett.\  {\bf 99}, 182003 (2007).

\bibitem{Vijande:2007fc} 
  J.~Vijande, E.~Weissman, N.~Barnea and A.~Valcarce,
  Phys.\ Rev.\ D {\bf 76}, 094022 (2007).

\bibitem{Dubnicka:2010kz} 
  S.~Dubnicka, A.~Z.~Dubnickova, M.~A.~Ivanov and J.~G.~Korner,
  Phys.\ Rev.\ D {\bf 81}, 114007 (2010).


\bibitem{Close:2003sg} 
  F.~E.~Close and P.~R.~Page,
  Phys.\ Lett.\ B {\bf 578}, 119 (2004).


\bibitem{Voloshin:2003nt} 
  M.~B.~Voloshin,
  Phys.\ Lett.\ B {\bf 579}, 316 (2004).
  
\bibitem{Wong:2003xk} 
  C.~-Y.~Wong,
  Phys.\ Rev.\ C {\bf 69}, 055202 (2004).


\bibitem{Swanson:2003tb} 
  E.~S.~Swanson,
  Phys.\ Lett.\ B {\bf 588}, 189 (2004).

\bibitem{Tornqvist:2004qy} 
  N.~A.~Tornqvist,
  Phys.\ Lett.\ B {\bf 590}, 209 (2004).

\bibitem{Voloshin:2004mh} 
  M.~B.~Voloshin,
  Phys.\ Lett.\ B {\bf 604}, 69 (2004).

\bibitem{AlFiky:2005jd} 
  M.~T.~AlFiky, F.~Gabbiani and A.~A.~Petrov,
  Phys.\ Lett.\ B {\bf 640}, 238 (2006).

\bibitem{Fleming:2007rp} 
  S.~Fleming, M.~Kusunoki, T.~Mehen and U.~van Kolck,
  Phys.\ Rev.\ D {\bf 76}, 034006 (2007).

\bibitem{Braaten:2007dw} 
  E.~Braaten and M.~Lu,
  Phys.\ Rev.\ D {\bf 76}, 094028 (2007).

\bibitem{Braaten:2007ft} 
  E.~Braaten and M.~Lu,
  Phys.\ Rev.\ D {\bf 77}, 014029 (2008).

\bibitem{Liu:2008fh} 
  Y.~-R.~Liu, X.~Liu, W.~-Z.~Deng and S.~-L.~Zhu,
  Eur.\ Phys.\ J.\ C {\bf 56}, 63 (2008).

\bibitem{Canham:2009zq} 
  D.~L.~Canham, H.~-W.~Hammer and R.~P.~Springer,
  Phys.\ Rev.\ D {\bf 80}, 014009 (2009).

\bibitem{Stapleton:2009ey} 
  E.~Braaten and J.~Stapleton,
  Phys.\ Rev.\ D {\bf 81}, 014019 (2010).

\bibitem{Lee:2009hy} 
  I.~W.~Lee, A.~Faessler, T.~Gutsche and V.~E.~Lyubovitskij,
  Phys.\ Rev.\ D {\bf 80}, 094005 (2009).

\bibitem{Gamermann:2009uq} 
  D.~Gamermann, J.~Nieves, E.~Oset and E.~Ruiz Arriola,
  Phys.\ Rev.\ D {\bf 81}, 014029 (2010).

\bibitem{Wang:2013kva} 
  P.~Wang and X.~G.~Wang,
  Phys.\ Rev.\ Lett.\  {\bf 111}, no. 4, 042002 (2013).


\bibitem{Kalashnikova:2005ui} 
  Yu.~S.~Kalashnikova,
  Phys.\ Rev.\ D {\bf 72}, 034010 (2005).

\bibitem{Suzuki:2005ha} 
  M.~Suzuki,
  Phys.\ Rev.\ D {\bf 72}, 114013 (2005).

\bibitem{Barnes:2007xu} 
  T.~Barnes and E.~S.~Swanson,
  Phys.\ Rev.\ C {\bf 77}, 055206 (2008).

\bibitem{Zhang:2009bv} 
  O.~Zhang, C.~Meng and H.~Q.~Zheng,
  Phys.\ Lett.\ B {\bf 680}, 453 (2009).

\bibitem{Matheus:2009vq} 
  R.~D'E.~Matheus, F.~S.~Navarra, M.~Nielsen and C.~M.~Zanetti,
  Phys.\ Rev.\ D {\bf 80}, 056002 (2009).

\bibitem{Kalashnikova:2009gt} 
  Yu.~S.~Kalashnikova and A.~V.~Nefediev,
  Phys.\ Rev.\ D {\bf 80}, 074004 (2009).

\bibitem{Ortega:2010qq} 
  P.~G.~Ortega, J.~Segovia, D.~R.~Entem and F.~Fernandez,
  Phys.\ Rev.\ D {\bf 81}, 054023 (2010).

\bibitem{Danilkin:2010cc} 
  I.~V.~Danilkin and Y.~.A.~Simonov,
  Phys.\ Rev.\ Lett.\  {\bf 105}, 102002 (2010).

\bibitem{Coito:2010if} 
  S.~Coito, G.~Rupp and E.~van Beveren,
  Eur.\ Phys.\ J.\ C {\bf 71}, 1762 (2011).

\bibitem{Coito:2012vf} 
  S.~Coito, G.~Rupp and E.~van Beveren,
  Eur.\ Phys.\ J.\ C {\bf 73}, 2351 (2013).

\bibitem{Ferretti:2013faa}
  J.~Ferretti, G.~Galat\`{a} and E.~Santopinto,
  Phys.\ Rev.\ C {\bf 88}, no. 1, 015207 (2013).

  
\bibitem{Chen:2013pya} 
  W.~Chen, H.~-y.~Jin, R.~T.~Kleiv, T.~G.~Steele, M.~Wang and Q.~Xu,
  Phys.\ Rev.\ D {\bf 88}, no. 4, 045027 (2013).

\bibitem{Takizawa:2012hy} 
  M.~Takizawa and S.~Takeuchi,
  PTEP {\bf 2013}, no. 9, 0903D01 (2013).

\bibitem{Takeuchi:2014mma} 
  S.~Takeuchi, M.~Takizawa and K.~Shimizu,
  Few Body Syst.\  {\bf 55}, 773 (2014).


\bibitem{Takeuchi:2008wc}
  S.~Takeuchi and K.~Shimizu,
  Phys.\ Rev.\  C {\bf 79}, 045204 (2009).

\bibitem{Newton}
R.\ G.\ Newton, {\it
Scattering Theory of Waves and Particles, 2nd ed.} 
(Springer-Verlag, New York, 1982),
Chap.\ 17.

\bibitem{Beringer:1900zz} 
  J.~Beringer {\it et al.}  [Particle Data Group Collaboration],
  Phys.\ Rev.\ D {\bf 86}, 010001 (2012).


\bibitem{Braaten:2005ai} 
  E.~Braaten and M.~Kusunoki,
  Phys.\ Rev.\ D {\bf 72}, 054022 (2005).

\bibitem{Li:2012cs} 
  N.~Li and S.~-L.~Zhu,
  Phys.\ Rev.\ D {\bf 86}, 074022 (2012).


\bibitem{Takizawa:2014nma} 
  M.~Takizawa, S.~Takeuchi and K.~Shimizu,
  Few Body Syst.\  {\bf 55}, 779 (2014).

\bibitem{Braaten:2004rw} 
  E.~Braaten and M.~Kusunoki,
  Phys.\ Rev.\ D {\bf 69}, 114012 (2004).

\bibitem{Eichten:2004uh} 
  E.~J.~Eichten, K.~Lane and C.~Quigg,
  Phys.\ Rev.\ D {\bf 69}, 094019 (2004).
  

\bibitem{Braaten:2004fk} 
  E.~Braaten, M.~Kusunoki and S.~Nussinov,
  Phys.\ Rev.\ Lett.\  {\bf 93}, 162001 (2004).


\bibitem{Braaten:2005jj} 
  E.~Braaten and M.~Kusunoki,
  Phys.\ Rev.\ D {\bf 72}, 014012 (2005).


\bibitem{Braaten:2004jg} 
  E.~Braaten,
  Phys.\ Rev.\ D {\bf 73}, 011501 (2006).

\bibitem{Bignamini:2009sk} 
  C.~Bignamini, B.~Grinstein, F.~Piccinini, A.~D.~Polosa and C.~Sabelli,
  Phys.\ Rev.\ Lett.\  {\bf 103}, 162001 (2009).

\bibitem{Artoisenet:2009wk} 
  P.~Artoisenet and E.~Braaten,
  Phys.\ Rev.\ D {\bf 81}, 114018 (2010).

\bibitem{Zanetti:2011ju} 
  C.~M.~Zanetti, M.~Nielsen and R.~D.~Matheus,
  Phys.\ Lett.\ B {\bf 702}, 359 (2011).


\bibitem{Terasaki:2009in} 
  K.~Terasaki,
  Prog.\ Theor.\ Phys.\  {\bf 122}, 1285 (2010).

\bibitem{Gamermann:2009fv} 
  D.~Gamermann and E.~Oset,
  Phys.\ Rev.\ D {\bf 80}, 014003 (2009).


\bibitem{Ablikim:2013mio} 
  M.~Ablikim {\it et al.}  [BESIII Collaboration],
  Phys.\ Rev.\ Lett.\  {\bf 110}, 252001 (2013).


\bibitem{Liu:2013dau} 
  Z.~Q.~Liu {\it et al.}  [Belle Collaboration],
  Phys.\ Rev.\ Lett.\  {\bf 110}, 252002 (2013).

%
\bibitem{Chen:2013coa} 
  D.~-Y.~Chen, X.~Liu and T.~Matsuki,
  Phys.\ Rev.\ D {\bf 88}, no. 3, 036008 (2013).

\bibitem{Belle:2011aa} 
  A.~Bondar {\it et al.}  [Belle Collaboration],
  Phys.\ Rev.\ Lett.\  {\bf 108}, 122001 (2012).

\bibitem{Krokovny:2013mgx} 
  P.~Krokovny {\it et al.}  [Belle Collaboration],
  Phys.\ Rev.\ D {\bf 88}, no. 5, 052016 (2013).




\bibitem{Anderson:1999ui}
  S.~Anderson {\it et al.} [ CLEO Collaboration ],
  Phys.\ Rev.\  {\bf D61}, 112002 (2000).


\bibitem{Achasov:2003ir}
  M.~N.~Achasov {\it et al.},
  Phys.\ Rev.\  D {\bf 68}, 052006 (2003).

\bibitem{Akhmetshin:2003zn}
  R.~R.~Akhmetshin {\it et al.}  [CMD-2 Collaboration],
  Phys.\ Lett.\  B {\bf 578}, 285 (2004).


\bibitem{GellMann:1962jt} 
  M.~Gell-Mann, D.~Sharp and W.~G.~Wagner,
  Phys.\ Rev.\ Lett.\  {\bf 8}, 261 (1962).



%
%
%
%
%
%
%
%

\end{thebibliography}
\end{document}